\documentclass[prx,twocolumn,twoside,superscriptaddress,notitlepage,floatfix]{revtex4-2}

\usepackage{setspace}
\usepackage[utf8]{inputenc}
\usepackage[T1]{fontenc}
\usepackage{amsmath}
\usepackage{amssymb}
\usepackage{stmaryrd}
\usepackage{mathrsfs}
\usepackage{braket}
\usepackage{xcolor}
\usepackage{graphicx}
\usepackage{bm}
\usepackage{csquotes}
\usepackage{siunitx}

\def\dd{\text{d}}
\def\ii{\text{i}}
\def\ee{\text{e}}

\def\RR{\mathbb{R}}
\def\ZZ{\mathbb{Z}}

\def\im{\text{i}}
\def\re{\text{r}}

\def\Im#1{\text{Im}#1}

\let\strong\textbf

\usepackage{hyperref}
\usepackage{bookmark}

\hypersetup{breaklinks=true,
  colorlinks=true,
  linkcolor=black,
  filecolor=black,
  urlcolor=black,
  citecolor=black,
  pdfborder=0 0 0,
}

\makeatletter
\let\@orig@make@capt@title\@make@capt@title
\def\@make@capt@title#1#2{\@orig@make@capt@title{{\bf #1}}{#2}}
\makeatother

\renewcommand{\eqref}[1]{\textup{{\normalfont(\ref{#1}}\normalfont)}}

\makeatletter\begin{document}

\title{Exceptional points in nonlinear and stochastic dynamics}

\author{Cheyne Weis}
\affiliation{James Franck Institute and Department of Physics, The University of Chicago, Chicago, IL 60637, USA}
\author{Michel Fruchart}
\affiliation{James Franck Institute and Department of Physics, The University of Chicago, Chicago, IL 60637, USA}
\affiliation{Gulliver, UMR CNRS 7083, ESPCI Paris PSL, 75005 Paris, France}
\author{Ryo Hanai}
\affiliation{Center for Gravitational Physics and Quantum Information, Yukawa Institute for Theoretical Physics, Kyoto University, Kyoto 606-8502, Japan}
\affiliation{Asia Pacific Center for Theoretical Physics, Pohang 37673, Korea}
\author{Kyle Kawagoe}
\affiliation{James Franck Institute and Department of Physics, The University of Chicago, Chicago, IL 60637, USA}
\affiliation{Kadanoff Center for Theoretical Physics, The University of Chicago, Chicago, IL 60637, USA}
\affiliation{Department of Physics and Department of Mathematics, The Ohio State University, Columbus, OH 43210, USA}
\author{Peter B. Littlewood}
\affiliation{James Franck Institute and Department of Physics, The University of Chicago, Chicago, IL 60637, USA}
\author{Vincenzo Vitelli}
\affiliation{James Franck Institute and Department of Physics, The University of Chicago, Chicago, IL 60637, USA}
\affiliation{Kadanoff Center for Theoretical Physics, The University of Chicago, Chicago, IL 60637, USA}

\begin{abstract}
We study a class of bifurcations generically 
occurring in dynamical systems with non-mutual couplings ranging from models of coupled neurons to predator-prey systems and non-linear oscillators.
In these bifurcations, extended attractors such as limit cycles, limit tori, and strange attractors merge and split in a similar way as fixed points in a pitchfork bifurcation.
We show that this merging and splitting coincides with the coalescence of covariant Lyapunov vectors with vanishing Lyapunov exponents, generalizing the notion of exceptional points to non-linear dynamical systems.
We distinguish two classes of bifurcations, corresponding respectively to continuous and discontinuous behaviors of the covariant Lyapunov vectors at the transition. 
We outline some physical consequences of generalized exceptional points on the dynamics of the system, including non-reciprocal responses, the destruction of isochrons, and enhanced sensitivity to noise.
We illustrate our results with concrete examples from neuroscience, ecology, and physics.
When applied to interpret existing experimental observations, our analysis suggests a simple explanation for the non-trivial phase delays observed in the population dynamics of plankton communities and the recently measured statistics of rotation reversals for a solid body immersed in a Rayleigh-Bénard convection cell.
\end{abstract}

\maketitle

Linear systems, ranging from sound and light waves to non-interacting bosonic quantum systems, can be thought of as a collection of harmonic oscillators. The dynamics of these systems is summarized by their so-called normal modes, and the corresponding oscillation frequencies. These are the eigenmodes and eigenvalues of the linear operator (dynamical matrix or Hamiltonian) describing the system.
Open systems are typically described by non-Hermitian operators, or in simpler cases real-valued asymmetric matrices~\cite{Shankar2022,Ashida2020,Nassar2020,Bergholtz2021,Clerk2022,Fruchart2023}. These mathematically capture dissipation and active driving, as well as the possible non-mutual (or non-reciprocal) couplings between the relevant degrees of freedom or fields (i.e., the action of A on B is different from the action of B on A). Of special interest are points in parameter space where such operators or matrices become non-diagonalizable and at least two eigenvectors coalesce -- they are technically known as exceptional points~\cite{Kato1984,Ashida2020}. Despite their name, exceptional points are rather common: a harmonic oscillator at critical damping is a simple example~\cite[\S~25]{landaumechanics}. Exceptional points typically mark the transition between travelling or oscillatory solutions and exponentially decaying dynamics. When studying non-reciprocal systems, one must face the fact that non-linearities are often crucial since the linearized dynamics of these driven, active or excitable media is unstable, leading to phase transitions between time-dependent states. 
The study of these transitions is tantamount to investigating bifurcations of the underlying dynamical system~\cite{StrogatzNL,Manneville2010,Murray2013,Golubitsky1985b,Golubitsky1988,Golubitsky2002,Crawford1991}.

\begin{figure}[t]
\includegraphics[width=8cm]{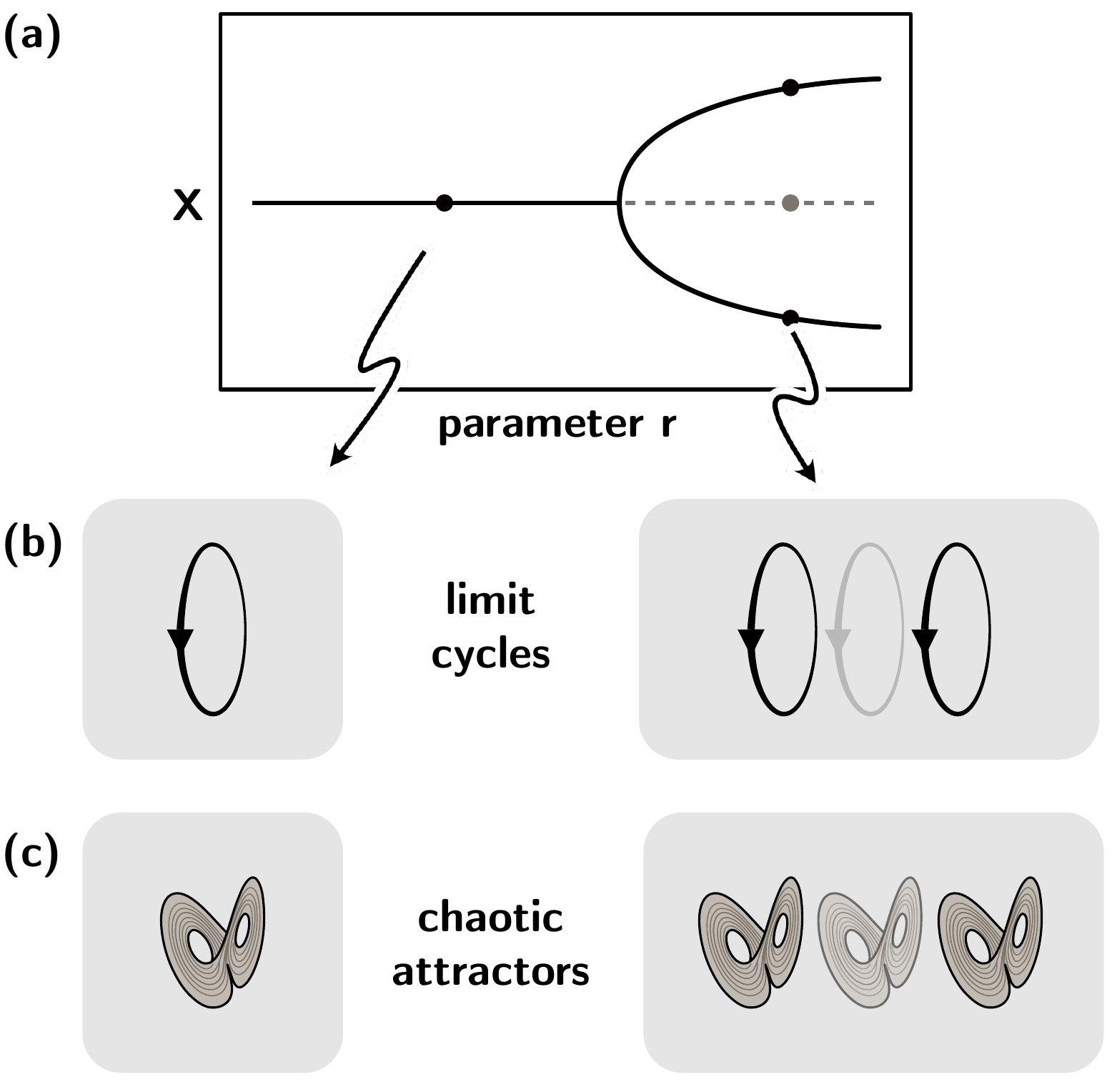}
    \caption{\label{figure_coalescence_attractors}
    \strong{Pitchfork bifurcation of attractors and Lyapunov vectors.}
    In a pitchfork bifurcation (panel a), a stable fixed point bifurcates into an unstable fixed point (dashed line) plus two stable fixed points when a control parameter $r$ reaches a critical value.
    This can be extended to generic attractors, such as limit cycles (panel b), higher-dimensional tori, or chaotic attractors (panel c), in which unstable attractors (repellers) are drawn with a lower opacity.
    }
\end{figure}

One of the simplest examples of a bifurcation is an elastic beam buckling under a load~\cite{StrogatzNL,Timoshenko1961}. The vertical beam is straight below a critical load at which it can curve either to the left or to the right.
Mathematically, this phenomenon is described by a pitchfork bifurcation (Fig.~\ref{figure_coalescence_attractors}a): it describes the appearance of the left- and right-buckled state of the beam that bifurcate from the unbuckled state at the critical load.
In the pitchfork bifurcation, the steady states of the system are fixed points. The unbuckled state is symmetric with respect to left-right inversion, while the buckled states break this symmetry, and therefore occur in conjugate pairs that coalesce at the bifurcation point.

In this article, we analyse certain classes of bifurcations in which the states coalescing at the bifurcation point are dynamical steady states (called attractors) as opposed to the two fixed points corresponding to the static shapes of the buckled beam.
These coalescing steady states can be limit cycles (closed orbits corresponding to a periodic evolution in time, see Fig.~\ref{figure_coalescence_attractors}b), limit tori, or strange attractors (more complicated structures that generally correspond to a chaotic evolution, see Fig.~\ref{figure_coalescence_attractors}c). The underlying dynamical systems need not  possess any symmetry besides time-translation invariance, i.e. they need not be equivariant under a continuous symmetry group, unlike the cases considered in \cite{Fruchart2021} and references therein. We show that the coalescence of attractors is accompanied by a generalized exceptional point defined in terms of so-called Lyapunov exponents and covariant Lyapunov vectors~\cite{Pikovsky2016,Eckmann1985}, that serve as generalized normal modes. The generalized exceptional point we investigate is an exact tangency of two covariant Lyapunov vectors with vanishing Lyapunov exponents (see appendix section~\ref{appendix_finite_LE} for the case of finite Lyapunov exponents).
These mathematical objects capture the dynamics of perturbations around the attractor, generalizing to extended attractors of non-linear systems the notion of exceptional points familiar from linear non-Hermitian physics.

This mathematical approach allows one to explore the rich phenomenology that arises as a result of this coalescence of attractors across various domains of science.
In the remainder of this paper, we develop a general theory for the coalescence of dynamical attractors in non-reciprocal dynamical systems based on the behavior of the covariant Lyapunov vectors at the bifurcation. We illustrate this behavior in several different physical models describing predator-prey systems, neural dynamics, coupled Hopf oscillators and chaotic dynamics. Next, we explore consequences of the combination of generalized exceptional points with noise or external perturbations. Finally, we highlight how our analysis can be used to explain existing experimental observations ranging from plankton dynamics~\cite{Beninca2008} to stochastic rotation reversals of rigid bodies in convection cells~\cite{Wang2023}.

\section{Generalized exceptional points in nonlinear dynamics}

\subsection{Linearized dynamics and exceptional points} 

Start with a dynamical system
\begin{equation}
    \label{ds}
    \dot{X} = f(X).
\end{equation}
in which $X(t) \in \mathbb{R}^N$ is a vector (the dot represents the time derivative, and $f$ is a function $\mathbb{R}^N \to \mathbb{R}^N$ defining the dynamical system). 
Consider now two nearby states that evolve by following the same dynamical system.
At long times, will these states be closer or farther away? 

Around a fixed point of the dynamical system $X_0$, defined by $f(X_0) = 0$, the answer is given by the Jacobian
\begin{equation}
    \label{jacobian}
    J_{a b} = \frac{\partial f_a}{\partial X_b}
\end{equation}
evaluated at the fixed point $X_0$ of interest.
The difference $\delta X(t)$ between the perturbed trajectory and the unperturbed one~\footnote{Technically, $\delta X(t)$ is a tangent vector, i.e. an element of the tangent space to the unperturbed trajectory at point $X(t)$. For a fixed point, $X(t) = X_0$.} evolves according to
\begin{equation}
\label{eom_perturbation_notime}
\delta \dot{X} = J \delta X.
\end{equation}
Hence, perturbations along the eigenvector ${c}_i$ of $J$ with eigenvalue $\lambda_i + \ii \omega$ grows or decays as $\ee^{\lambda_i t}$.

\begin{figure}[t]
    \centering
    \includegraphics[width=8cm]{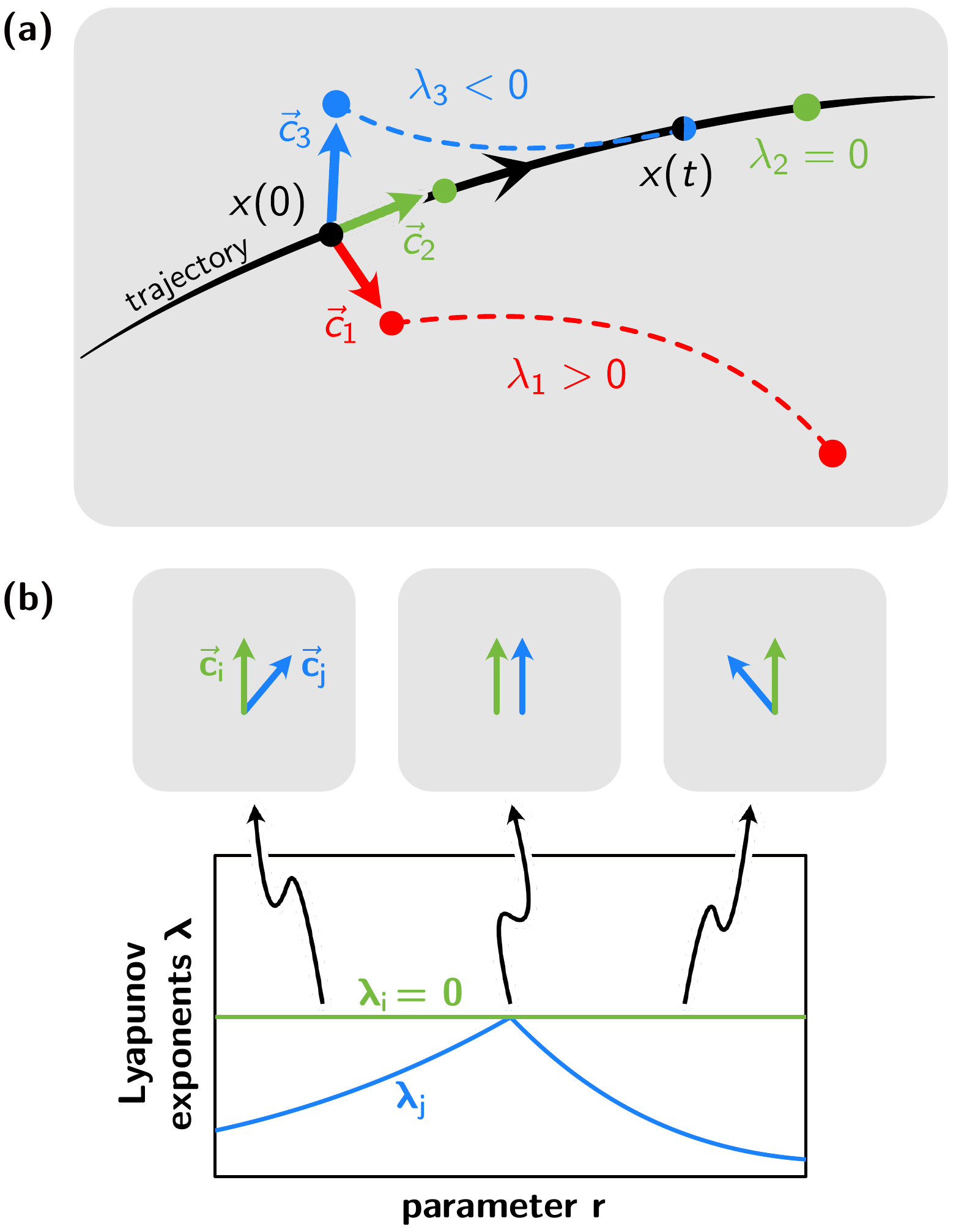}
    \caption{\label{figure_lyapunov_coalescence}
    \strong{Coalescence of Lyapunov exponents and covariant vectors at the bifurcation.}
    In panel a, we provide a schematic picture of Lyapunov exponents $\lambda_i$ and (covariant) Lyapunov vectors ${c}_i$.
    Consider an initial condition $x(0)$ for a dynamical system $\dot{x} = f(x)$. It evolves into $x(t)$. A slightly perturbed initial condition $x(0) + \epsilon \delta x$ ($\epsilon \ll 1$) evolves under the same dynamical system. Decomposing the perturbation $\delta x$ over the covariant Lyapunov vectors ${c}_i(0)$, we can predict whether at long times, the perturbation will grow ($\lambda > 0$, red), decay ($\lambda < 0$, blue) or stay finite ($\lambda = 0$, green).
    In panel b, we show the evolution of Lyapunov exponents and covariant Lyapunov vectors through the bifurcation represented in Fig.~\ref{figure_coalescence_attractors}a. One Lyapunov exponent $\lambda_i$, corresponding to the motion along the attractor ${c}_j$, always vanishes. At the transition, another Lyapunov exponent $\lambda_j$ goes to zero, while the corresponding Lyapunov vector ${c}_j$ becomes parallel to  ${c}_i$.
    }
\end{figure}

To draw this conclusion, we have implicitly assumed that the Jacobian $J$ can be diagonalized. 
However, not all matrices are diagonalizable: a counter-example is the non-diagonalizable matrix
\begin{equation}
    \label{EP}
    J = J_{\text{EP}} = \begin{pmatrix}
    0 & 1 \\
    0 & 0
    \end{pmatrix}.
\end{equation}
These matrices are somewhat unusual: they typically occur at codimension-two points in parameter space called exceptional points~\cite{Kato1984} (for instance, they would be isolated points in a two-dimensional parameter space~\cite{Seyranian2005}). 
When one approaches an exceptional point in parameter space, two eigenvectors of the matrix become more and more collinear, and end up parallel to each other at the exceptional point.
The particular form in Eq.~\eqref{EP} is called a Jordan block~\cite{Lang2010}.
Despite occurring at isolated points, these non-diagonalizable matrices do play an important role when they appear in the description of a physical or biological system. 

Physically, the main feature of the matrix Eq.~\eqref{EP} is that it encodes a non-reciprocal interaction between the two degrees of freedom (i.e., vector components) on which it operates: as $J_{1 2} = 1$, the degree of freedom 2 has an effect on 1, while 1 has no effect on 2 because $J_{2 1} = 0$. 
Mathematically, a crucial feature of this matrix is that it is non-normal~\cite{Trefethen2005}. A matrix $M$ is normal when it is unitarily diagonalizable. In contrast, non-normal matrices can have non-orthogonal eigenvectors, and even reach an exceptional point in which two eigenvectors become parallel, like ${c}_i$ and ${c}_j$ in Fig.~\ref{figure_lyapunov_coalescence}b (middle).

In the context of dynamical systems, the non-diagonalizable matrix Eq.~\eqref{EP} appears as (a block in) the Jacobian of any dynamical system undergoing a so-called Bogdanov-Takens bifurcation~\cite{Kuznetsov2004}. This is a codimension-two bifurcation that occurs in systems ranging from fluid mechanics to ecology and neuroscience~\cite{Kuznetsov2004,Bazykin1998,Izhikevich2007,Renardy1999}. It can be seen as the intersection of a Hopf bifurcation and a saddle-node bifurcation. 

Deviations from normality produce striking physical consequences related to transient growths and to an enhanced sensitivity to fluctuations and to boundary conditions~\cite{Trefethen2005,Shankar2022,Bergholtz2021,Ashida2020,Jaeger1997,Biancalani2017,Chajwa2020}.
For instance, consider an initial perturbation $\delta X(0) = (\delta X_1^0,\delta X_2^0)$ evolving according to Eq.~\eqref{eom_perturbation_notime} with the Jacobian $J=J_{\text{EP}}$ in Eq.~\eqref{EP}. Based on the (vanishing) eigenvalues of $J_{\text{EP}}$ alone, one may expect that the perturbation will have a constant magnitude. Yet, an explicit solution 
\begin{equation}
    \delta X(t) = \begin{pmatrix}
    X_1^0 + X_2^0 \, t \\
    X_2^0
    \end{pmatrix}
\end{equation}
shows that $\delta X$ actually growth with time, until the anomalous polynomial growths $\delta X_1(t) \sim X_2^0 \, t$ is stopped by non-linearities. When the eigenvalue $\lambda$ associated to the Jordan block is finite and negative ($J = J_{\text{EP}} + \lambda \text{Id}$ with $\lambda < 0$), the polynomial growth observed above still takes place, but it is eventually stopped by an exponential prefactor $\ee^{\lambda t}$. 
In addition to the polynomial growth, note that a perturbation initially along the direction $\delta X_2$ gradually moves along direction $\delta X_1$. The converse does not occur: perturbations along $\delta X_1$ stay along this direction. This is non-reciprocity in action. 
Both the anomalous growth and the non-reciprocal evolution of perturbations arise not only at exceptional points, but also in their neighborhood, where the Jacobian is strongly non-normal.

\subsection{Lyapunov exponents and covariant Lyapunov vectors}

When the dynamical system has attractors with a spatial extent in configuration space (like a limit cycle), the Jacobian at a single point is insufficient to indicate if the system has the characteristics of an exceptional point.
As before, perturbations $\delta X$ about an unperturbed trajectory $X_0(t)$ evolve according to 
\begin{equation}
\label{eom_perturbation}
\delta \dot{X} = J(X_0(t)) \, \delta X.
\end{equation}
in which the Jacobian now depends on time through the unperturbed trajectory $X_0(t)$.

Formally, this linear differential equation is solved by defining the evolution operator as the time-ordered exponential\footnote{Formally, the evolution operator $U(t,t_0)$ is defined as the unique solution of the Cauchy problem \eqref{eom_perturbation} with $\delta X(t_0) = \text{Id}$, with $\text{Id}$ the identity matrix.}
\begin{equation}
    \label{U_Texp}
    U(t, t_0) = T \exp\left(
    \int_{t_0}^{t} J(X_0(\tau)) \dd \tau
    \right),
\end{equation}
very much like in quantum mechanics~\cite{SakuraiModernQM,LeBellac}.
The evolution operator can thus be decomposed as
\begin{equation}
    \label{evo_op_nondegenerate}
    U(t, t_0) = \sum_{i}   \ket{c_i(t)} 
\ee^{\int_{t_0}^{t}\lambda_{i}(\tau) \dd\tau}
    \bra{\tilde{c}_i(t_0)}.
\end{equation}
in which ${c_i}(t)$ are called covariant Lyapunov vectors (CLVs), $\lambda_i(t')$ are instantaneous growth rates associated to the CLVs, and $\tilde{c}_i$ are called dual (or adjoint) CLVs \cite{Pikovsky2016,Kuptsov2012}.
At long times, the time averages of the instantaneous growth rates converge towards the Lyapunov exponents
\begin{equation}
    \lambda_i = \lim_{t \to \infty} \frac{1}{t} \int_{0}^{t} \lambda_i(t') \dd t'.
\end{equation}
The decomposition corresponding to \eqref{evo_op_nondegenerate} is known as an Oseledets splitting.
We emphasize that the CLVs are not necessarily orthogonal to each other (namely, $\braket{c_i(t),c_j(t)} \neq \delta_{ij}$).
The dual CLVs generalize left eigenvectors, while normal CLVs generalize right eigenvectors, so $\braket{\tilde{c}_i(t),c_j(t)} = \delta_{ij}$.

An arbitrary perturbation $\delta X(0)$ at time $t=0$ then evolves as~\cite{Schubert2015,Gaspard1995}
\begin{equation}
    \label{evo_nondegenerate}
    \delta X(t) = U(t,0) \delta X(0) = \sum_{i} a_i \ee^{\int_0^t \lambda_i(t') \dd t'} c_i(t)
\end{equation}
in which the $a_i$ are the coefficients obtained by projecting the initial perturbation $\delta X(0)$ on the CLVs $c_i(0)$. 
Infinitesimal perturbations either decay exponential ($\lambda_i < 0$), grow exponentially ($\lambda_i > 0$), or stay constant ($\lambda_i = 0$) at long times, as illustrated in Fig.~\ref{figure_lyapunov_coalescence}a.
Perturbations along the covariant Lyapunov vector ${c}_i$ grow or decay with the Lyapunov exponent $\lambda_i$ (Fig.~\ref{figure_lyapunov_coalescence}a).
The Lyapunov exponents generalize the (real part of the) eigenvalues of the Jacobian at a fixed point, while the covariant Lyapunov vectors generalize the eigenvectors~\footnote{We refer the reader to \cite{Trevisan1998,Huhn2019}, that includes a discussion of degenerate cases.}.
For more details about Lyapunov exponents and covariant Lyapunov vectors and applications, we refer the readers to Refs.~\cite{Eckmann1985,Pikovsky2016,Ginelli2007,Kuptsov2012,Wolfe2007,Ginelli2013,Takeuchi2013,Yang2010,Froyland2013,Trevisan1998,Noethen2019,Wiesel1993,Wiesel1993b,Hejazi2017,Voth2002}.
Please note that the CLVs $c_i(t)$ depend on time, because they have to follow the unperturbed trajectory.
Indeed, most of the complexity of the time evolution Eq.~\eqref{eom_perturbation} is crucially packed into the time-dependence of the CLVs.

Although CLVs are generically nonorthogonal, the normalization of the vectors can often be chosen such that $\braket{\tilde{c}_i, c_j} = \delta_{i j}$. Doing so is \emph{almost} always possible: the only exception are generalized exceptional points, that we discuss in the next section.

\subsection{Generalized exceptional points}
Equations~\eqref{evo_nondegenerate} and \eqref{evo_op_nondegenerate} are valid most of the time, but not always: they must encompass the simple example of a constant non-diagonalizable Jacobian such as Eq.~\eqref{EP}. This feature can extend to non-constant Jacobians.
In general, one has to write
\begin{equation}
    \label{Lyapunov_evolution_operator}
    U(t, t_0) = \sum_{i j} \ket{c_i(t)} 
    \Lambda_{i j}(t)
    \bra{\tilde{c}_j(t_0)}
\end{equation}
in which $\Lambda_{i j}(t)$ is composed of Jordan blocks, and hence is not necessarily diagonal~\cite{Arnold1999,GonzalezTokman2013}. 
This situation is the generalization of an exceptional point. It occurs when two Lyapunov exponents are equal and the corresponding CLVs, say $c_{1}(t)$ and $c_{2}(t)$, become identical (Fig.~\ref{figure_lyapunov_coalescence}b)~\footnote{Two remarks are in order. First, it is possible to have equal Lyapunov exponents with different CLVs (a trivial example consists in two identical decoupled dynamical systems, considered as a single dynamical system). Conversely, two identical CLVs must have the same Lyapunov exponent (because they are the same vector; this is a tautology).}. At the singularity, the evolution operator cannot be diagonalized anymore and assumes a Jordan block form. In the simplest case of a Jordan block of size two~\footnote{In the main text, all the examples we consider correspond to a Jordan block of size two. The case of higher-order singularities is discussed in appendix section \ref{appendix_tori_higher}.},
\begin{equation}
    \label{jordan_Lambda}
    \Lambda(t) \simeq
    \begin{pmatrix}
    \Lambda_{1}(t) & \Lambda_{12}(t) \\
    0 & \Lambda_{1}(t) 
    \end{pmatrix}
\end{equation}
where $\Lambda_{1}(t) = \ee^{\int_{t_0}^{t}\lambda_{1}(\tau) \dd\tau}$.
This can be seen as a generalization of Eq.~\eqref{EP}. A similarity transformation $J \mapsto P^{-1} J P$ for an arbitrary invertible matrix $P$ leaves the matrix $J$ in Eq.~\eqref{EP} non-diagonalizable. Similarly, the presence of a generalized exceptional point where two CLVs are parallel is invariant under a non-linear change of coordinate of the dynamical system (see Appendix B of Ref.~\cite{Yang2010} for how CLVs transform under such a change of coordinates).
Because of the covariance of the CLVs with respect to the time evolution, two CLVs that are exactly parallel at one point of the trajectory remain exactly parallel all along the trajectory~\footnote{In general, the angles between CLVs depend on time (equivalently, on the base point on the trajectory). However, two CLVs that are exactly parallel at one point are also exactly parallel along the whole trajectory. In order to illustrate the presence of a tangency in the examples, we either average the angles $\theta_{i j}$ between the CLVs $\vec{c}_i$ and $\vec{c}_j$ over the trajectory, or, in the case of (quasi)periodic dynamics we evaluate them at a distinguished point (e.g. where an arbitrary coordinate is maximal). See Appendix \ref{app_plankton} (in particular Fig.~\ref{figure_model_plankton}m) for an example.}.

The case in which two CLVs are almost identical has been known as \emph{near tangencies} in the literature \cite{Kuptsov2012,Yang2009,Takeuchi2011,Sharafi2017,Xu2016,Bosetti2010,Vannitsem2016}. 
In this work, we focus on \emph{exact} tangencies (i.e., generalized exceptional points).
It is natural to expect near tangencies in the neighborhood of generalized tangencies. This is indeed what we observe in several cases. As we shall see, this does not always happen: perhaps surprisingly, the behavior of CLVs can be discontinuous.

\subsubsection{Generalized EPs with vanishing Lyapunov exponent}
\label{zero_CLV}

The main focus of this work is the case of exact tangencies (generalized EPs) where the involved CLVs have vanishing Lyapunov exponents. These tangencies correspond to bifurcations in the underlying dynamical system. 

We have implicitly assumed that the dynamical system in Eq.~\eqref{ds} is autonomous, namely, that $f$ does not explicitly depend on time. As a consequence, the dynamical system Eq.~\eqref{ds} always has a vanishing (local and global) Lyapunov exponent $\lambda_* = 0$ corresponding to the CLV $\vec{c}_* \equiv f$ tangent to the trajectory at each point~\cite{Pikovsky2016}. This can be shown directly by taking the time derivative of Eq.~\eqref{ds} and using the chain rule, yielding $(d/dt) \dot{X} = J(X(t)) \, \dot{X}$. One can interpret this fact as a consequence of the time-translation invariance of the dynamical system, that is spontaneously broken by the solutions.

\subsubsection{Generalized EPs with finite Lyapunov exponent}

It is also possible to find tangencies  between CLVs with finite Lyapunov exponents that coincide with a change in the transient behavior of the system rather than a change in the nature of the attractors. In other words, they do not correspond to a bifurcation of the dynamical system. A finite Lyapunov exponent CLV tangency can be demonstrated in a linear system $\dot{x} = Ax$ with a fixed point~\footnote{Consider for instance the matrix
\begin{equation}
    J(\epsilon) = \begin{pmatrix}
        \sigma & \epsilon \\
        1 - \epsilon & \sigma
    \end{pmatrix}.
\end{equation}
This matrix has exceptional points $\epsilon = 0$ and $\epsilon = 1$.
Consider now the linear dynamical system $\dot{X} = J(\epsilon) X \equiv f_{\epsilon}(X)$ for $X \in \mathbb{R}^2$.
It has a unique fixed point at $X = 0$.
In the case of a fixed points, the CLVs are the eigenvectors of the Jacobian (here $J(\epsilon)$), so there is indeed a tangency of the CLVs at $\epsilon = 0$ (and at $\epsilon = 1$). 
The fixed point $X=0$ does change nature at $\epsilon = 0$ from a node (for $\epsilon = 0^-$, where the streamlines of the vector field are essentially straight), to a focus (for $\epsilon = 0^+$, where the streamlines are spiraling). Nevertheless, the dynamical systems at $\epsilon = 0^{\pm}$ are topologically equivalent (see for instance Example 2.1 in Ref.~\cite{Kuznetsov2004}), so there is no bifurcation.
}. When a parameter is tuned such that two eigenvalues of $A$ coalesce at a finite value to become a complex conjugate pair, the fixed point will change from a node to a focus changing the transient dynamics. Other examples are discussed in appendix section~\ref{appendix_finite_LE}.

\section{Coalescence of attractors and generalized exceptional points}

\subsection{Parity-breaking bifurcations}
\label{sec_parity_breaking}

As a first example, consider the so-called parity-breaking bifurcation~\cite{Knobloch1995}
\begin{subequations}
\label{parity_breaking}
\begin{align}
    \dot{\phi} &= w \\
    \dot{w} &= r w - w^{3}.
\end{align}
\end{subequations}
The variable $\phi$ can be seen as an angular variable representing the motion of a point on a circle of fixed radius, or as a real variable representing the motion of a point on the line~\footnote{Up to a non-linear change of variable, the circle and line can be mapped to arbitrary closed and open curves.}.
The variable $w$ is an additional variable that undergoes a pitchfork bifurcation when the parameter $r$ changes sign (neither $w$ nor $r$ are related to the radius of the circle). At the same time, the full dynamical system \eqref{parity_breaking} undergoes a parity-breaking bifurcation (also known as drift-pitchfork bifurcation).
This bifurcation occurs in pattern formation~\cite{Coullet1985b,Coullet1989,Malomed1984,Douady1989,Brachet1987,Pan1994,Bensimon1989}, fluid dynamics~\cite{Xin1998,Thiele2004}, excitable media~\cite{Kness1992,Krischer1994}, coupled lasers~\cite{Hassan2015,Clerkin2014,Soriano2013}, synchronization~\cite{Hong2011,Hong2011b,Hong2014,Fruchart2021}, biological tissues~\cite{Leonetti2006}, driven-dissipative condensates~\cite{Hanai2019,Hanai2020}, active matter and other collective phenomena~\cite{Fruchart2021,Saha2020,You2020,FrohoffHulsmann2023,Ophaus2018}.
When $r<0$, $\phi$ is constant, so the system simply has a fixed point.
When $r > 0$, a limit cycle appears. Its frequency $\pm \sqrt{r}$ depends on the initial condition $w(t=0)$, which fixes the sign $\pm = \operatorname{sign}(w(\infty))$. Hence, the sense of rotation (clockwise or counterclockwise) of the limit cycle in the parity-breaking bifurcation is not determined in advance: it depends on the initial conditions.
This is in contrast with the case of a Hopf bifurcation, in which the sense of rotation is always the same for a given set of parameters.

Let us compute the Jacobian of the dynamical system defined by Eq.~\eqref{parity_breaking}. Evaluated at $w=0$, it is
\begin{equation}
    J = \begin{pmatrix}
    0 & 1 \\
    0 & r
    \end{pmatrix}.
\end{equation}
This matrix becomes non-diagonalizable at the bifurcation point $r=0$. In other words, the bifurcation occurs at an exceptional point of the Jacobian~\footnote{It is interesting to compare the parity-breaking bifurcation to the so-called Bogdanov-Takens bifurcation~\cite{Kuznetsov2004}, typically associated with a non-diagonalizable Jacobian. Despite having the same linear part (corresponding to the non-diagonalizable Jacobian), their non-linear parts do not match. This is because the dynamical system \eqref{parity_breaking} does not satisfy the genericity conditions that are usually assumed to obtain the normal form of the Bogdanov-Takens bifurcation (equations (BT.0-3) in Theorem 8.4 of Ref.~\cite{Kuznetsov2004}, for instance).
The parity-breaking can then be thought as a degenerate version of a Bogdanov-Takens bifurcation (see Ref.~\cite{Kuznetsov2005} and references therein for other similar situations).}.

\subsection{Pitchfork of attractors} 
\label{pitchfork_attractors}

The parity-breaking bifurcation introduced in the previous paragraph describes a transition from a fixed point to two limit cycles (plus an unstable fixed point). In a rotating frame of reference, this is equivalent to a bifurcation from a single limit cycle to two different limit cycles (plus an unstable limit cycle).
This suggests that a generalized notion of exceptional point could appear when the Jacobian is not constant along trajectories, even if it is difficult or impossible to do the equivalent of rotating the reference frame. This could occur for limit cycles, but also more complex attractors. In order to analyze this situation, we generalize Eq.~\eqref{parity_breaking} to produce a pitchfork of attractors in which a given non-linear attractor splits or merges into attractors of the same nature.

We start with a dynamical system producing the desired attractor, of the form
\begin{equation}
    \dot{x} = f(x)
\end{equation}
in which $x \in R^M$.
This dynamical system is controlled by a 1D system exhibiting a pitchfork bifurcation as follows:
\begin{subequations}
\label{eq_pitchfork_attractors}
\begin{align}
    \dot{x} &= f(x) + g(x,w)  \\
    \dot{w} &= r w - w^{3}. \label{pitchfork}
\end{align}
\end{subequations}
The simplest coupling consists in making the parameters of $f$ depend on the value of $w$ (a bit like in a parametric oscillator)~\footnote{Let $p$ be the vector of parameters needed to parameterize $f(x; p)$. We can choose a function $\pi(w)$ and set $p=\pi(w)$. By choosing a reference parameter $p_0$ and setting $g(x,w) = f(x; \pi(w)) - f(x; p_0)$, and $f(x) = f(x; p_0)$, we go back to the case discussed in the main text.}. 
The properties of the attractor described by $f$ are then tuned through $w$. Because there are two possible (stable) values of $w$ on the right of the pitchfork bifurcation, there are also two different attractors, whose properties gradually become identical when we approach the pitchfork bifurcation.

In this paper, we focus on pitchforks of attractors as described by Eq.~\eqref{eq_pitchfork_attractors}. One can also consider a generalization of Eq.~\eqref{eq_pitchfork_attractors} in which the pitchfork bifurcation in Eq.~\eqref{pitchfork} is replaced with another bifurcation, such as a saddle-node or a transcritical bifurcation, in the same way as the so-called drift-transcritical bifurcation \cite{Ophaus2018,FrohoffHulsmann2023,FrohoffHulsmann2021,Ophaus2021} can be obtained from the drift-pitchfork bifurcation \eqref{parity_breaking}. See for instance \cite{Kuznetsov2004} and references therein for a discussion on bifurcations of limit cycles. This leads to other ways of merging/splitting attractors, that can also be accompanied with generalized exceptional points. 

In addition, we note that the all the bifurcations we consider can arise in spatially extended systems, in which Eq.~\eqref{eq_pitchfork_attractors} (or variants) describes the dynamics on an invariant manifold. Simple examples can be obtained by adding a diagonal diffusive term to the normal form \eqref{eq_pitchfork_attractors}. Other examples include drifting defects and localized states~\cite{Knobloch2015,FrohoffHulsmann2023,Ophaus2021,Brusch2000}.

\section{Examples}

\subsection{Pitchfork of limit cycles}
\label{pflc}

\begin{figure*}
    \centering
    \includegraphics[width=0.75\paperwidth]{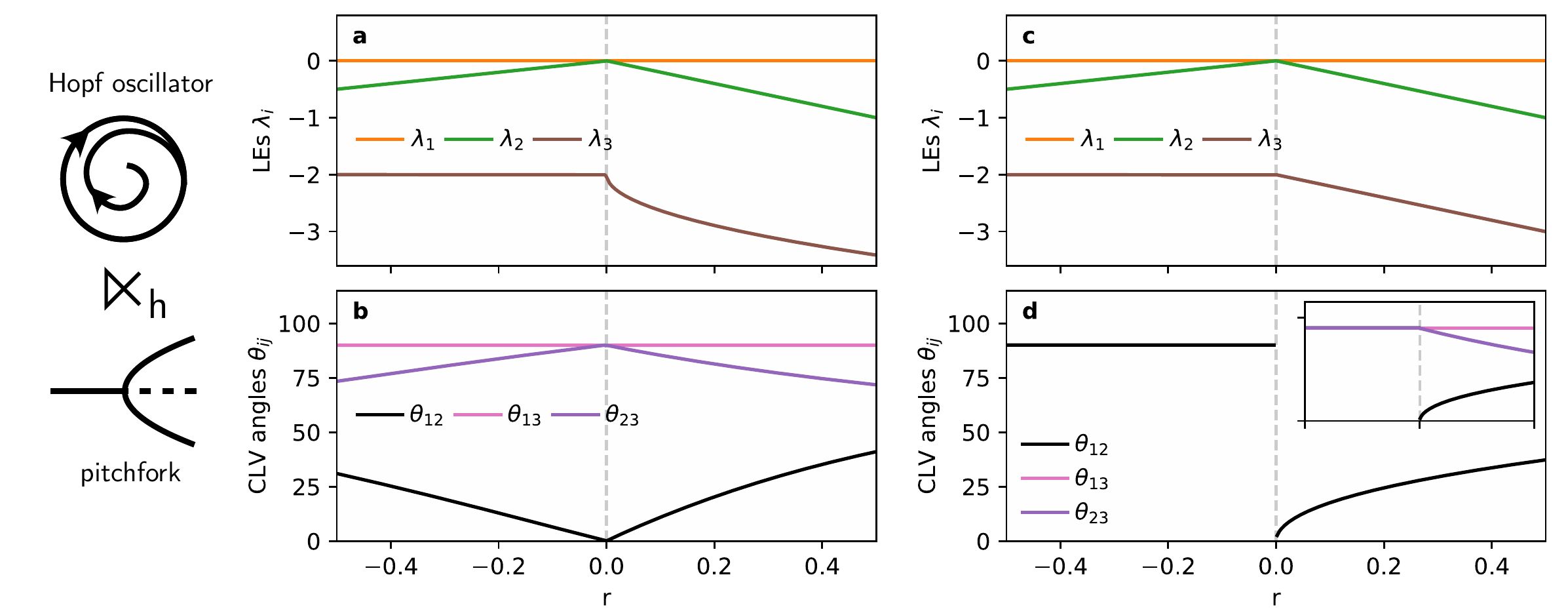}
	\caption{\strong{Pitchfork of limit cycles.}
    We consider a Hopf oscillator controlled by a pitchfork bifurcation, as in Eq.~\eqref{hopf_pitchfork}, leading to a pitchfork bifurcation of limit cycles.
	The Lyapunov exponents and the angles $\theta_{i j}$ between the covariant Lyapunov vectors are plotted as a function of the control parameter $r$, with two choices of the coupling function: (a,b) $h(w) = w$ and (c,d) $h(w) = w^2$.
	The angles $\theta_{ij}$ (panels b,d) is the angle between the CLVs ${c}_i$ and ${c}_j$ corresponding to $\lambda_i$ and $\lambda_j$ (panels a,c).
	The dashed gray line indicates the bifurcation point, where the Lyapunov exponents $\lambda_1$ and $\lambda_2$ both vanish and the corresponding CLVs align.
	In panel b, the behavior of the CLVs is continuous through the bifucation. In panel d, in contrast, the angle $\theta_{1 2}$ (black line) jumps from a finite value to zero when $r \to 0^{-}$. For clarity, we have only plotted $\theta_{12}$ in panel d; the other angles are shown in the inset.
	We have set $\alpha = 1+\ii$, $\beta = -1$, $\gamma = 1 + \ii$ in Eq.~\eqref{hopf_pitchfork}. 
	}
	\label{figure_pitchfork_hopf}
\end{figure*}

Let us illustrate these ideas when $f$ is the normal form of a Hopf oscillator. 
In this case, it is convenient to write $x=(x_1,x_2)$ as a complex number $z = x_1 + \ii x_2$, and the general form Eq.~\eqref{eq_pitchfork_attractors} becomes
\begin{subequations}
\label{hopf_pitchfork}
\begin{align}
    \dot{z} &= (\alpha + \beta \,|z|^2) \, z + \gamma h(w) z \label{hp_z} \\
    \dot{w} &= r w - w^{3} \label{hp_w}
\end{align}
\end{subequations}
where $\alpha$, $\beta$, $\gamma$ are complex, while $r$ and the arbitrary function $h(w)$ are real. 
Here, adding the coupling $g(z,w) = \gamma h(w) z$ is equivalent to replacing the coupling $\alpha$ with $\alpha'(w) = \alpha + \gamma h(w)$. 
Let us analyze the dynamical system Eq.~\eqref{hopf_pitchfork}. 
First, Eq.~\eqref{hp_w} shows that $w = 0$ when $r<0$ and $w = \pm \sqrt{r}$ when $r>0$, irrespective of the value of $z$.
Then, writing $z = R(t) \ee^{\ii \phi(t)}$ and $\alpha = \alpha_\re + \ii \, \alpha_\im$ (similar for $\alpha'$, $\beta$ and $\gamma$), we find that 
\begin{subequations}
\label{pitchfork_lc_polar}
\begin{align}
    \dot{R} &= \alpha'_\re R + \beta_\re R^3 \\
    \dot{\phi} &= \alpha'_\im + \beta_\im R^2.
\end{align}
\end{subequations}
Hence, a limit cycle with radius $R = \sqrt{-\alpha'_\re(w)/\beta_\re}$ and angular frequency $\Omega = \alpha_\im'(w) - \beta_\im/\beta_\re \alpha'_\re(w)$ exists when the inside of the square root is positive (the period of the oscillation is $T=2\pi/\Omega$). Otherwise, there is a stable fixed point $R = 0$.
The radius and frequency of the limit cycle depend explicitly on $w$. Hence, on the symmetry-breaking side of the pitchfork bifurcation ($r > 0$), there are two different limit cycles whose properties depend on the value of $w = \pm \sqrt{r}$, provided that $h(w) \neq h(-w)$. 
The system mimics a traditional pitchfork bifurcation, but with stable/unstable fixed points replaced with stable/unstable limit cycles. 
Let us tune the parameters of the Hopf bifurcation so that a stable limit cycle exists \footnote{Here, we  assume that a limit cycle occurs on both sides of the bifurcation. This is not necessarily the case. Setting $\alpha = \beta = 0$ and $h(w) = w$, we find that Eq.~\eqref{pitchfork_lc_polar} reduces to the normal form of the parity-breaking bifurcation \eqref{parity_breaking}, in which a fixed point exists on one side, and two limit cycles on the other side.}.
For $r<0$, there is a single limit cycle invariant under the transformation $w \to - w$. For $r>0$, there are two stable limit cycles that are mapped to each other by the transformation $w \to - w$, plus an unstable cycle that is mapped onto itself. 
(We refer to appendix section \ref{appendix_Z2} for a discussion on symmetries.)
At $r=0$, a pitchfork bifurcation of limit cycles occurs (see also Refs.~\cite{Kuznetsov2004,Nikolaev1998a,Nikolaev1998b,Nikolaev1995} and for examples Refs.~\cite{Abshagen2005,Pedersen2022,Sherman1994,Rohm2018,Marques2013,Bacic2020,Willms2017}), in which the two stable limit cycles and the unstable one merge into one stable cycle, as depicted in Fig.~\ref{figure_pitchfork_hopf}c. 
In this bifurcation, the radius of the limit cycle is irrelevant. Therefore, it can be ignored to focus on the reduced dynamics 
\begin{subequations}
\label{limit_cycle_parity_breaking}
\begin{align}
    \dot{\phi} &= \omega_0 + \omega_1 \, w \\
    \dot{w} &= r w - w^{3}.
\end{align}
\end{subequations}
similar to Eq.~\eqref{parity_breaking}.
As the bifurcation is approached, the limit cycles are pushed against each other, the distance between the cycles shrinks. 
One can expect that this confinement may constrain the dynamics of small perturbations close to the attractors to be tangent to the cycles. As we now show, this expectation can be made precise using Lyapunov exponents and covariant vectors.

In the case of limit cycles, the Lyapunov exponents and covariant vectors can be computed using Floquet theory. More precisely, the real part of the Floquet exponents are the Lyapunov exponents, while the corresponding Floquet eigenvectors span the same spaces as the covariant Lyapunov vectors~\cite{Trevisan1998,Huhn2019}. 
The dynamical system Eq.~\eqref{hopf_pitchfork} is simple enough to do this calculation analytically: this is done in the Methods.
We find that the Lyapunov exponents are $0$, $r - 3 w^2$ and $2 \beta_\re R^2$ with corresponding covariant vectors at time $t=T$ (see Appendix section \ref{explicit_computation_floquet} for a derivation and for the expression at arbitrary times)
\begin{align}
\!\!\!\!
    \psi_1 = \begin{pmatrix}
    0 \\ 1 \\ 0
    \end{pmatrix}
    \quad\!
    \psi_2 = \frac{1}{N_2} \begin{pmatrix}
    0 \\ - \gamma_\im \, R \, h'(w) \\ 2 r
    \end{pmatrix}
    \quad\!
    \psi_3 = \begin{pmatrix}
    1 \\ 0 \\ 0
    \end{pmatrix}
\end{align}
when $\alpha_\re = 0$ and $\beta_\im = 0$, and where $N_2$ is a normalization factor (the general case is similar, but the expressions are longer: see appendix section \ref{explicit_computation_floquet}).
In these expressions, $R$ and $w$ are shorthands for their value on the current attractor as a function of the system parameters, and the vectors are written in the basis of perturbations $\delta X = (\delta x_1, \delta x_2, \delta w)$, see appendix section \ref{explicit_computation_floquet} for details.

Consider first the case in which $h(w) = w$, so $h'(w) = 1$ (Fig.~\ref{figure_pitchfork_hopf}a top). At the bifurcation, $\psi_2 \to (0,1,0)$ when $r \to 0$, independent of whether the limit is taken from above or below. 
Near the bifurcation, we have essentially $\psi_2 \sim (0,1,2r)$, so the angle between $\psi_1$ and $\psi_2$ goes to zero: the two covariant Lyapunov vectors become parallel.

Consider now the case in which $h(w) = w^2/2$, so $h'(w) = w$ (Fig.~\ref{figure_pitchfork_hopf}a bottom).
When we approach the bifurcation from positive values $r \to 0^{+}$, the stable branches of solutions are $w = \pm \sqrt{r}$, so $\psi_2 \sim (0,\sqrt{r},r) \to (0,1,0)$. As previously, the two vectors $\psi_1$ and $\psi_2$ become identical. However, when we approach the bifurcation from negative values ($r \to 0^{-}$), the stable branch is $w = 0$ and $\psi_2 \to (0,0,1)$, so the two covariant Lyapunov vectors do not become parallel in a continuous way. The coalescence of covariant Lyapunov vectors still occurs, but there is a discontinuity in the Lyapunov vectors at the bifurcation.
As we shall now see, this analytically tractable toy model exemplifies a situation that generically occurs in asymmetrically coupled oscillators.

\begin{figure*}
    \centering
    \includegraphics[width=0.9\textwidth]{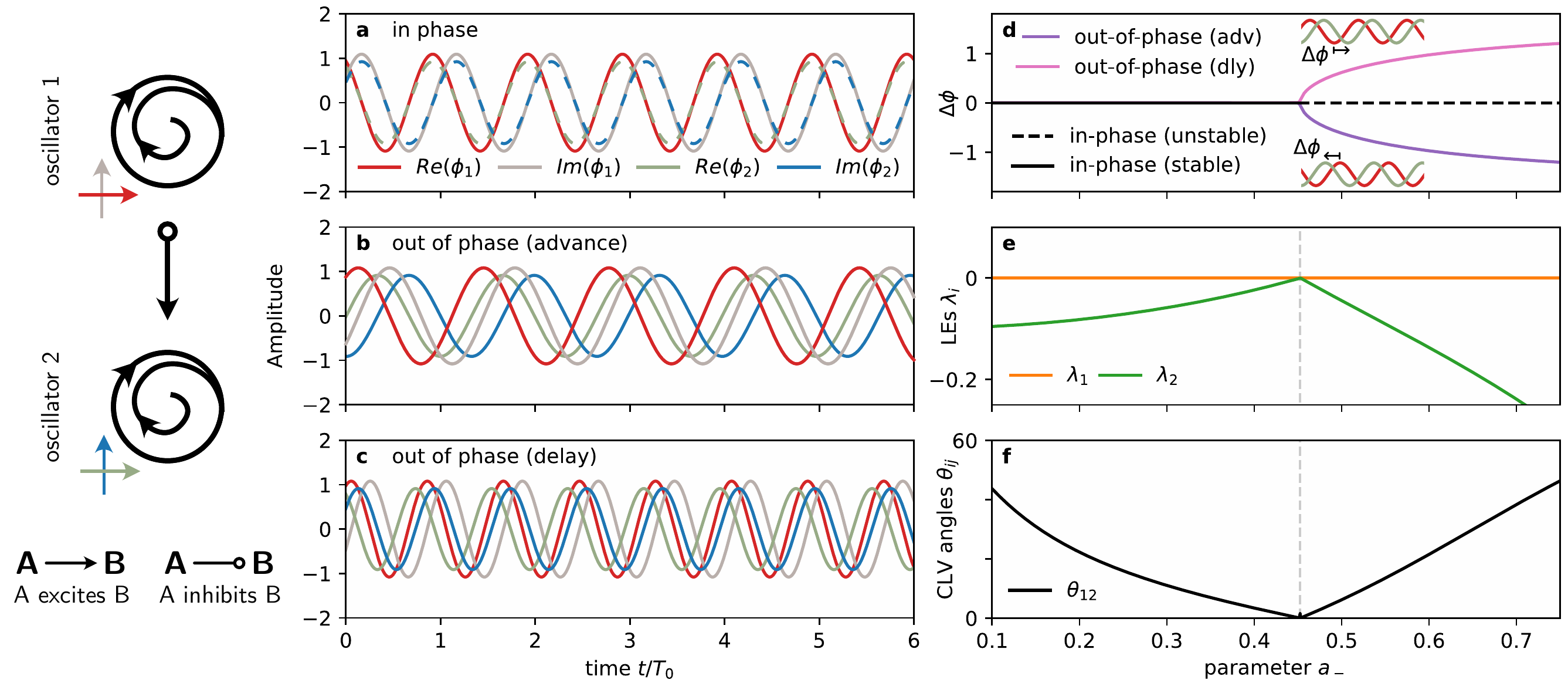}
    \caption{\strong{Coupled non-linear oscillators.}
    We simulate the coupled nonlinear oscillators given in Eq.~\eqref{eq:coupledhopf}.
    Depending on the value of the parameter $a_{-}$, we observe different behavior illustrated by the time series in panels a-c. When $a_{-} < a_{-}^{\text{c}} \approx \num{0.22625}$, the two oscillators exhibit synchronized oscillations corresponding to a single limit cycle (panel a). 
    When $a_{-} > a_{-}^{\text{c}}$, the two oscillators become out of phase (panels b and c). Depending on the initial condition, the phase difference $\Delta \phi$ between the oscillators $z_1(t)$ and $z_2(t)$ can either be positive (panel c) or negative (panel d). Hence, there are two stable limit cycles. The phase difference is plotted as a function of $a_{-}$ in panel d.
    At the bifurcation point $a_{-} = a_{-}^{\text{c}}$, 
    a generalized exceptional point occurs, as shown in panels e-f. In panel e, the largest Lyapunov exponents $\lambda_1$ and $\lambda_2$ are plotted.
    We always have $\lambda_1 = 0$ because the dynamical system is autonomous. At the bifurcation point $a_{-} = a_{-}^{\text{c}}$, the second Lyapunov exponent $\lambda_2$ also vanishes. At the same time, the angle $\theta_{12}$ between the CLVs ${c}_1$ and ${c}_2$ vanishes (panel f), showing that a generalized exceptional point occurs. Here, the CLVs are continuous across the bifurcation.
    We have set $a=\omega=-b =1, a_{+}=0.05$. In panel a, $a_{-}$ is varied from \num{0.05} to \num{0.375}. 
    In panel a, $a_{-} = 0.3$. 
    In panel b-c, $a_{-} = 0$. 
}  \label{fig:hopf}
\end{figure*}

\subsection{Coupled nonlinear oscillators} 
\label{coupled_nonlinear_oscillators}
We now show that coupling two nonlinear oscillators in a non-reciprocal way typically leads to a pitchfork bifurcation of limit cycles. To do so, we consider the dynamical system
\begin{equation}
\!\!\!\!
    \begin{bmatrix}
        \dot{z}_{1} \\
        \dot{z}_{2}
    \end{bmatrix}
    = 
    \begin{bmatrix}
        (a+i\omega)+b|z_{1}|^{2} & a_{12} \\
        a_{21} & (a+i\omega)+b|z_{2}|^{2}
    \end{bmatrix}
    \begin{bmatrix}
        z_{1}\\
        z_{2}
    \end{bmatrix}
    \label{eq:coupledhopf}
\end{equation}
consisting of two copies of the normal form of a Hopf bifurcation (Hopf oscillators) for the complex variables $z_n$ coupled by the terms $a_{1 2}$ and $a_{2 1}$.
Depending on the value of $a_{+}=[a_{12}+a_{21}]/2$ and $a_{-}=[a_{12}-a_{21}]/2$, the Hopf oscillators can be in an aligned, antialigned, or chiral state. When the coupling of the oscillators is sufficiently nonreciprocal, the aligned/antialigned state transitions to the chiral state, and a bifurcation occurs such that the steady state of the dynamical system will have a finite angle between the oscillators, which approaches 90 degrees as $a_{-}$ becomes large. There are two such steady states, one where $z_{1}$ chases $z_{2}$ as the system oscillates, and one where $z_{2}$ chases $z_{1}$. The symmetry of the system is spontaneously broken, determining which Hopf oscillator leads and which follows. The transition from the aligned to chiral state occurs at an exceptional point, where one Lyapunov exponent becomes zero at the bifurcation, and the corresponding Lyapunov vectors become parallel at all points on the limit cycle, as shown in Fig.~\ref{figure_lyapunov_coalescence}.
The behavior of the model \eqref{eq:coupledhopf} is phenomenologically similar to the pitchfork of limit cycles with $h(w) = w$ analyzed in the previous section (compare Fig.~\ref{figure_pitchfork_hopf}a-b with Fig.~\ref{fig:hopf}e-f). 

Writing $z_n = r_n \ee^{\ii \phi_n}$ in polar form, performing an adiabatic elimination of $r_n$ (see e.g. Ref.~\cite{Haken1983}) assuming $a_\pm \ll a$, we end up with
\begin{subequations}
\begin{align}
    \partial_t \Delta \phi &= 2 \left( a_{+} - a_{-}^2/a \cos \Delta \phi \right) \sin \Delta \phi \\
    \partial_t \phi &= \omega + a_{-} \sin \Delta\phi
\end{align}
\end{subequations}
in which $\Delta \phi \equiv \phi_2 - \phi_1$ and $\phi = [\phi_1 + \phi_2]/2$.
A series expansion for small $\Delta \phi$ and a rescaling leads to the normal form Eq.~\eqref{limit_cycle_parity_breaking}.
Crucially, this allows us to identify the control variable $w$ in the normal form Eq.~\eqref{limit_cycle_parity_breaking} to the dephasing $\Delta \phi$ between the oscillators (namely, $w \propto \Delta \phi$). The two symmetric non-zero solutions $\Delta \phi_{\pm}$ of $\partial_t \Delta \phi = 0$ correspond to the two limit cycles that emerge at the bifurcation. This is a generic feature, that we will find in other examples of coupled oscillators (Secs.~\ref{sec_prey_predator} and \ref{sec_neurons}). 

\begin{figure*}
    \centering
    \includegraphics[width=\textwidth]{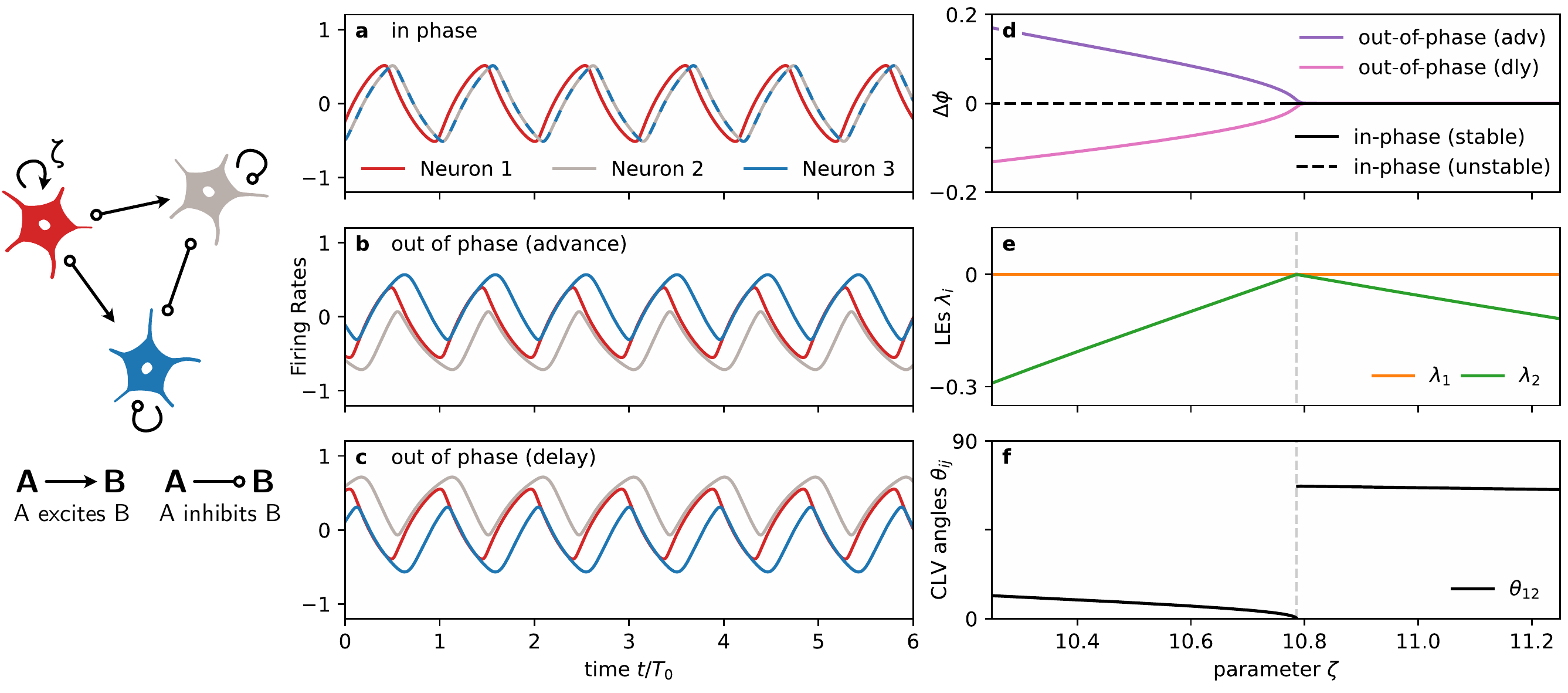}
	\caption{\textbf{Wilson-Cowan model.} 
	Panels a-c show time series
	of the Wilson-Cowan model in Eq. \eqref{eq:wceq} for one excitatory neuron ($i=1$) coupled to two inhibitory neurons ($i=2,3$). When $\zeta > \zeta_{\text{c}} \approx \num{10.786}$, the firing rates of the two inhibitory neurons are in-phase oscillations corresponding to a limit cycle (panel a).
	When $\zeta < \zeta_{\text{c}}$, the two inhibitory neurons become out of phase (panels b and c), and the phase difference $\Delta \phi$ depends on the initial condition, as there are two stable limit cycles. The phase difference is plotted as a function of $\zeta$ in panel d.
    The two limit cycles correspond to phase advance (adv, purple) or delay (dly, pink).
In panel e, the largest Lyapunov exponents $\lambda_1$ and $\lambda_2$ are plotted ($\lambda_1 \equiv 0$ because the dynamical system is autonomous). At the bifurcation $\zeta = \zeta_{\text{c}}$, the second Lyapunov exponent $\lambda_2$ and the angle $\theta_{12}$ between the CLVs ${c}_1$ and ${c}_2$ vanish (panel f). This is a generalized exceptional point. 
	The angle $\theta_{12}$ is discontinuous through the transition: it stays constant when $\zeta \to \zeta_{\text{c}}$ from the right. 
	We have set
$A_{1 1} = \zeta$, $A_{2 1} = A_{3 1} = \num{8}$ in which $\zeta \geq 0$ can vary (for the excitatory neuron $j=1$), 
$A_{1 2} = \num{-4.5}$, $A_{2 2} = \num{0}$, $A_{3 2} = \num{-4}$, and
$A_{1 3} = \num{-7.5}$, $A_{2 3} = \num{-4}$, $A_{3 3} = \num{0}$ (for the inhibitory neurons $j=2,3$) and $h_i = 0$ for all neurons.
In panel a, $\zeta = 11$. In panels b-c, $\zeta = 9$.
Times are normalized by the period $T_0$ at $\zeta_{\text{c}}$.
	}
	\label{fig:wilsoncowan}
\end{figure*}

\subsection{Wilson-Cowan neurons} 
\label{sec_neurons}
We next consider a simplified model of neural network dynamics, the Wilson-Cowan model~\cite{Wilson1972,PrezCervera2019}.
The model describes the dynamics of the average firing rates $x_i$ of coupled neurons, which evolve according to
\begin{equation}
    \dot{x}_{i} = -x_{i} + \tanh\left(\sum_j A_{i j} x_{j} + h_{i}\right)
    \label{eq:wceq}
\end{equation}
where the indices $i,j=1,\dots,N$ label neurons, $h_{i}$ is an external forcing, and $A_{ij}$ contains the couplings between the neurons.
It is assumed that each neuron $j$ is either excitatory, in which case it increases the firing rates of the neurons it influences ($A_{i j} \geq 0$ for all $i$) or inhibitory, in which case it decreases it ($A_{i j} \leq 0$ for all $i$).
The coupling between an excitatory neuron and an inhibitory neuron is therefore non-reciprocal: more precisely, $A_{i j}$ and $A_{j i}$ have opposite signs.
Here, we consider the case of one excitatory plus two inhibitory Wilson-Cowan neurons and choose to use the coupling $A_{1 1} = \zeta$ as the bifurcation parameter.
Here, the angle between CLVs generally depends on the position on the limit cycle, because it is not a circle, except if the angle is zero (i.e., if two CLVs are parallel). 
To obtain comparable quantities for different values of the parameters, we arbitrarily select a particular point on the cycle (the one that maximizes one of the coordinates) for each value of $\zeta$.
The three limit cycles coalesce at a critical value $\zeta_{\text{c}}$ where a bifurcation occurs. The bifurcation is marked in Fig.~\ref{fig:wilsoncowan} by a dotted line, where two Lyapunov exponents combine, while the angle between the corresponding CLVs goes to zero.
The behavior of the Wilson-Cowan model \eqref{eq:wceq} is phenomenologically similar to the pitchfork of limit cycles with $h(w) = w^2$ analyzed in the previous section (compare Fig.~\ref{figure_pitchfork_hopf}c-d with Fig.~\ref{fig:wilsoncowan}e-f). 
This suggests that a slow manifold reduction of Eq.~\eqref{eq:wceq} would produce Eq.~\eqref{hopf_pitchfork} near the bifurcation.

The merging and splitting of limit cycles has been suggested as a toy model of perceptual bistability for temporally periodic stimuli~\cite{PrezCervera2019,Rankin2015,Pressnitzer2006}.
A simple example consists in a sequence of tones A and B with different frequencies repeated in ABA... patterns: the same sound can be heard as two separate sequences A-A-A-... and B-B-B-... with different periods, or as a single ABA-ABA-... sequence, and one can switch from one perception to the other and back. 
You can hear a demonstration in Ref.~\cite{perceptual}. There is only one signal, but our brains can interpret it in different ways. 
These two different perceptions are distinguished by whether the tones are perceived to be synchronized or not. This information can be encoded in the in-phase and out-of-phase states of two coupled oscillators, which could therefore provide a mechanism for perceptual bistability~\cite{PrezCervera2019}. 
In this context, the coalescence of two attractors corresponds to the confusion between two percepts, that become indistinguishable at the bifurcation. 

\subsection{Rosenzweig-MacArthur prey-predator model}
\label{sec_prey_predator}
The notion of non-reciprocity is perhaps nowhere as striking as in ecology: predators eat prey, but prey rarely eat predators.
Based on the results of section \ref{coupled_nonlinear_oscillators}, we expect that a non-reciprocal coupling may affect the phase coordination of the species. This phase coordination is believed to influence biodiversity in prey-predator and other consumer-resource models~\cite{Vandermeer2006,Chesson2000,Armstrong1980,Beninca2009}. Essentially, the idea is that in-phase synchronization of, say, two predators opens a periodic opportunity for a third one to enter the system. However, the invading species may modify the phase coordination of the two original species, potentially breaking the conditions for invasion~\cite{Vandermeer2006}. This motivates a precise analysis of the phase coordination before any invasion.

\begin{figure*}
    \centering
    \includegraphics[width=\textwidth]{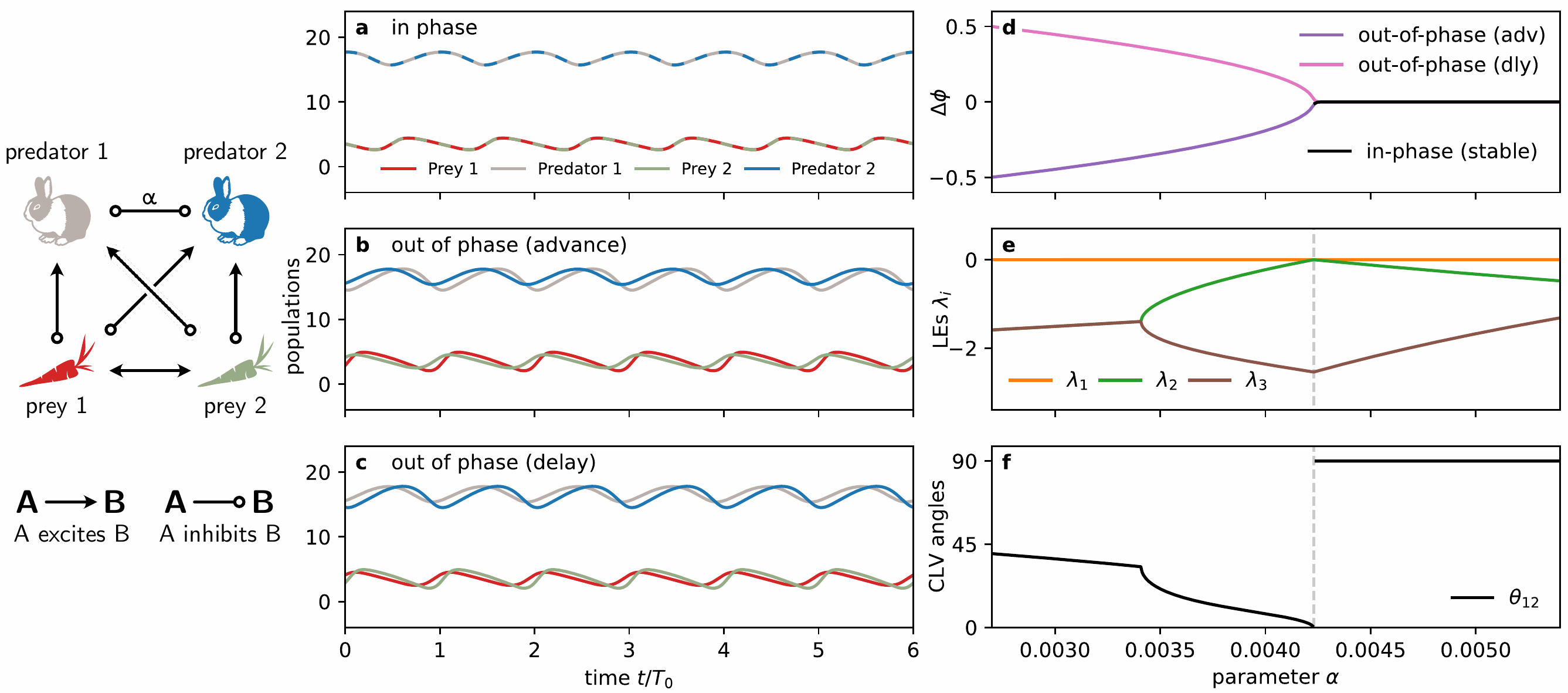}
	\caption{\label{figure_ecology}
	\strong{Rosenzweig-MacArthur prey-predator model.}
	We simulate the Rosenzweig-MacArthur model in Eq.~\eqref{eq:ecology}, with two prey and two predators. 
	The size of the populations of prey and predators evolve periodically in time (panels a-c). When $\alpha > \alpha_{\text{c}} \approx \num{0.00423}$, the prey are in phase with each other, and the predators in phase with each other (panel a). When $\alpha < \alpha_{\text{c}}$, there is a dephasing $\Delta \phi$ between the two prey (and the two predators), positive or negative depending on the initial condition: two stable attractors are present.
	Panel d shows the phase difference between the two prey  as a function of $\alpha$ (black line for the in-phase state of panel a; purple and pink lines for the out-of-phase states of panels b-c, respectively phase advance (adv) or delay (dly)).
	At $\alpha = \alpha_{\text{c}}$ (grey dashed lines), 
	a generalized exceptional point occurs.
	At the bifurcation,  both the Lyapunov exponent $\lambda_2$ (panel e) and the angle $\theta_{12}$ between the CLVs ${c}_1$ and ${c}_2$ (panel f) vanish. The angle $\theta_{12}$ is discontinuous at $\alpha_{\text{c}}$.
We have set $s=12$, $r = 12$, $k = 15.6$, $u_0 = 8$, $\epsilon=0.1$, and $\alpha$ is varied from \num{0.0027} to \num{0.0054}.
	In panel a-c, the initial conditions are $\vec{x}(0) =(u_1(0),v_1(0),u_2(0),v_2(0)) = {2.8, 16.2, 4.9, 14.7}, $ and $\vec{x}(0) = {4.9, 14.7, 2.8, 16}$. In panel a, $\alpha = 0.0054$. In panels b-c, $\alpha = 0.0024$.
    For technical reasons, we didn't track the unstable limit cycle.
	}
\end{figure*}

Here, we consider two coupled versions of a minimal but realistic predator-prey model called the Rosenzweig-MacArthur model~\cite{Turchin2013,Maynard1978,Murray2013,Rosenzweig1963}.
In our toy model, two kinds of prey and two kinds of predators are present. The prey are kind and hence they collaborate with each other and increase their respective populations in a mutualistic way.  In contrast, the predators are unkind, so they compete and kill each other.
This model is described by the coupled differential equations

\begin{equation}
    \begin{split}
    \dot{u}_1 &= r (u_0 - u_1) u_1 - k \frac{u_1 v_1}{1 + u_1} + \epsilon r (u_0 - u_2) u_2 \\
    \dot{v}_1 &= k \frac{u_1 v_1}{1 + u_1} - s v_1 - \alpha v_2 v_1 \\
    \dot{u}_2 &= r (u_0 - u_2) u_2 - k \frac{u_2 v_2}{1 + u_2} + \epsilon r (u_0 - u_1) u_1 \\
    \dot{v}_2 &= k \frac{u_2 v_2}{1 + u_2} - s v_2 - \alpha v_1 v_2
    \end{split}
	\label{eq:ecology}
\end{equation}
where $u_1$ and $u_2$ are the numbers (or concentrations) of prey, while $v_1$ and $v_2$ are the numbers of predators. The coefficient $r$ measures the natural growth of prey ($u_0$ sets the capacity of the environment), $s$ measures the natural decline of predators, and $k$ measures predation, which reduces the number of prey while increasing the number of predators. Finally, $\epsilon$ measures the collaboration between prey and $\alpha$ measures the direct competition between predators.
Again, we observe that the Lyapunov vectors become parallel at the bifurcation.
The behavior of the Rosenzweig-MacArthur model \eqref{eq:ecology} is phenomenologically similar to the pitchfork of limit cycles with $h(w) = w^2$ analyzed in the previous section (compare Fig.~\ref{figure_pitchfork_hopf}c-d with Fig.~\ref{figure_ecology}e-f). 

\begin{figure*}
    \centering
    \includegraphics[width=\textwidth]{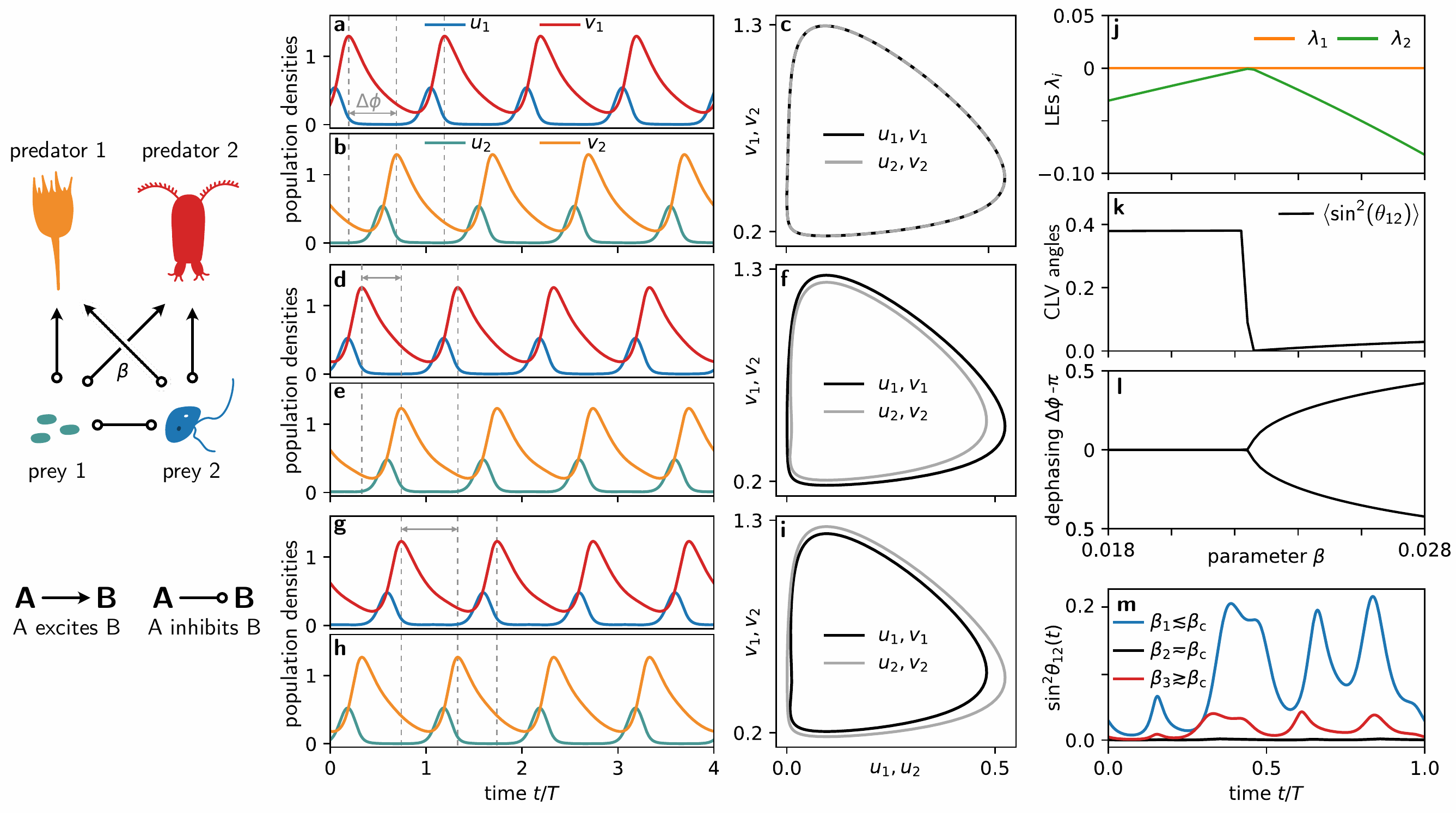}
    \caption{\label{figure_model_plankton}
    \strong{Model of plankton food web.}
    We simulate a set of coupled Rosenzweig-MacArthur-like equations given by Eq.~\eqref{rma_pfw} and used in Ref.~\cite{Beninca2009} to model laboratory experiments on a plankton food web isolated from the Baltic Sea \cite{Beninca2008}. 
    This model describes the population densities of prey $u_i(t)$ and of predators $v_k(t)$, and involves a parameter $\beta$ that quantifies the coupling through predation of the two coupled prey-predator systems.
    Below a critical value $\beta_{\text{c}} \simeq \num{0.02275}$, a single limit cycle exists (panels a-c). Two stable limit cycles exist above the critical value, as shown in panels d-f and g-i, respectively.
    At the bifurcation,  both the Lyapunov exponent $\lambda_2$ (panel j) and the angle $\theta_{12}$ between the CLVs ${c}_1$ and ${c}_2$ (panel k) vanish. The angle $\theta_{12}$ is discontinuous from the left at $\beta_{\text{c}}$.
    Parameter values and initial conditions are given in the text of App.~\ref{app_plankton}, after Eq.~\eqref{rma_pfw}.
    }
\end{figure*}

Before the bifurcation (right side in Fig.~\ref{fig:wilsoncowan}d-f), the two species of prey are synchronized together (and the two kinds of predators are synchronized together). After the bifurcation (left side in Fig.~\ref{fig:wilsoncowan}d-f), this is not the case, and the two different limit cycles correspond to different sequences of the local maxima of the different populations (e.g., $u_1, u_2, v_1, v_2$ versus $u_2, u_1, v_2, v_1$). This is illustrated in Fig.~\ref{figure_ecology}.
Again, the splitting of the limit cycle leads to a finite phase delay between the two populations, that are neither in phase nor in anti-phase. In the steady-state, the dephasing takes two possible values corresponding to the two stable limit cycles, depending on which populations is in advance. While our model is a simplified version of what could take place in realistic situations, we expect this phenomena to occur in nature.
For instance, non-trivial phase delays (neither zero nor $\pi$) have been observed in a laboratory experiment probing the food web structure of a plankton community isolated from the Baltic Sea \cite{Beninca2008,Beninca2009}. In this case, the preys are phytoplankton species (picocyanobacteria and nanoflagellates), while the predators are zooplankton (rotifers and calanoid copepods). In Ref.~\cite{Beninca2009}, this situation is modelled using coupled Rosenzweig-MacArthur-like equations with more realistic couplings~\cite{Beninca2009,Vandermeer2004,Vandermeer1993}. In Appendix \ref{app_plankton}, we show that this model also exhibits a generalized exceptional point marking the bifurcation between a single limit cycle with populations in anti-phase, and two stable limit cycles with a non-trivial phase delay, as summarized in Fig.~\ref{figure_model_plankton}. Our analysis of the model of Ref.~\cite{Beninca2009} suggests the existence of two different limit cycles. The non-trivial dephasings observed in the experiment is indirect evidence supporting this statement. In principle, it could be directly verified experimentally. We note that it would have consequences on the correlations between species' abundances. In particular, two different results could be obtained for the same values of the couplings. In section \ref{consequences}, we will show that the proximity to a generalized exceptional point typically leads to an increased effect of noise: it is therefore intriguing to ask what would be the consequences of stochastic fluctuations~\cite{Xue2017} on the ecological model.

\subsection{Pitchfork of Lorentz chaotic attractors}
\begin{figure}[t]
    \centering
    \includegraphics[width=8cm]{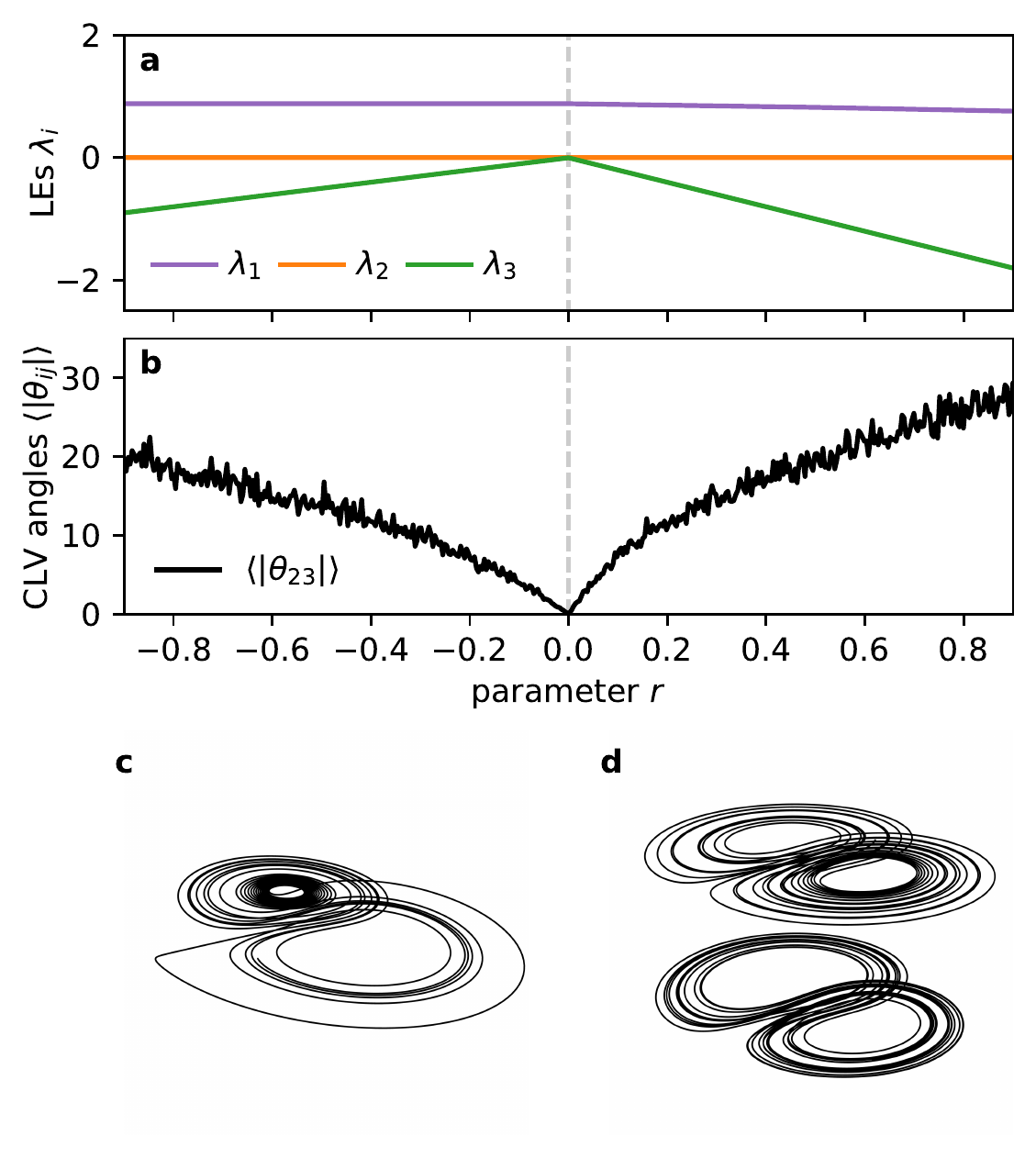}
	\caption{\strong{Pitchfork of chaotic attractors.}
	Following the general procedure of Eq.~\eqref{eq_pitchfork_attractors}, we construct a pitchfork of chaotic attractors by controlling the Lorenz system~\cite{Lorenz1963,Ott2002} by a pitchfork. The full dynamical system is given in Eq.~\eqref{eq:chaos}.
	The Lyapunov exponents are plotted with respect to the bifurcation parameter $r$ in panel a.
	The angle between the covariant Lyapunov vectors ${c}_2$ and ${c}_3$, averaged over the trajectory, is plotted in panel b.
	On one side of the bifurcation (for $r<0$), a single chaotic attractor is present (panel c). On the other side ($r>0$), two chaotic attractors coexist (panel d), in addition to a chaotic repeller (not pictured in panel d).
	At the bifurcation ($r=0$), the Lyapunov exponents $\lambda_2$ and $\lambda_3$ (orange and green curves in panel a) vanish, while the angle between the corresponding CLVs also vanishes (panel b).
	We have set $\sigma = 10$, $\beta = 8/3$, $\rho = 28$, $\gamma = 5$ in Eq.~\eqref{eq:chaos}.
    Equation \eqref{eq:chaos} is numerically integrated using \texttt{DifferentialEquations.jl} \cite{Rackauckas2017}.
    The LEs and CLVs are computed using the algorithms of Refs.~\cite{Benettin1980,Benettin1980b,Ginelli2007,Ginelli2013}.
	The average angles are computed by randomly choosing points along the chaotic attractor.
	We have set $r = \num{-1}$ in panel c and
	$r = \num{100}$ in panel d. 
	} 
	\label{fig:lorenz_EP}
\end{figure}

To conclude this section, we go beyond limit cycles and consider the symmetric coalescence of more complex attractors. As an example, we now discuss a pitchfork bifurcation of chaotic attractors. (We also refer to appendix section \ref{appendix_tori_higher} for a discussion on limit tori.)
To do so, we take $f$ in Eq.~\eqref{eq_pitchfork_attractors} to be a Lorenz attractor, which is a prototypical example of chaos~\cite{Lorenz1963,Ott2002}. Accordingly, we consider~\footnote{The model \eqref{eq:chaos} does not directly describes coupled chaotic attractors. The parallels between this situation and the case of limit cycles suggest that a bifurcation accompanied by a coalescence of covariant Lyapunov attractors could occur in coupled chaotic attractors, maybe in relation with their synchronization~\cite{Boccaletti2002,Pecora1990}. However, this is not analyzed in this work.
Relations between Lyapunov vectors and attractor merging crises~\cite{Grebogi1982,Grebogi1983,Grebogi1987}, that occur when attractors collide with each other have also been put forward~\cite{Beims2016,Tantet2017}.}
\begin{equation}
    \begin{split}
\label{eq:chaos}
        \dot{x} &= \sigma(y - x) + \gamma w \\
        \dot{y} &= x(\rho - z) - y + \gamma w \\
        \dot{z} &= xy - \beta z + \gamma w \\
        \dot{w} &= rw - w^{3}
\end{split}
\end{equation}
where we have chosen a linear coupling  $g(\vec{x}, w) = \gamma \, w$.
Due to the non-periodicity of a chaotic attractor, there is no simple way to select a distinguished point where to compute the CLV for each choice of $r$, as we did for limit cycles.
Instead, we compute the average $\langle|\theta_{ij}|\rangle$ over the attractor of the angles $\theta_{ij}$ between the CLVs ${c}_i$ and ${c}_j$.
As the CLVs are covariant with respect to the dynamics, they are parallel all along the attractor if they are parallel at one point. Figure~\ref{fig:lorenz_EP}a-b shows that the CLVs become parallel at the bifurcation point $r=0$ where the chaotic attractors merge. 
On one side of the bifurcation, a single chaotic attractor is present (Fig.~\ref{fig:lorenz_EP}c), while two of them coexist on the other side (Fig.~\ref{fig:lorenz_EP}d).

\section{Physical consequences and their experimental manifestations}
\label{consequences}

In this section, we discuss the physical consequences of generalized exceptional points. First, we discuss how a system close to a generalized EP responds to small perturbations in Sec.~\ref{atoms_nr}. The key point is that this response is not reciprocal (Sec.~\ref{atoms_nr}). This occurs irrespective of the value of the Lyapunov exponent (finite or not) at the generalized exceptional point.  
When the corresponding Lyapunov exponent vanishes, this mechanism may lead to a pileup of perturbations along a certain direction in tangent space. This occurs because the non-reciprocity of the response is combined with the softening (critical slowing down) typical of criticality near a bifurcation~\cite{Munoz2018}.
In particular, this can lead to the destruction of isochrons, in which points in the bassin of a limit cycle do not have a well-defined asymptotic phase (Sec.~\ref{no_isochrons}). We also consider the effect of noise on a system near a generalized exceptional point (Sec.~\ref{lota_noise}). We find that the flatness of the response leads to an increased sensitivity to noise (Sec.~\ref{noise_flat}). In addition, on one side of the bifurcation, the presence of multiple attractors with different properties leads to the generation of telegraph noise from the white noise to which the full system is submitted (Sec.~\ref{telegraph_noise}). As an example, this conversion of white noise into telegraph noise has visible experimental consequences for a solid body rotating in a convection cell, see Sec.~\ref{free_rotation}.

\subsection{Generalized EPs as atoms of non-reciprocity}
\label{atoms_nr}

\begin{figure}[t]
    \centering
    \includegraphics[width=8cm]{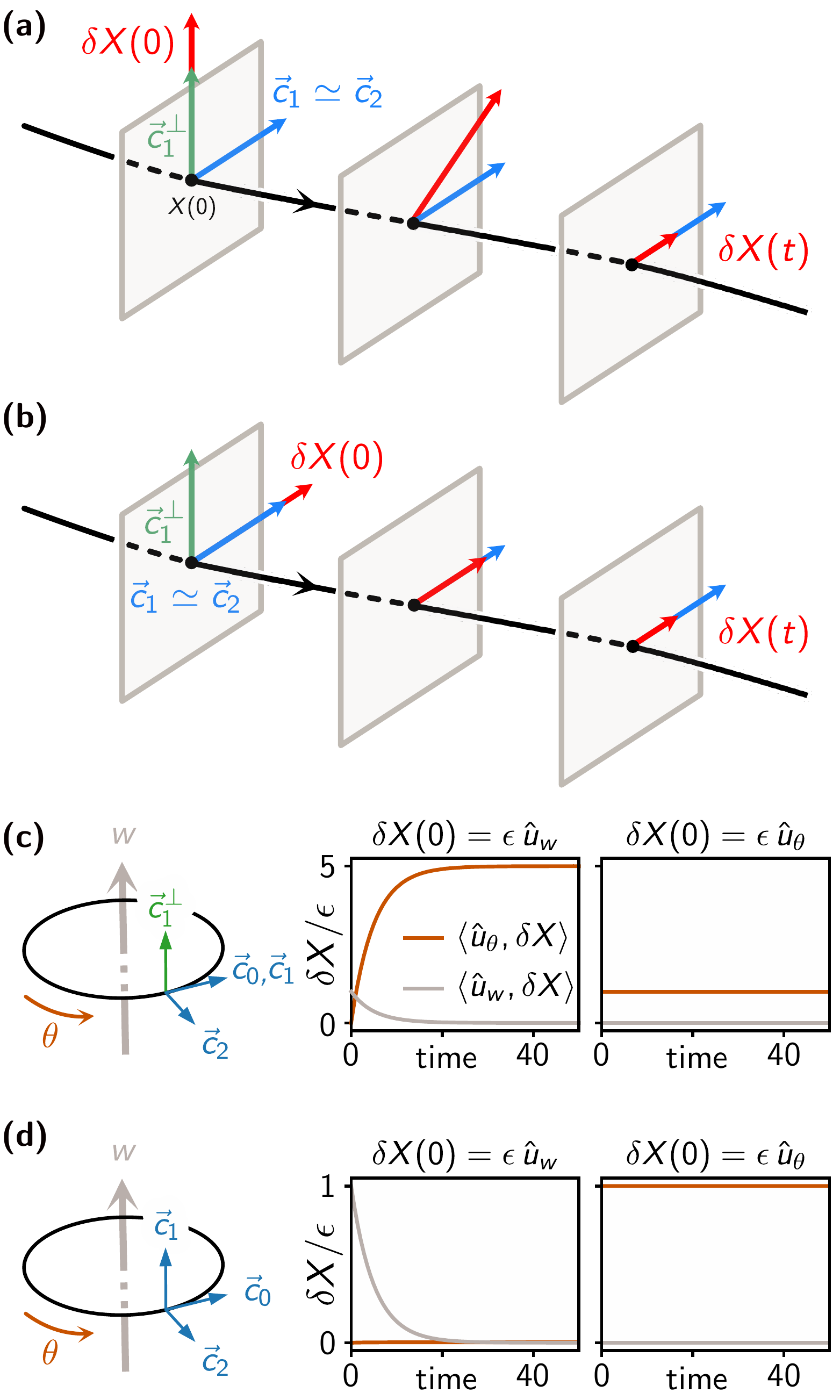}
    \caption{\label{figure_evolution_perturbation}
    \strong{Effect of a generalized exceptional point on perturbations.}
    (a-b) When two CLVs $\vec{c}_1$ and $\vec{c}_2$ (in blue) are approximately parallel, there is a \enquote{missing} direction $\vec{c}_1^\bot$ (in green) in tangent space (light gray square), approximately orthogonal to $\vec{c}_1$ and $\vec{c}_2$ \cite{NoteCaptionNRP}.
    In panels a-b, we sketch the evolution $\delta X(t)$ of a perturbation $\delta X(0)$ (in red) along the trajectory (black) in this situation. The relaxation of perturbations is non-reciprocal.
    In panel a, the perturbation initially along $\vec{c}_1^\bot$ gradually aligns with $\vec{c}_1$.
    In panel b, the perturbation initially along $\vec{c}_1$ stays along this direction.
    The evolution of the amplitude depends on the Lyapunov exponents.
    (c-d) The non-reciprocal relaxation described in panels a-c relies on the angles between CLVs. This allows us to distinguish between the continuous and discontinuous behaviors discussed in Sec.~\ref{pflc}.
    We simulate Eq.~\eqref{hopf_pitchfork} for (c) the continuous case $h(w) = w$ (like in Fig.~\ref{figure_pitchfork_hopf}a,b) and (d) the discontinuous case $h(w) = w^2$ on the discontinuous side of the bifurcation (like in Fig.~\ref{figure_pitchfork_hopf}c,d). 
    We compare a trajectory on the limit cycle to a perturbed trajectory: a perturbation $\Delta X(t=0) \sim \epsilon$ along $w$ is added as an initial condition, in the limit where $\epsilon \to 0$. The non-reciprocal relaxation does not occur when the CLVs are orthogonal (panel d).
    We have taken $r = \num{-0.2}$, $\alpha = 1 + \ii$, $\beta = -1$, $\gamma = 1 + \ii$, and $\epsilon = \num{0.001}$.
    }
\end{figure}

The tangency of two CLVs $\vec{c}_1$ and $\vec{c}_2$ leads to a non-reciprocal coupling between perturbations about the corresponding directions. 
This arises irrespective of whether the corresponding Lyapunov exponents are finite or not. 
Far away from tangencies, when $\vec{c}_1$ and $\vec{c}_2$ are approximately orthogonal, a perturbation $\delta X$ initially along $\vec{c}_1(0)$ essentially stays along $\vec{c}_1(t)$ (same for $\vec{c}_2$). 
On the other hand, when $\vec{c}_1 \simeq \vec{c}_2$, a non-reciprocal response emerges.
To illustrate this, let us consider a situation where a perturbation along the direction $\vec{c}_1^\bot$ orthogonal to the two approximately parallel CLVs $\vec{c}_1 \simeq \vec{c}_2$, as represented in Fig.~\ref{figure_evolution_perturbation}. 
In such a case,
it will eventually align with $\vec{c}_1$ (Fig.~\ref{figure_evolution_perturbation}a). The converse is not true: a perturbation along $\vec{c}_1$ stays along $\vec{c}_1$ (Fig.~\ref{figure_evolution_perturbation}b). 
In Fig.~\ref{figure_evolution_perturbation}c-d, we compare the fate of infinitesimal perturbations of a trajectory of Eq.~\eqref{hopf_pitchfork} in the two cases of Fig.~\ref{figure_pitchfork_hopf}, corresponding to continuous and discontinuous behaviors of the CLVs. As expected, there is no mode interconversion in the discontinuous case, because the CLVs are orthogonal on this side of the bifurcation. This illustrates that the covariant Lyapunov vectors, not the Lyapunov exponents~\footnote{Lyapunov exponent are asymptotic quantities describing the growth rate at large times. By definition, they do not pick up the transient effects we discuss here. These can be captured more easily by so-called finite-time Lyapunov exponents and similar quantities. It is remarkable that the long-time CLVs are still a partial proxy for some transient effects.}, are at the origin of the non-symmetric mode interconversion.

\subsection{EP-induced destruction of isochrons}
\label{no_isochrons}

Points starting in the basin of attraction of a limit cycle eventually end up on the limit cycle. One may ask what is the eventual dephasing between two such points. The answer to this question is provided by the notion of isochrons~\cite{Kuramoto1984,Winfree2001,Guckenheimer1975,Winfree1974,Mauroy2012,Mauroy2013,Mauroy2016}. 
Under certain conditions, the notion of isochron can be extended to transient dynamics such as fixed points or excitable systems~\cite{Mauroy2012,Mauroy2013,Ichinose1998,Shirasaka2017,Wilson2015}, heteroclinic orbits~\cite{Shaw2012}, or chaotic attractors~\cite{Rosenblum1996,Schwabedal2012,Tonjes2022}.
The main idea is that the $n$-dimensional basin of attraction of the limit cycle can be decomposed into slices invariant under one period of evolution.
These slices, called isochrons, essentially consists of all the points that eventually have a certain asymptotic phase.
In more technical terms, isochrons provide a foliation of the basin of attraction by a one-parameter family of $(n-1)$-dimensional hypersurfaces parameterized by the asymptotic phase $\phi \in S^1$.
Isochrons have been used to analyze the response and controllability of oscillators, including their transient dynamics, excitability, and synchronization~\cite{Himona2022,Mauroy2013,Shirasaka2017}. They are also important tools to perform model reduction~\cite{Nakao2014,Kuramoto1984}.

In this section, we show that the presence of a generalized exceptional point can lead to the destruction of isochrons. As we have seen in section \ref{zero_CLV}, the CLV $c_* = \dot{X}$ with Lyapunov exponent $\lambda_* = 0$  associated with time-translation invariance is, by definition, tangent to the limit cycle.
Away from (generalized) exceptional points, it can be shown that the hyperplane tangent to the isochron at position $X$ forms the eigenspace spanned by the $n-1$ CLVs (or Floquet eigenvectors) with finite Lyapunov exponents (see for instance Ref.~\cite[\S~3.4]{Kuramoto1984}). 

\begin{figure}[t]
    \centering
    \includegraphics[width=8cm]{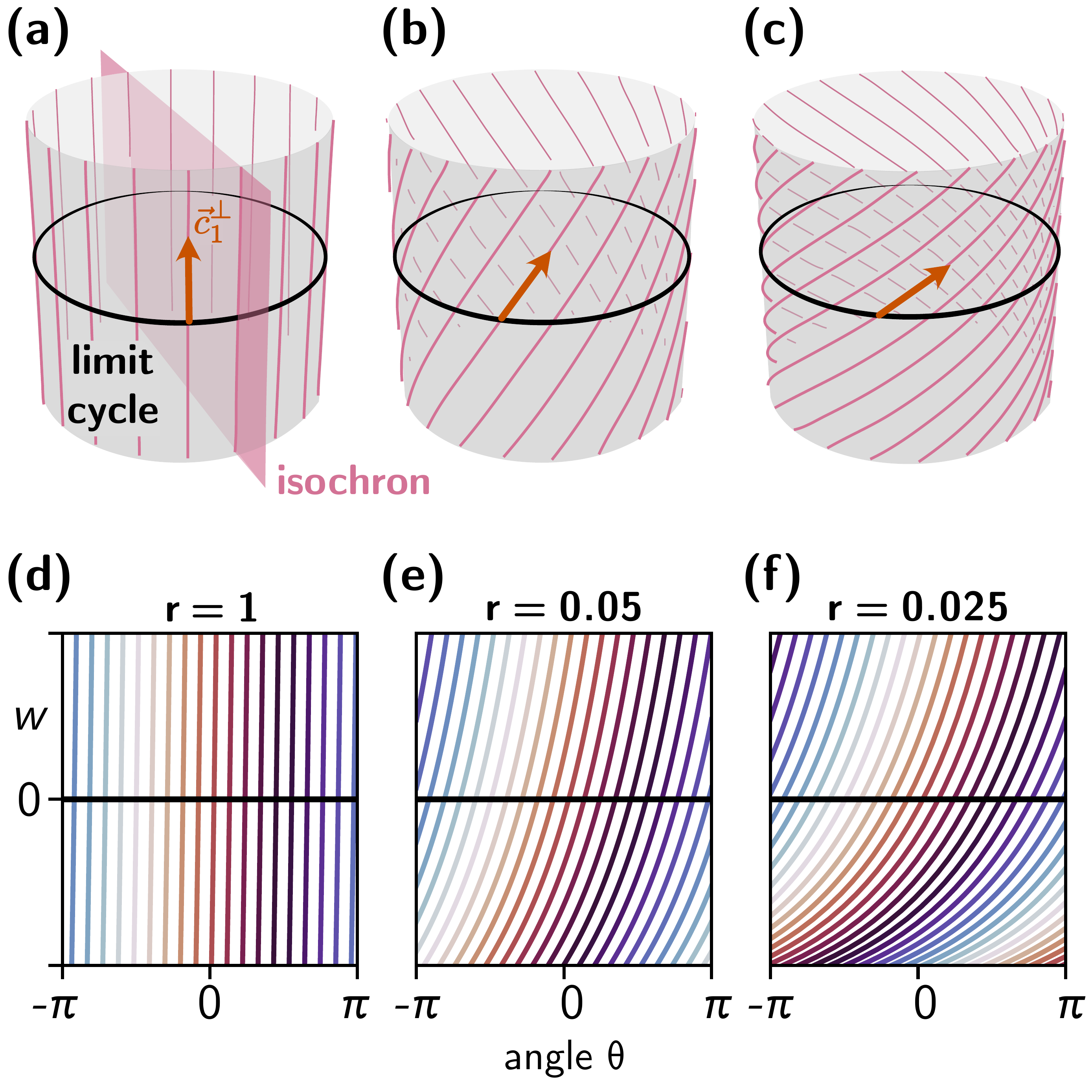}
    \caption{\label{figure_isochrons_schematic}
    \strong{Isochrons and Floquet vectors.}
    In panels a-c, we schematically show intersections of the isochrons with a cylinder (in purple) in the normal form Eq.~\eqref{noisy_parity_breaking} when approaching a generalized exceptional point (panels a,b,c, correspond to increasing closeness to the generalized EP). The limit cycle is drawn in black. The CLV $c_2$ that coalesces with $c_*$ is drawn in red.
    One full isochron (a hypersurface of codimension one) is also represented in purple in panel a.
    In panels d-f, we show slices of the isochrons of the dynamical system in Eq.~\eqref{hopf_pitchfork} computed numerically.
    The limit cycle located at $w=\sqrt{r}$ is shown as a thick black line.
    As the generalized exceptional point at $r=0$ is approached (from d to f), isochrons tilt towards the limit cycle.
    In the limit $r \to 0$, the isochrons are not transversal to the limit cycle anymore and cease to be well-defined.
    Colors label different isochrons, and the $w$ axis ranges from $\sqrt{r} - \num{0.1}$ to $\sqrt{r} + \num{0.1}$.
    Isochrons are computed using the method of Ref.~\cite{Mauroy2012}. 
    We have set $\alpha = 1 + \ii$, $\beta = -1$, $\gamma = \ii$, $h(w) = w$, and (d) $r=\num{1}$, (e) $r=\num{0.05}$, (f) $r=\num{0.025}$.
    }
\end{figure}

As a CLV with finite exponent tends to align with $c_*$, the isochrons become more and more tangent to the limit cycle, as represented schematically in Fig.~\ref{figure_isochrons_schematic}a-c.
At an exceptional point, the isochrons are not well-defined anymore. Equivalently, the asymptotic phase is not well-defined \footnote{In Ref.~\cite{Guckenheimer1975}, it is proven that isochrons exist under the hypothesis that the limit cycle is hyperbolic (this means that the only Floquet multiplier on the unit circle corresponds to the Floquet eigenvector $\vec{c}_*$ with multiplier $\mu = 1$, see Ref.~\cite{Kuznetsov2004}). Non-hyperbolic limit cycles typically have no well-defined isochrons, but this is not necessarily true~\cite{Castejon2013,Freire2007}. Conversely, limit cycles without well-defined isochrons are necessarily non-hyperbolic. This is because the hyperbolicity is defined in terms of Floquet multipliers (or Lyapunov exponents), while the existence or absence of isochrons is related to the Floquet eigenvectors (or CLVs). Namely, the isochrons cannot be well-defined when another CLV aligns with $\vec{c}_*$. For instance, an exactly solvable planar system without isochrons can be found in Ref.~\cite{Demir2007}, in which the monodromy matrix [Eq~. (13) in the reference] is a Jordan block of size two. We also refer to Refs.~\cite{Tantet2020,Tantet2019,Chekroun2020} for a discussion of the effect of noise on an oscillator with twisted isochrons from the point of view of so-called Ruelle-Pollicott resonances.}. 
Figure~\ref{figure_isochrons_schematic}d-f shows isochrons computed using the method of Ref.~\cite{Mauroy2012} for the dynamical system defined by Eq.~\eqref{hopf_pitchfork}.
The destruction of isochrons has potentially drastic consequences: even tiny perturbations lead to unbounded dephasings at long times~\cite{Demir2007}.

\subsection{Effect of noise}
\label{lota_noise}

We now show that adding noise near a generalized exceptional point can produce striking effects: noise is amplified with an arbitrarily large factor, and colored noise is generated from white noise.
More precisely, on one side of the bifurcation, a random telegraph noise is generated as the system jumps between the two stable attractors heralded by the generalized EP.
We illustrate these effects in the case of limit cycles.

To do so, we now consider a noisy version of the normal form in Eq.~\eqref{parity_breaking} (and Eqs.~\eqref{hopf_pitchfork} and \eqref{pitchfork_lc_polar})
\begin{subequations}
\label{noisy_parity_breaking}
\begin{align}
    \dot{w} &= r w - w^3 + \eta_w(t) \\
    \dot{\phi} &= \omega_0 + \omega_1 \, w + \eta_\phi(t)
\end{align}
\end{subequations}
in which $\omega_0$ and $\omega_1$ are characteristic frequencies, and $\eta_w$ and $\eta_\phi$ are Gaussian noises satisfying $\braket{\eta_i(t)} = 0$ and 
$\braket{\eta_i(t)\eta_j(t')} = 2 T_i \delta_{i j} \delta(t-t')$ ($i,j$ label $w, \phi$).
We assume that $\omega_0 = 0$ (the $\omega_0 \neq 0$ can be reduced to this one by going in a comoving frame).
When $\omega_1 = 0$, the two equations are decoupled. The random variable $\phi$ is the Wiener process corresponding to $\eta_\phi$, so it has zero mean and its variance scales as $t$. 

Now, let us consider the case in which $\omega_1 \neq 0$. As the equation for $w$ is fully decoupled from the $\phi$ equation, we can proceed in two steps. First, note that the variable $w$ describes a Brownian particle in the potential $V(w) = - [r w^2/2 - w^4/4]$.
We consider separately the cases where $r<0$ (Sec.~\ref{noise_flat}) and $r>0$ (Sec.~\ref{telegraph_noise}).

\subsubsection{Enhancement of noise by generalized EPs}
\label{noise_flat}

We first consider the situation where $r \leq 0$. In this case, the potential has a single minimum at $w=0$. In a nutshell, the noise on $w$ is converted into noise in $x$ with an arbitrarily large (or small) conversion ratio (for instance, proportional to $\omega_1$ in Eq.~\eqref{noisy_parity_breaking}). Approaching the generalized exceptional point ($r \to 0$), the potential landscape for $w$ becomes flatter, affecting the properties of the noise.
What are the consequences on fluctuations of the variable of interest $\phi$? Neglecting the additional noise $\eta_\phi$, a simple calculation (see Appendix \ref{app_noise}) shows that $\braket{\phi}^2 = 0$ and that in the limit where $|r| \to 0$, the variance of $\phi(t)$ at long times is
\begin{equation}
    \label{phase_diffusion}
    \braket{\phi^2} \sim A \omega_1^2 \, t
\end{equation}
in which $A \simeq \num{0.97499}$ is a prefactor of order one.
By analogy with particle diffusion, the slope $D$ in the relation $\braket{\phi^2} \sim D \, t$ is called a phase diffusion coefficient~\cite{Kuramoto1984}. 
We have seen that here, $D = A \omega_1^2$.
Notably, the effective phase diffusion coefficient does not depend on $T_w$. As a point of comparison, we find that $D = \omega_1^2 (2 T_w/r^2)$ when $r < 0$ is large enough to neglect the non-linear term (in this limit, $w$ is an Ornstein-Uhlenbeck process).
Note that the phase diffusion coefficient can be arbitrarily large in both cases, due to the coefficient $\omega_1$. 
The fact that the phase diffusion coefficient $D$ is independent of the strength of the noise $T_w$ may look paradoxical, as it suggests that a finite $D$ would persist with no noise at all. The paradox is only apparent: the asymptotic diffusive regime only takes place after a transient with characteristic time $T_c \sim 1/\sqrt{T_w}$ which diverges as $T_w \to 0$ (see Appendix \ref{app_noise}) \footnote{If we assume that the phase correlations are eventually diffusive (namely, $\braket{\phi(t_0+t)\phi(t_0)} \sim D t$), these results can be guessed by dimensional analysis. 
To do so, let us consider the SDEs $\dd w = r w \dd t - b w^3 + \sigma \dd W$ and $\dd \phi = \omega_1 w$ in which $W$ is a Wiener process. 
We attribute independent dimensions $[w] = \mathsf{W}$ and $[\phi] = \Phi$ to the variables $w$ and $\phi$ (here, square brackets label the dimension of a quantity). In addition, the dimension symbol of time is $\mathsf{T}$. The parameters are $[r] = \mathsf{T}^{-1}$, $[b] = \mathsf{T}^{-1} \mathsf{W}^{-2}$, $[\omega_1] = \Phi \mathsf{W}^{-1} \mathsf{T}^{-1}$, and $[\sigma] = \mathsf{W} \mathsf{T}^{-1/2}$ (recall that $[\dd W] = \sqrt{[\dd t]}$). In addition, the diffusion coefficient we are looking for has dimension $[D] = \Phi^2 \mathsf{T}^{-1}$.
First, assume $b=0$ (and $r \neq 0$); dimensional analysis yields that the only combination of parameters with the appropriate dimension is $[D] = [\omega_1^2 \sigma^2/r^2]$, so $D \sim \sigma^2 \sim T_{w}$.
Second, assume $r=0$ (and $b \neq 0$); in the same way, dimensional analysis yields $[D] = [\omega_1^2/b]$ which does not depend on the noise strength $\sigma^2$ (i.e. on the temperature $T_w$). 
A similar analysis can be performed when $\dd \phi = \omega_1^{(2)} w^2$ (this corresponds to the case of Sec.~\ref{pflc} with discontinuous CLVs); in this case, we find that dimensionally, $D \sim [\omega_1^{(2)}]^2 \sigma b^{-3/2} \sim \sqrt{T_w}$. We can therefore expect a significant increase of fluctuations near the bifurcation compared to the case far from the exceptional point where $r \gg b$ at low $T_w$. 
The general case in which both $r$ and $b$ are finite is not fully determined by dimensional analysis.
We refer to Ref.~\cite{Shmakov2023} for a more complete scaling theory.}.

\subsubsection{EP-induced telegraphic noise}
\label{telegraph_noise}

When $r > 0$, the potential $V$ is a double-well. When the noise is small enough, the particle spends a long periods in each well, separated by quick jumps between the wells. This intermittent process is approximately described by the Ahrenius-Kramers-Eyring theory and can be approximated by a telegraph process~\footnote{
More precisely, the time evolution of $w$ can be approximated as a discrete-state continuous-time Markov process with two states $w_{\pm}$ (corresponding to the two minima of $U(w)$) with transition rates $1/\tau^*$ between the states following the master equation
\begin{equation}
    \partial_t 
    \begin{pmatrix}
    p_+ \\
    p_-
    \end{pmatrix}
    =
    \begin{pmatrix}
    - \lambda_{-+} & \lambda_{+-} \\
    \lambda_{-+} & -\lambda_{+-}
    \end{pmatrix}
    \begin{pmatrix}
    p_+ \\
    p_-
    \end{pmatrix}
\end{equation}
in which $\lambda_{i j} = \lambda_{j\to i}$ is the rate of transition from $i$ to $j$, given by $\lambda_{+-} = \lambda_{-+} = 1/\tau^*$ as the double-well is symmetric.
As the transition rates are constant, the distribution of waiting times is exponential, reproducing Eq.~\eqref{expdisttimes}. 
The telegraph noise is also known as a burst noise, a Markovian dichotomous noise, or a Kac process. It is related to a Poisson process in the sense that the number of changes between the two states in an interval of duration $\tau$ is Poissonian. See Refs.~\cite{vanKampen1992,Horsthemke2006,Bena2006} for details. 
We also note that a superposition of a large number of random telegraph processes can produce a $1/f$ noise \cite{Weissman1988,Kogan1996}.
}. 
In this approximation, $w(t) = \pm 1$ for random durations $\tau_{\pm}=\tau$ (the double-well is symmetric) following an exponential distribution with probability density function (pdf)
\begin{equation}
    \label{expdisttimes}
    p(\tau) = \frac{1}{\tau^*} \, \ee^{-\tau/\tau^*}
\end{equation}
where 
\begin{equation}
    \tau^* \equiv \mathbb{E}[\tau] = \frac{2\pi}{|V''(w^*)| \, V''(w_{\pm})} \ee^{[V(w^*) - V(w_{\pm})] / T}
\end{equation}
is given by the Eyring–Kramers-Arrhenius formula
in the limit where $T \to 0$ (see \cite{Hanggi1990,Berglund2011,Bovier2004,Bovier2005,Bouchet2016} and references therein), in which $w^*$ is the position of the barrier (maximum of $V$) and $w_{\pm}$ is the position of the well (minimum of $V$), see Fig.~\ref{figure_noise_telegraph}a. Here, with $V(w) = - r w^2/2 + w^4/4$, we find
\begin{equation}
    \tau^* = \frac{\pi}{r^2} \, \ee^{r^2/4 T}
\end{equation}

The telegraph noise generated by this process is a non-Gaussian, colored noise that tends to drive out of equilibrium the system on which it is applied~\cite{vanKampen1992,Horsthemke2006,Bena2006,Hanggi1995,Weiss2002,Masoliver2017,Kutner2017,Tailleur2008,Touzo2023}.
On both sides of the bifurcation, the amplitude of the noise is proportional to the magnitude of the function $g$ in Eq.~\eqref{eq_pitchfork_attractors} which can in principle be arbitrarily large or small.

The fate of $\phi$ can be understood by first ignoring the noise $\eta_\phi$. In this case, $\dot{\phi} = \omega_1 w$.
As a consequence, two effects occur.
First, $\phi$ is composed of branches with positive and negative slopes $\omega_1 w_{\pm}$, obtained by integrating the random telegraph signal. 
The corresponding signal is highly correlated at short times, becoming diffusive only at long times.
Second, the standard deviation of $\phi$ is proportional to $\omega_1$, so it can be orders of magnitude larger than the standard deviation of the white noise $\eta_w$ acting on the variable $w$.
Both effects can be directly observed in numerical simulations of the stochastic differential equation \eqref{noisy_parity_breaking}. In Fig.~\ref{figure_noise_telegraph}, we compare (i) in panel e the case where $\omega_1 \equiv 0$ (so there is only a Gaussian white noise) and (ii) in panel f the case where $\omega_1 \gg T_\phi$ and  (so the EP-induced telegraphic noise is large compared to the pre-existing Gaussian noise). 
In case (i), $\phi$ is directly a Wiener process (i.e. a Brownian motion). Therefore, it is scale invariant and diffusive (as $\braket{\phi^2} \propto t$). Note $\braket{\phi} = 0$.
In the case (ii), $\phi(t)$ is a succession of upward and downward slopes with values $\pm w_{\text{c}} = w_\pm = \pm \sqrt{r}$, extending over durations distributed exponentially, following Eq.~\eqref{expdisttimes}. 

When the measurement time scales are small compared to the characteristic time $\tau^*$ of reversals, this succession of upward and downward slopes can be directly observed. In the next paragraph, we discuss an experimental example in fluid mechanics that we model by this mechanism. When the measurement time scale is large compared to $\tau^*$, the process we described has indirect consequences. For instance, one can measure the frequency of oscillation of the system. This is what happens in the case of coupled lasers, in which the consequence is a two-color laser characterized by a power spectrum with two peaks, as it has been experimentally observed in Ref.~\cite{Hassan2015} (see also Refs.~\cite{Clerkin2014,Soriano2013} and SI Sec.~XII of Ref.~\cite{Fruchart2021}).

\subsubsection{Experimental example: free rotation of a solid body in a convection cell}
\label{free_rotation}

\begin{figure*}
\includegraphics[width=16cm]{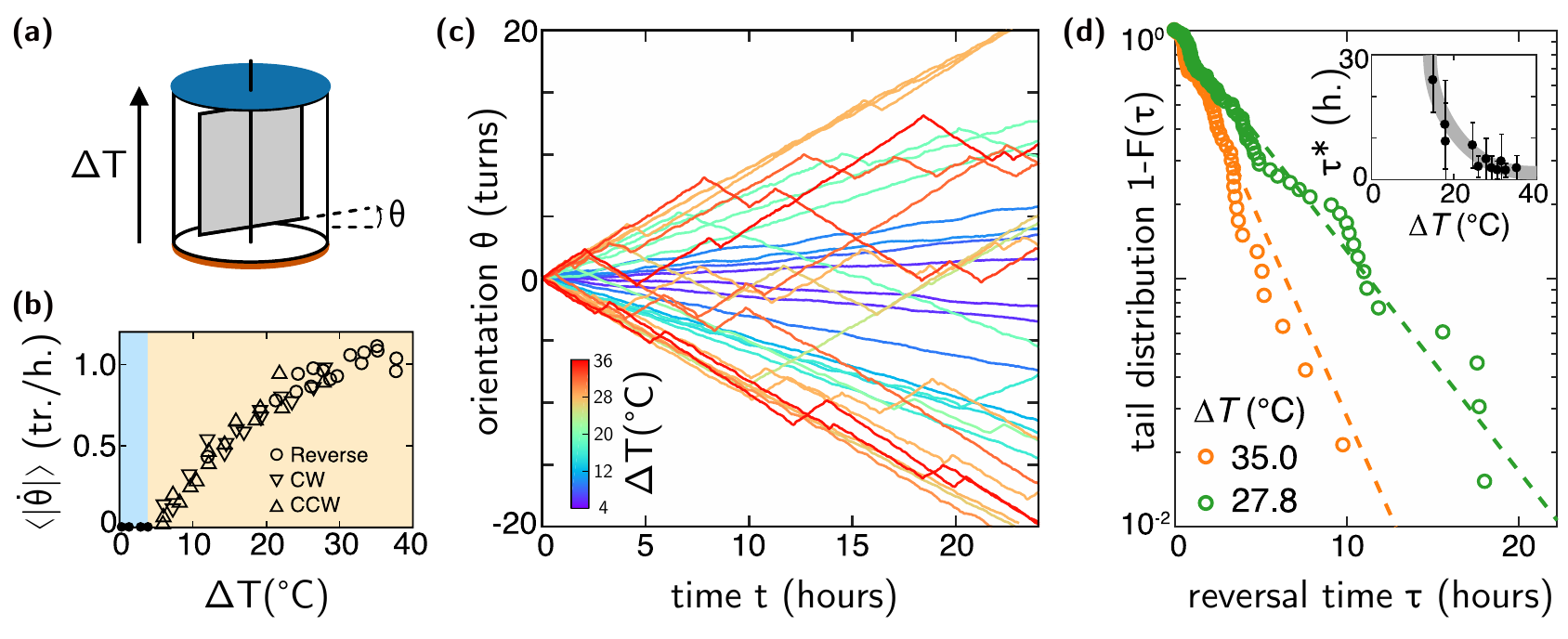}
    \caption{\label{figure_convection_experiment}
    \strong{Free rotation of a solid body in Rayleigh-Bénard convection.}
    We reproduce from Ref.~\cite{Wang2023} figures summarizing their experimental results.
    (a) Description of the system: a temperature gradient $\Delta T$ is applied to a cylindrical convection cell containing a solid body freely rotating about the axis of the cylinder. 
    The angular position $\theta$ of the body with respect to a fixed axis is recorded.
    (b) When $\Delta T$ is increased, the system bifurcates from a state with no average rotation to a state with finite average rotation speed. The magnitude of the rotation speed is determined by $\Delta T$ (it is approximately a square root of the distance to threshold) but its sign is random.
    (c) Observation of time series of the angle $\theta$ in the high-$\Delta T$ regime shows that there are frequent reversals of the rotation: the system switches between clockwise and counterclockwise motion.
    (d) The distribution of sojour times in the (counter)clockwise states is approximately exponential, as evidenced by the tail distribution ($F(\tau)$ is the cumulative distribution; the probability density function is $p(\tau) = F'(\tau)$).
    Dashed lines are fits by $F(\tau) = 1 - \exp(-\tau/\tau^*)$.
    Panels b-d are adapted from \cite{Wang2023}.
    }
\end{figure*}

As an illustration, we now compare in Fig.~\ref{figure_convection_experiment} the predictions of the normal form Eq.~\eqref{noisy_parity_breaking} to recent experimental results from Ref.~\cite{Wang2023}. In this work, a plate is immersed in a cylindrical convection cell. The plate can freely rotate about the axis of the cylinder. A temperature gradient $\Delta T$ is imposed between the top and bottom of the cell, leading to Rayleigh-Bénard convection. The rotation of the plate due to its interaction with the fluctuating flow is then monitored through its angle $\theta$ with a fixed axis. At low temperature gradients, the average rotation rate $\braket{|\dot{\theta}|}$ vanishes. 
As $\Delta T$ passes a critical value $\Delta T_{\text{c}}$, the average rotation rate $\braket{|\dot{\theta}|}$ becomes finite. In this regime, reversals in the sign of the rotation rate are observed, and are reported to follow a Poisson process.

\begin{figure}
\includegraphics[width=\columnwidth]{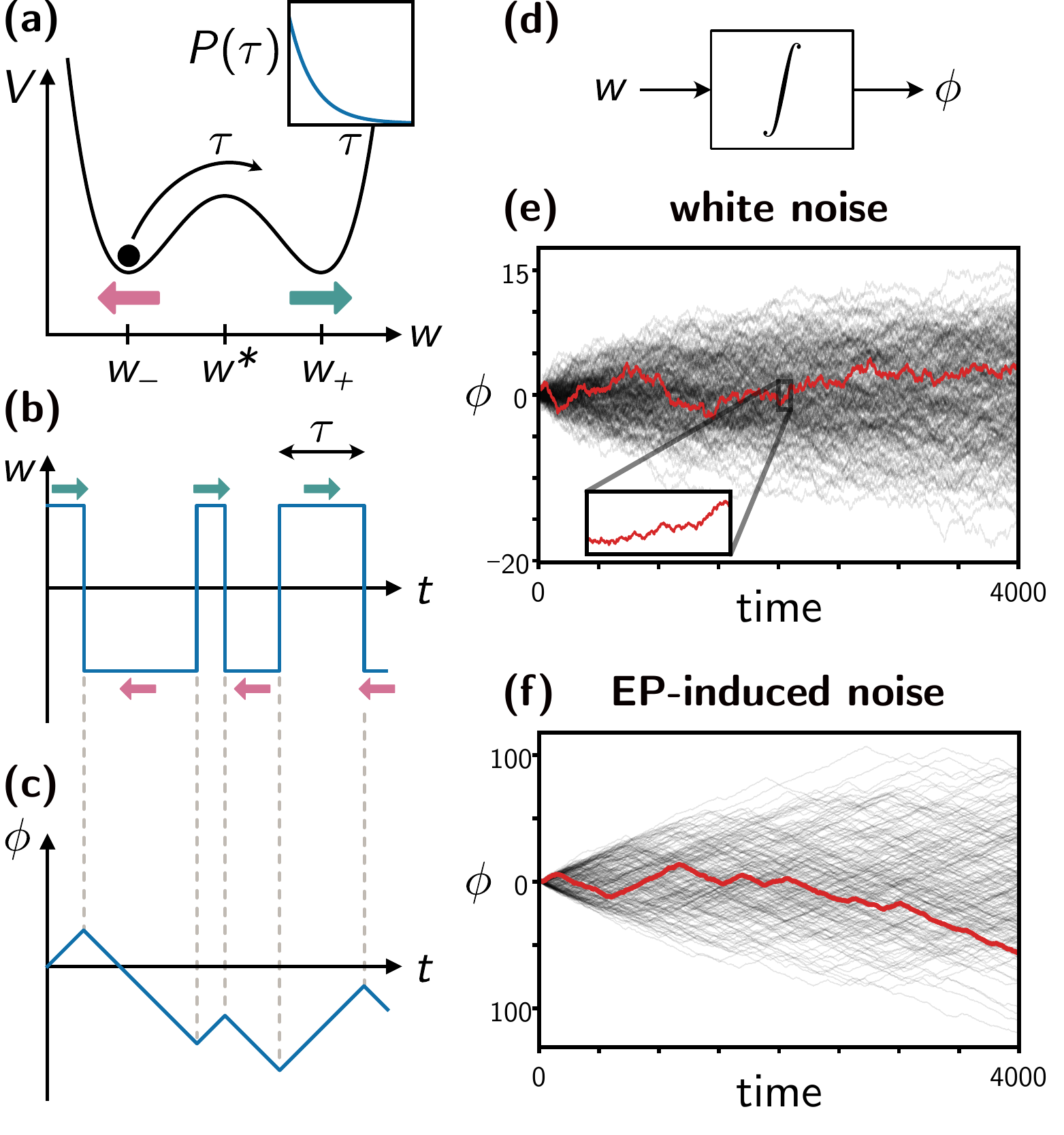}
    \caption{\label{figure_noise_telegraph}
    \strong{EP-induced telegraph noise.}
    Applying a weak noise produces stochastic jumps between attractors.
    Here, this is illustrated in the simple case of Eq.~\eqref{noisy_parity_breaking}.
The two stable limit cycles are labelled by the minima $w_\pm$ of a double potential $U(w)$ (panel a).
    The sojourn times $\tau$ in each attractor are random variables, approximately following an exponential distribution (inset) described by the Ahrenius-Kramers-Eyring theory.
    As a consequence, $w$ can be approximated as a random telegraph process (panel b).
    The state of the system then drifts along the attractor at a speed determined by $w(t)$. In panel c, we have shown a schematic of the position $\phi$ on the attractor (e.g. the phase of the limit cycle) as a function of time, in the simple case where the angular velocity is directly proportional to $w$ (see Eq.~\eqref{noisy_parity_breaking}).
    Schematically, $\phi$ is obtained by integrating $w$ (panel d).
    In panels e-f, we compare the results of numerical integrations of Eqs.~\eqref{noisy_parity_breaking}) for different realizations of the noise, with $\omega_1 = 0$ (panel d) and $\omega_1 \neq 0$ (panel e).  
    An arbitrary trajectory has been highlighted in red.
    We have set $r = \num{0.22}$, $\omega_0 = 0$, $\omega_1 = \num{0.1}$, $T_w = T_\phi = \num{0.1}$.
    Here, we do not assume that $\phi$ is periodic.
}
\end{figure}

Rather than performing a first-principles model of this experimental situation and trying to reduce it, we make a guess and directly identify $w = (\Delta T - \Delta T_{\text{c}})/\Delta T_{\text{c}}$ in Eq.~\eqref{noisy_parity_breaking}. Our simple model then reproduces the experimentally observed (i) square-root-like behavior of $\braket{|\dot{\theta}|}$ with $\Delta T$, (ii) the two classes of behavior in time of $\theta$ (compare panels e and f of Fig.~\ref{figure_noise_telegraph} with the two regimes [in blue and red, respectively] in panel c of Fig.~\ref{figure_convection_experiment}) and (iii) the Poissonian distribution of the reversal times. 

  \section{Conclusion and perspectives}

In this article, we have shown that the tangency of covariant Lyapunov vectors provides a generalization of exceptional points to arbitrary non-linear systems. 
This identification suggests intriguing parallels between classical dynamical systems~\cite{Yang2009,Takeuchi2011} and quantum many-body systems~\cite{Luitz2019}.
An exact tangency occurs for the covariant Lyapunov vectors associated with vanishing Lyapunov exponents in a class of bifurcations describing the splitting and merging of extended invariant sets. In turn, these bifurcations generically occur in systems with asymmetric couplings. 
We show that this tangency of covariant Lyapunov vectors, dubbed a generalized exceptional point, can occur both continuously and discontinuously. 

In linear systems, exceptional points possess distinctive properties~\cite{Miri2019,Trefethen2005,Shankar2022,Bergholtz2021,Ghatak2020,Baconnier2022,Tlusty2021,Wiersig2020} including an enhanced sensitivity to perturbations~\cite{Trefethen2005} as well as chiral mode conversion when dynamically encircling the exceptional point~\cite{Miri2019,Doppler2016,Hassan2017,Dembowski2004}. To what extent these properties extend to non-linear systems? 
We have analyzed the effect of noise and perturbations in systems with generalized exceptional points. In addition to displaying a non-reciprocal response to small perturbations, these systems exhibit an enhanced sensitivity to noise. For instance, in the case of limit cycles, isochrons become tangent to the limit cycle and are ill-defined at the bifurcation.

\begin{acknowledgments}
V.V. acknowledges support from the Simons Foundation, the Complex Dynamics and Systems Program of the Army Research Office under grant W911NF-19-1-0268, the National Science Foundation under grant DMR-2118415 and the University of Chicago Materials Research Science and Engineering Center, which is funded by the National Science Foundation under award no. DMR-2011854. 
M.F. acknowledges support from a MRSEC-funded Kadanoff–Rice fellowship (DMR-2011854), the National Science Foundation under grant DMR-2118415, and the Simons Foundation. M.F. thanks Daniel S. Seara for useful discussions and A. Mauroy for sharing code computing isochrons.
R.H. was supported by an appointment to the JRG Program at the APCTP through the Science and Technology Promotion Fund and Lottery Fund of the Korean Government.
CW was partially supported by NSF-MPS-PHY award 2207383.
\end{acknowledgments}

\clearpage

\onecolumngrid

\begin{center}
{ \bf Appendix}
\end{center}

\setcounter{section}{0}

\section{Computation of the Lyapunov exponents and vectors using Floquet theory}
\label{explicit_computation_floquet}

In this Appendix, we compute analytically the Lyapunov exponents and covariant Lyapunov vectors (also known as Floquet vectors) for the very simple system Eq.~\eqref{hopf_pitchfork} using Floquet theory~\cite{Floquet1883,Lyapunov1907,Teschl2012}. To do that, we evaluate the evolution operator in Eq.~\eqref{U_Texp} directly by diagonalizing the Jacobian in a rotating frame.

\subsection{Floquet theory}
\label{floquet_basics}

In this section, we recall basics of Floquet theory~\cite{Floquet1883,Lyapunov1907,Teschl2012}.
In the main text, we have introduced the evolution operator $U(t,t_0)$ in Eq.~\eqref{U_Texp}, defined as the unique solution of the Cauchy problem Eq.~\eqref{eom_perturbation} with $U(t_0,t_0) = \text{Id}$, where $\text{Id}$ is the identity matrix. 
In the case of periodic orbits, one can show that 
\begin{equation}
    U(t+T, t_0+T) = U(t, t_0).
\end{equation}
In order to keep notations simple, we now set the origin of times so that $t_0 = 0$ and write $U(t) = U(t,t_0=0)$. 
As a consequence of the property $U(t+T, t_0+T) = U(t, t_0)$, we have 
\begin{equation}
    U(t + n T) = U(t) [U(T)]^n
\end{equation}
for integer $n$. This suggests that the long-time evolution of the system is mainly controlled by the matrix $U(T)$, called the Floquet operator or monodromy matrix. This can be made precise by writing
\begin{equation}
    U(t) = V(t) \ee^{t F}
\end{equation}
in which $F = (1/T) \log U(T)$ and $V(t) = U(t) \ee^{-t F}$ satisfies $V(t+T) = V(t)$. 
This decomposes the evolution operator in a periodic part $V(t) = V(t+T)$ and a non-periodic part related to $U(T)$. The eigenvalues $\mu_i$ of $U(T)$ are called Floquet multipliers, and we also defined the Floquet exponents $s_i$ such that $\mu_i = \ee^{s_i}$ (the exponents are related to the eigenvalues of $F$ by a factor $T$).

In the case of a periodic orbit, the Lyapunov exponents $\lambda_i$ are given by $\lambda_i = \log |\mu_i| = \text{Re}(s_i)$, while the covariant Lyapunov vectors are directly the eigenvectors of $U(T)$ (in case of degeneracies, the Floquet eigenvectors at least span the same spaces as the covariant Lyapunov vectors)~\cite{Trevisan1998,Huhn2019}. 

\subsection{Floquet analysis of the Hopf-pitchfork system}

We start from Eq.~\eqref{hopf_pitchfork} and compute the Jacobian \eqref{jacobian}.
The variables are $X_1 = \text{Re}(z) \equiv x$, $X_2=\text{Im}(z) \equiv y$, and $X_3 = w$. Matrices and vectors follow the ordering $(X_1, X_2, X_3)$.
Perturbations $\delta X$ satisfy
\begin{equation}
    \delta \dot{X} = J(X(t)) \delta X
\end{equation}
A direct computation shows that the Jacobian is 
\begin{equation}
    J(X) = 
    \begin{pmatrix}
    \alpha_\re' + \beta_\re (3x^2 + y^2) - \beta_\im 2 x y
    & 
    -\alpha_\im' + \beta_\re 2 x y - \beta_\im (x^2+3y^2)
    & 
    [\gamma_\re x
    - 
    \gamma_\im y] h'
    \\
    \alpha_\im' + \beta_\im(3x^2+y^2) + \beta_\re 2 x y 
    &
    \alpha_\re' + \beta_\im 2 x y + \beta_\re (x^2 + 3 y^3)
    &
    [\gamma_\im x
    +
    \gamma_\re y] h'
    \\
    0 & 0 &
    r - 3 w^2
    \end{pmatrix}
\end{equation}
in which $h' = \dd h/\dd w$.

Evaluating the Jacobian on the limit cycle solutions with radius $R$ and angular frequency $\Omega$ (the expression of which is given as a function of the parameters in the main text), we find
\begin{equation}
    J(t) = J(X(t)) = 
    \begin{pmatrix}
    2 R^2 [\beta_\re c^2 - \beta_\im c s]
    & 
    - \Omega + 2 R^2 [- \beta_\im s^2 + \beta_\re c s]
    & 
    R (\gamma_\re c - \gamma_\im s) h'
    \\
    \Omega + 2 R^2 [ \beta_\im c^2 + \beta_\re c s] &
    2 R^2 [ \beta_\re s^2 + \beta_\im c s] 
    &
    R (\gamma_\im c + \gamma_\re s) h'
    \\
    0 & 0 & r - 3 w^2
    \end{pmatrix}
\end{equation}
where we have temporarily set $c = \cos(\Omega t)$ and $s = \sin(\Omega t)$, and in which $R$ and $\Omega$ are functions of the parameters of the system.

We then perform a time-dependent unitary change of basis
\begin{equation}
    X \to X' = V(t) X
\end{equation}
with
\begin{equation}
    V = \begin{pmatrix}
    \cos(\Omega t) & \sin(\Omega t) & 0\\
    -\sin(\Omega t) & \cos(\Omega t) & 0 \\
    0 & 0 & 1 
    \end{pmatrix}
\end{equation}
to go to the frame rotating at the frequency $\Omega$ about the $w$-axis.
In the rotating frame, perturbations $\delta X' = V(t) \delta X$ follow
\begin{equation}
    \label{eom_dXprime}
    \delta\dot{X}' 
    = \dot{V} \delta X + V \delta \dot{X} 
    = (\dot{V} V^{-1} + V J V^{-1}) \delta X'
\end{equation}
so the Jacobian transforms as
\begin{equation}
    J(t) \to J'(t) = \dot{V} V^{-1} + V J V^{-1}. 
\end{equation}
Here, this expression reduces to
\begin{equation}
    J' = 
    \begin{pmatrix}
    2 R^2 \beta_\re & 0 & R \gamma_\re h' \\
    2 R^2 \beta_\im & 0 & R \gamma_\im h' \\
    0 & 0 & r - 3 w^2
    \end{pmatrix}.
\end{equation}
As $J'$ does not depend on time, the solution of Eq.~\eqref{eom_dXprime} is
\begin{equation}
    \delta X'(t) = \ee^{t J'} \delta X'(0)
\end{equation}
Hence, going back to the original reference frame,
\begin{equation}
    \delta X(t) = U(t) \delta X(0)
\end{equation}
in which
\begin{equation}
    U(t) = V^{-1}(t) \ee^{t J'} V(t)
\end{equation}

As $V(T) = 1$, we find that the Floquet operator is $U(T)=\ee^{T J'}$. Hence, the eigenvectors of $U(T)$ are identical to those of $J'$, and the eigenvalues $\mu_i$ of $U(T)$ (Floquet multipliers) are obtained from the eigenvalues $\epsilon_i$ of $J'$ as $\mu_i = \ee^{T \epsilon_i}$. A direct calculation gives
\begin{align}
    \label{psis_floquet}
    \psi_1 = \begin{pmatrix}
    0 \\
    1 \\
    0
    \end{pmatrix},
    \qquad
    \psi_2 = \frac{1}{N_2} \, \begin{pmatrix}
    {h'} R {\gamma_\re} (r-3 w^2) \\
    {h'} R \left({\gamma_\im} r+2 R^2 ({\beta_\im} {\gamma_\re}-{\beta_\re} {\gamma_\im})-3 {\gamma_\im} w^2\right) \\
    (r-2 {\beta_\re} R^2-3 w^2) (r-3 w^2)
    \end{pmatrix},
    \qquad
    \psi_3 = \frac{1}{\sqrt{\beta_\re^2+\beta_\im^2}} \, \begin{pmatrix}
    \beta_\re \\
    \beta_\im \\
    0
    \end{pmatrix}
\end{align}
in which $N_2$ is a normalization factor. The corresponding eigenvalues are
\begin{align}
\epsilon_1 = 0,
\qquad
\epsilon_2 = r - 3 w^2,
\qquad
\epsilon_3 = 2 R^2 \beta_\re.
\end{align}

Please note that the $\psi_a$ in Eq.~\eqref{psis_floquet} are the CLVs evaluated at $t=0,T,\dots$ (namely, $c_{a}(0) = \psi_a$).
The CLVs at arbitrary time $c_{a}(t) \propto U(t) \psi_a$ are obtained by applying the evolution operator. In the case of periodic orbits, it is convenient to choose the phase to obtained time-periodic CLVs $c_{a}(t) = \ee^{-i t \epsilon_a} U(t) \psi_a = c_{a}(t+T)$.

We now look for the limits of $\psi_i$ when $r \to 0$.
We have $w = \pm \sqrt{r}$ when $r \to 0^{+}$ (from above), while $w = 0$ when $r \to 0^{-}$ (from below).
The eigenvectors $\psi_1$ and $\psi_3$ do not depend on $r$, so we focus on $\psi_2$. 

\strong{First case.} Assume $h'(0) \neq 0$. 
In this case, we find that as $r \to 0$ (from above or below), $\psi_2 \to (0, 1, 0)$ provided that ${\beta_\im} {\gamma_\re}-{\beta_\re} {\gamma_\im} \neq 0$. 
(When this last condition is not met, $\psi_2 \to (1,h'(0) R \gamma_\im, - 2 \beta_\re R^2)/N_2$.)

\strong{Second case.}  When $h'(w) = w^\alpha$, the situation is more subtle. 
For $r < 0$, we always have $h' = 0$, so
$\psi_2 \to (0, 0, 1)$
as $r \to 0^{-}$.
For $r > 0$, $h' = r^{\alpha/2}$ and 
\begin{align}
    \psi_2 = \frac{1}{N_2} \, \begin{pmatrix}
    r^{\alpha/2} R {\gamma_\re} (-2 r) \\
    r^{\alpha/2} R \left({\gamma_\im} r+2 R^2 ({\beta_\im} {\gamma_\re}-{\beta_\re} {\gamma_\im})-3 {\gamma_\im} r \right) \\
    4 r (r + {\beta_\re} R^2)
    \end{pmatrix}
    \sim
    \frac{1}{N_2}
    \begin{pmatrix}
    C_1 r^{\alpha/2+1}  \\
    C_2 r^{\alpha/2} \\
    C_3 r + 4 r^2
    \end{pmatrix}
\end{align}
in which $C_i$ are constants, $N_2$ ensures that the vector is normalized, and we have retained only the most important terms as $r \to 0$. In particular, $C_2 = 2 R^3 ({\beta_\im} {\gamma_\re}-{\beta_\re} {\gamma_\im})$ (we assume that this does not vanish) and $C_3 = -2 \beta_\re R^2$.
When $\alpha < 2$, $\psi_2 \to (0,1,0)$.
When $\alpha = 2$ and $C_3 \neq 0$, $\psi_2 \to (0, C_2, C_3)/N_2$.
When $\alpha > 2$ and $C_3 \neq 0$, $\psi_2 \to (0,0, 1)$.
When $C_3 = 0$, we have $\psi_2 \to (0,1,0)$ for $\alpha < 4$, $\psi_2 \to (0,C_2,4)/N_2$ for $\alpha = 4$, and $\psi_2 \to (0,0,1)$ for $\alpha > 4$.

The conclusion of this calculation is that, except in higher-codimension submanifolds, we have $\psi_2 \to (0,1,0) = \psi_1$ (i.e., the Floquet vectors become parallel at the bifurcation) as $r \to 0^+$ provided that $h(w) \sim w^{\alpha+1}$ with $0 \leq \alpha < 2$. Besides, $\psi_2 \to (0,1,0) = \psi_1$ as $r \to 0^-$ when $h'(0) \neq 0$. In this case, the Floquet vectors become parallel from both sides, while in the other cases their behavior at the bifurcation is discontinuous.

\section{\texorpdfstring{$\ZZ_2$}{Z2} symmetries}
\label{appendix_Z2}
While no continuous symmetry is required for the tangency of covariant Lyapunov vector to happen (besides time translation invariance), some of the systems discussed in the main text have a discrete $\ZZ_2$ symmetry that interchanges the two stable attractors and leaves the unstable attractor (repellers) unchanged.
In some cases, these symmetries are expressed as a time-dependent transformation. In this section, we discuss the general structure of these symmetries, and show how they emerge in the systems analyzed in the main text.

\subsection{Standard and time-dependent symmetries}

Consider a dynamical system
\begin{equation}
    \label{si_sym_general_ds}
    \dot{X} = f(X)
\end{equation}
in which $X(t) \in \mathbb{R}^N$ and a group $G$ equipped with a group action $\rho : G \times \mathbb{R}^N \to \mathbb{R}^N$. We use the shorthand $g \cdot X = \rho(g, X)$ to refer to the action of a group element $g \in G$ on a vector $X \in \RR^N$. 
The dynamical system is said to be equivariant with respect to the group $G$ (more precisely, the group action $\rho$) when $f(g \cdot X) = g \cdot f(X)$ for all $g \in G$ and $X \in \RR^N$.
This guarantees that the equation of motion for the transformed variables $\tilde{Y} = g \cdot X$ are identical to the equations of motion for the original variables $X$.
(In this appendix, all tilded variables are transformed variables.)
Namely, Eq.~\eqref{si_sym_general_ds} implies that $\dot{\tilde{X}} = f(\tilde{X})$.

Consider now a time-dependent \enquote{group action}, so that the transformed variable is now $\tilde{X}(t) = \rho(t, g, X(t)) \equiv g(t) \cdot X(t)$. As an example, let us take a state $X \in \RR^2$ represented (for convenience) as a complex number $z = X_1 + \ii X_2$ and the transformation $z(t) \mapsto \tilde{z}(t) = \ee^{\ii \omega t} z^*(t)$ (the star denotes complex conjugation). Applying this transformation two times is gives back $z(t)$, so we can interpret it as defining a (time-dependent) representation of the group $\ZZ_2$\footnote{To be more precise, let $\pm 1$ represent the two elements of $\ZZ_2$ (viewed multiplicatively), then we define $\rho(t, -1, z(t)) = \ee^{\ii \omega t} z^*(t)$ and $\rho(t, +1, z(t)) =  z(t)$.}. 

\subsection{Systems analyzed in the main text}

The pitchfork bifurcation (of fixed points) $\dot{w} = r w - w^3$ has a $\ZZ_2$ symmetry where the transformation is simply $\tilde{w} = -w$. Several of the systems analyzed in the main text have a similar $\ZZ_2$ symmetry that either maps a steady state of the system to itself or interchanges two steady states. 
\begin{itemize}
    \item The pitchfork of attractors Eq.~\eqref{pitchfork} is invariant under the transformation $w \mapsto \tilde{w} = -w$ when $g(x,w)$ is even in $w$ [i.e. $g(x, -w) = g(x,w)$], with $x$ left invariant by the transformation (i.e. $\tilde{x} = x$). 
    \item The Wilson-Cowan  Eq.~\eqref{eq:wceq} is invariant under $\tilde{x}_i=-x_i$ when $h_i = 0$.
    \item The predator-prey model in Eq.~\eqref{eq:ecology} is invariant under the transformation $\tilde{u}_1 = u_2, \tilde{v}_1 = v_2,\tilde{u}_2 = u_1, \tilde{v}_2 = v_1$ (because we have chosen identical parameters for the two prey-predator subsystems). 
\end{itemize}
In all of these systems, the symmetry acts as a reflection interchanging the stable states, and leaving the unstable state invariant.

\medskip

Other systems do not posses such a simple symmetry, but do exhibit a time-dependent $Z_2$ symmetry. 
\begin{itemize}
    \item Consider the pitchfork of limit cycles Eq.~\eqref{hopf_pitchfork}.
When $h(w) = w$, the frequency of the limit cycle depends on $w$, making a simple reflection ($\tilde{w} = - w$, $\tilde{z} = z$) insufficient: this transformation is not a symmetry, as it can be verified explicitly. Nevertheless, Eq.~\eqref{hopf_pitchfork} has a time-dependent symmetry under certain conditions on the parameters. Consider the transformation
\begin{equation}
    \begin{split}
        z \mapsto \tilde{z} &= \ee^{\ii 2 \Im(\alpha) t} z^* \\
        w \mapsto \tilde{w} &= -w \\
    \end{split}
    \label{trans_NF}
\end{equation}
As in the case of a single pitchfork, the equation of motion for $w$ is equivalent to the $\tilde{w}$ equation of motion.
Let us now consider the other equation of motion. From the definition \eqref{trans_NF} of $\tilde{z}$, we get
\begin{equation}
    \dot{\tilde{z}} = (\dot{z}^* + \ii 2 \Im(\alpha) z^*) \ee^{\ii 2 \Im(\alpha) t}.
\end{equation}
We start with the equation of motion Eq.~\eqref{hopf_pitchfork} for $z$ and complex conjugate it to obtain
\begin{subequations}
\begin{align}
    \dot{z}^* = (\alpha^* + \beta^* \,|z|^2) \, z^* + \gamma^* h(w) z^*
\end{align}
\end{subequations}
where we have used that $h(w) \in \RR$. Hence,
\begin{align}
    \dot{\tilde{z}} 
    &= [ (\alpha^* + \beta^* \,|z|^2) \, z^* + \gamma^* h(w) z^* + \ii 2 \Im(\alpha) z^*] \ee^{\ii 2 \Im(\alpha) t} \\
    &= [ (\alpha^* + \beta^* \,|z|^2) + \gamma^* h(w) + \ii 2 \Im(\alpha)]  \tilde{z} \\
    &= [ (\alpha^* + \beta^* \,|\tilde{z}|^2) \pm \gamma^* h(\tilde{w}) + \ii 2 \Im(\alpha)]  \tilde{z}
\end{align}
where we have replaced $h(w)$ by $\pm h(\tilde{w}) = \pm h(-w) = h(w)$, where the $\pm$ depends on whether $h$ is odd or even (cases where $h$ is neither odd nor even are not considered). We have also used that $|\tilde{z}|^2 = |z|^2$.
Assuming in addition that $\beta \in \RR$ and that $\pm \gamma^* = \gamma$, we end up with 
\begin{align}
    \dot{\tilde{z}} 
    &= [ (\alpha^* + \ii 2 \Im(\alpha) + \beta \,|\tilde{z}|^2) + \gamma h(\tilde{w})]  \tilde{z} \\
    &= [ (\alpha + \beta \,|\tilde{z}|^2) + \gamma h(\tilde{w})]  \tilde{z}
\end{align}
Hence, we have shown that the equation of motion for $(\tilde{z}, \tilde{w})$ is identical to the equation of motion for $(z, w)$ provided that $\beta \in \RR$ and $\gamma^* h(w) = \gamma h(-w)$.

\item Similarly, the equations \eqref{eq:coupledhopf} describing coupled non-linear oscillators are invariant under the transformation
\begin{equation}
    \begin{bmatrix}
        \tilde{z}_1  \\
        \tilde{z}_2
    \end{bmatrix}
    = 
    \ee^{\ii 2 \omega t}
    \begin{bmatrix}
        {z}_1^*  \\
        {z}_2^*
    \end{bmatrix}
    \label{trans_CH}
\end{equation}
First, we have 
\begin{equation}
    \begin{bmatrix}
        \dot{\tilde{z}}_1  \\
        \dot{\tilde{z}}_2
    \end{bmatrix}
    =
    \ee^{\ii 2 \omega t}
    \begin{bmatrix}
        \dot{z}^*_1 + \ii2\omega z^*_1 \\
        \dot{z}^*_2 + \ii2\omega z^*_2.
    \end{bmatrix}
\end{equation}
Complex conjugating Eq.~\eqref{eq:coupledhopf} yields
\begin{equation}
\!\!\!\!
    \begin{bmatrix}
        \dot{z}_{1}^* \\
        \dot{z}_{2}^*
    \end{bmatrix}
    = 
    \begin{bmatrix}
        (a-\ii\omega)+b|z_{1}|^{2} & a_{12} \\
        a_{21} & (a-\ii\omega)+b|z_{2}|^{2}
    \end{bmatrix}
    \begin{bmatrix}
        z_{1}^*\\
        z_{2}^*
    \end{bmatrix}
\end{equation}
where we have assumed that $a$, $\omega$, $b$, $a_{12}$, $a_{21}$ are real. The two previous equations combine to give
\begin{equation}
    \begin{bmatrix}
        \dot{\tilde{z}}_1  \\
        \dot{\tilde{z}}_2
    \end{bmatrix}
    =
    \ee^{\ii 2 \omega t}
    \begin{bmatrix}
        (a-\ii\omega + \ii2\omega)+b|z_{1}|^{2} & a_{12} \\
        a_{21} & (a-\ii\omega + \ii2\omega)+b|z_{2}|^{2}
    \end{bmatrix}
    \begin{bmatrix}
        z_{1}^*\\
        z_{2}^*
    \end{bmatrix}
    =
    \begin{bmatrix}
        (a+\ii\omega)+b|\tilde{z}_{1}|^{2} & a_{12} \\
        a_{21} & (a+\ii\omega)+b|\tilde{z}_{2}|^{2}
    \end{bmatrix}
    \begin{bmatrix}
        \tilde{z}_{1}\\
        \tilde{z}_{2}
    \end{bmatrix}
\end{equation}
Therefore, the equations of motion are equivalent in both the $\vec{z}$ and $\tilde{\vec{z}}$ coordinate systems: Eq.~\eqref{trans_CH} is a time-dependent symmetry of Eq.~\eqref{eq:coupledhopf} [when $a,\omega,b,a_{12},a_{21} \in \RR$].
\end{itemize}

The time-dependent symmetry transformations can be understood as the systems being invariant under complex conjugation in the rotating frame. 
The bifurcations of limit cycles analyzed in the main text fall in two classes:
\begin{itemize}
    \item systems with a time-independent $\ZZ_2$ symmetry (Wilson-Cowan model, Rosenzweig-MacArthur model), for which the CLVs are discontinuous at the bifurcation like Eq.~\eqref{hopf_pitchfork} with $h(w) = w^2$,
    \item systems with a time-dependent $\ZZ_2$ symmetry (coupled Hopf oscillators), for which the CLVs are continuous at the bifurcation, like Eq.~\eqref{hopf_pitchfork} with $h(w) = w$.
\end{itemize}
It is not yet known whether the relation between the two classes of symmetry and the behaviors of the CLVs is general.

\begin{figure}[t]
    \centering
    \includegraphics[width=8cm]{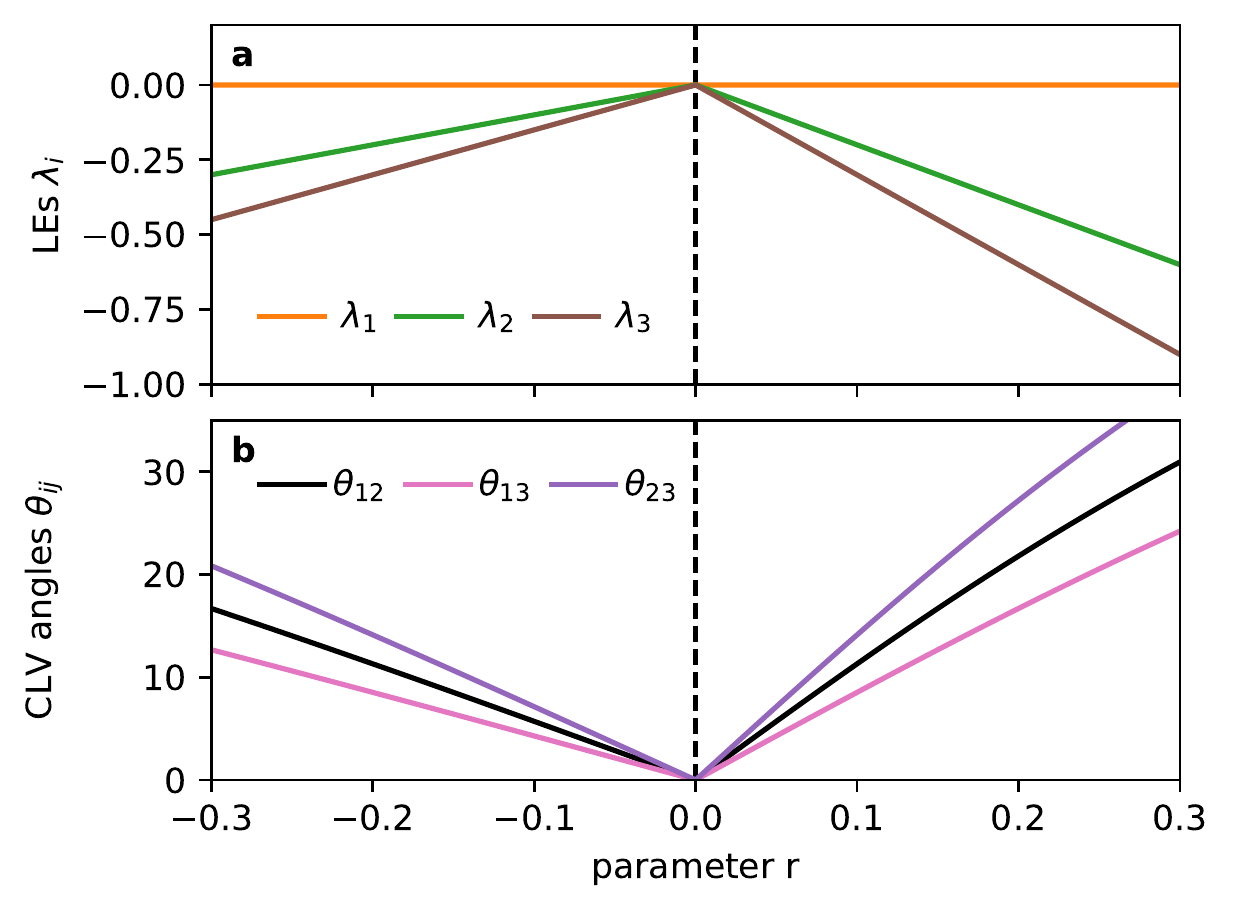}
	\caption{\strong{Higher-order degeneracies.}
	It is possible to have three (or more) covariant Lyapunov vectors becoming parallel at the same time.
	This generalizes higher-order exceptional point (a $n$-th order EP corresponds to $n$ vectors becoming parallel, and to a Jordan block of size $n$).
	Here, $c_1$, $c_2$ and $c_3$ all become parallel at $r=0$.
	Equation~\eqref{eq:torus_system} with parameters $M=2, N=1, \alpha =1+i, \beta = -1, \gamma_j = (i,2i), \tau_j = (1,\frac{2}{3})$ causes three CLVs to merge at $r=0$.
	}
	\label{fig:order2EP}
\end{figure}

\begin{figure}[t]
    \centering
    \includegraphics[width=8cm]{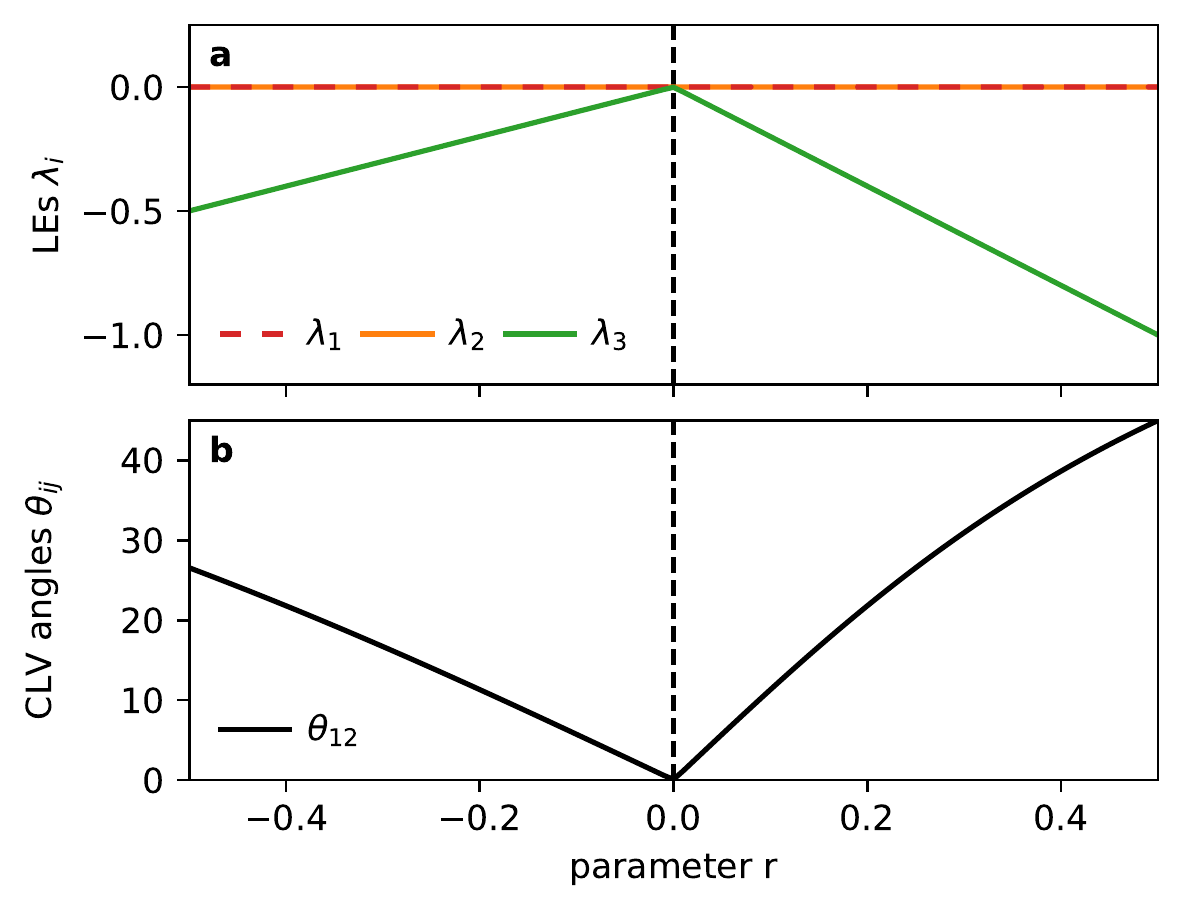}
	\caption{
	\strong{Coalescence of CLVs in limit tori.}
	In the main text, we have discussed the coalescence of CLVs in limit cycles (periodic evolutions). It is also possible to have a coalescence of CLVs in limit tori (a limit torus corresponds to a quasi-periodic evolution).
	This is illustrated using Eq.~\eqref{eq:torus_system} with parameters $M=1$, $N=2$, $\vec{\alpha} =(1+i,1/2+i\pi)$, $\vec{\beta} = (-1,-1)$, $\vec{\gamma} = (i,0)$, $\tau = 1$.
	The two highest Lyapunov exponents always vanish ($\lambda_1 = \lambda_2 = 0$, in red and orange in the top panel).
	At the critical value $r=0$ of the control parameter, a third Lyapunov exponent $\lambda_3$ (in green) vanishes. The angle between the corresponding CLV ${c}_3$ and the two-dimensional vector space spanned by the CLVs ${c}_1$ and ${c}_4$ vanishes at $r=0$ (bottom panel).
	}
	\label{fig:torusEP}
\end{figure}

\begin{figure}
    \centering
\includegraphics[width=8cm]{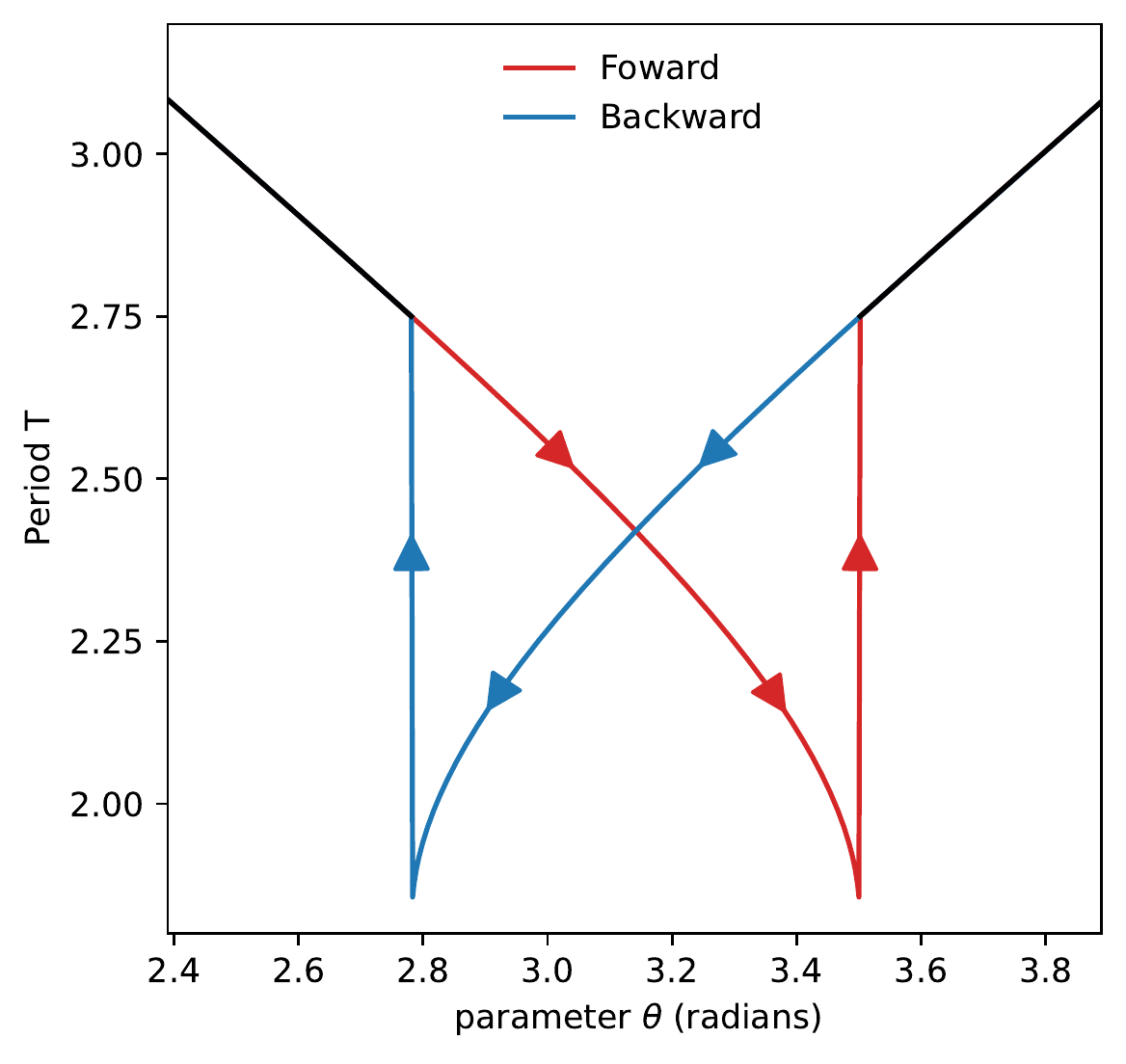}
    \caption{\label{figure_hysteresis}
    \strong{Hysteresis.}
    Hysteresis can be observed by adiabatically tuning the parameters around the exceptional point.
    We consider the Wilson-Cowan system with the same parameters as in Fig.~\ref{fig:wilsoncowan} of the main text. We set $\vec{h}= h(1,-1,1)$ and choose the path encircling the EP $(\zeta, h) = (10.786+2\cos(\theta), 0.5\sin(\theta))$.
    We then let $\theta$ evolve either the clockwise (forward) or counterclockwise (backward) direction, and monitor the state of the system through the period $T$ of the limit cycle.
    The paths coincide for part of the circle (black curves), but there is a region between $\theta \approx \num{2.75}$ where the trajectories of the system on the forward (red) and backward (blue) path are different, signaling hysteresis. 
    }
\end{figure}

\section{Pitchfork of tori and higher-order generalized exceptional points}
\label{appendix_tori_higher}
Higher-order generalized exceptional points and generalized exceptional points of limit tori can be constructed by modifying Eq.~\eqref{hopf_pitchfork} to include additional Hopf oscillators and pitchfork bifurcations. 
We consider $N$ Hopf oscillators indexed by $i=1,\dots,N$ and $M$ pitchforks indexed by $j=1,\dots,M$. The Hopf oscillators are controlled by the pitchforks as described by the dynamical system
\begin{equation}
    \begin{split}
        &\dot{z}_i = (\alpha_i + \beta_i \,|z_i|^2) \, z +  \gamma_{ik} \, h(w_{k})  \\
        &\tau_j\dot{w}_{j} = r_{i}w_{j} - w_{j}^{3}
    \end{split}
    \label{eq:torus_system}
\end{equation}
where $\gamma_{ik}$ is the matrix of couplings of each oscillator to each pitchfork, and an implicit sum over $k$ is taken. 
When $M=N=1$, the system reduces to Eq.~\eqref{hopf_pitchfork}. When $M=1$, $N>1$ (i.e. we add additional Hopf oscillators) and at least one Hopf oscillator is adequately coupled to the pitchfork, the system will undergo a generalized exceptional point at the bifurcation of $N$-tori (under the conditions derived in section \ref{explicit_computation_floquet} of the appendix), see Fig.~\ref{fig:torusEP}. As $r$ approaches zero, the CLV corresponding to the pitchfork will approach a vector in the degenerate vector space spanned by the zero-modes of each Hopf oscillator dependent on the exact choice of couplings. For the system in Eq. ~\eqref{eq:torus_system}, CLVs can be calculated by decomposing the system into Floquet problems for each Hopf oscillator, because the Hopf oscillators are decoupled from each other. 
By tuning multiple parameters, high-order generalized exceptional points can be found, where three or more CLVs align at once~\footnote{In linear algebra, an exceptional point of order $n$ corresponds to a Jordan block of size $n$. For instance, the canonical Jordan block of size three associated with the eigenvalue $\lambda$ has the form 
\begin{equation*}
\begin{pmatrix}
\lambda & 1 & 0 \\ 0 & \lambda & 1 \\ 0 & 0 & \lambda
\end{pmatrix}.
\end{equation*}
We refer to Refs.~\cite{Arnold1999,GonzalezTokman2013} for the equivalent notion in the case of covariant Lyapunov vectors. 
}. When $M>1$ and $N=1$ in Eq.~\eqref{eq:torus_system}, if each pitchfork is adequately coupled to the Hopf oscillator and each $r_{i}$ is simultaneously tuned to zero, then the CLV corresponding to each pitchfork will simultaneously align with the zero CLV of the Hopf oscillator producing an order-$M$ generalized exceptional point (Fig.~\ref{fig:order2EP}).
While these examples are obtained by merely considering a product of independent dynamics, we expect that they occur at (relatively high codimension) bifurcations in generic dynamical systems.

\section{Hysteresis}
\label{appendix_hysteresis}

In this Appendix, we show that a non-linear effect superficially similar to the dynamical encircling of exceptional points occurs when the symmetry between the attractors is explicitly broken.
This may provide a mechanism for memory formation~\cite{Keim2019} in which the memory is stored in a dynamical state (a limit cycle, or a more complex attractor).

Consider encircling a generalized exceptional point in one of the systems analyzed in the main text. To do so, we may add a term $b \, w^2$ to Eq.~\eqref{pitchfork}, so the pitchfork bifurcation becomes imperfect (like a paramagnetic/ferromagnetic transition under an external magnetic field). In this case, an effect superficially similar to the dynamical encircling of exceptional points occurs: the coalescence of attractors exhibits hysteresis. The effect is however fundamentally different: while the chiral mode conversion in linear systems is a purely dynamical effect, the number of steady-state solutions changes in a pitchfork of attractors when the parameters are varied. 

In figure~\ref{figure_hysteresis}, we show that hysteresis emerges in the Wilson-Cowan system by varying the parameters in a closed loop around the bifurcation point, clockwise and counterclockwise. The external fields $h_i$ break the $\ZZ_2$ symmetry of the Wilson-Cowan equations (see section \ref{appendix_Z2} of the appendix).

\section{Generalized exceptional points with non-zero Lyapunov exponents}
\label{appendix_finite_LE}

\begin{figure}
    \centering
    \includegraphics[width=8cm]{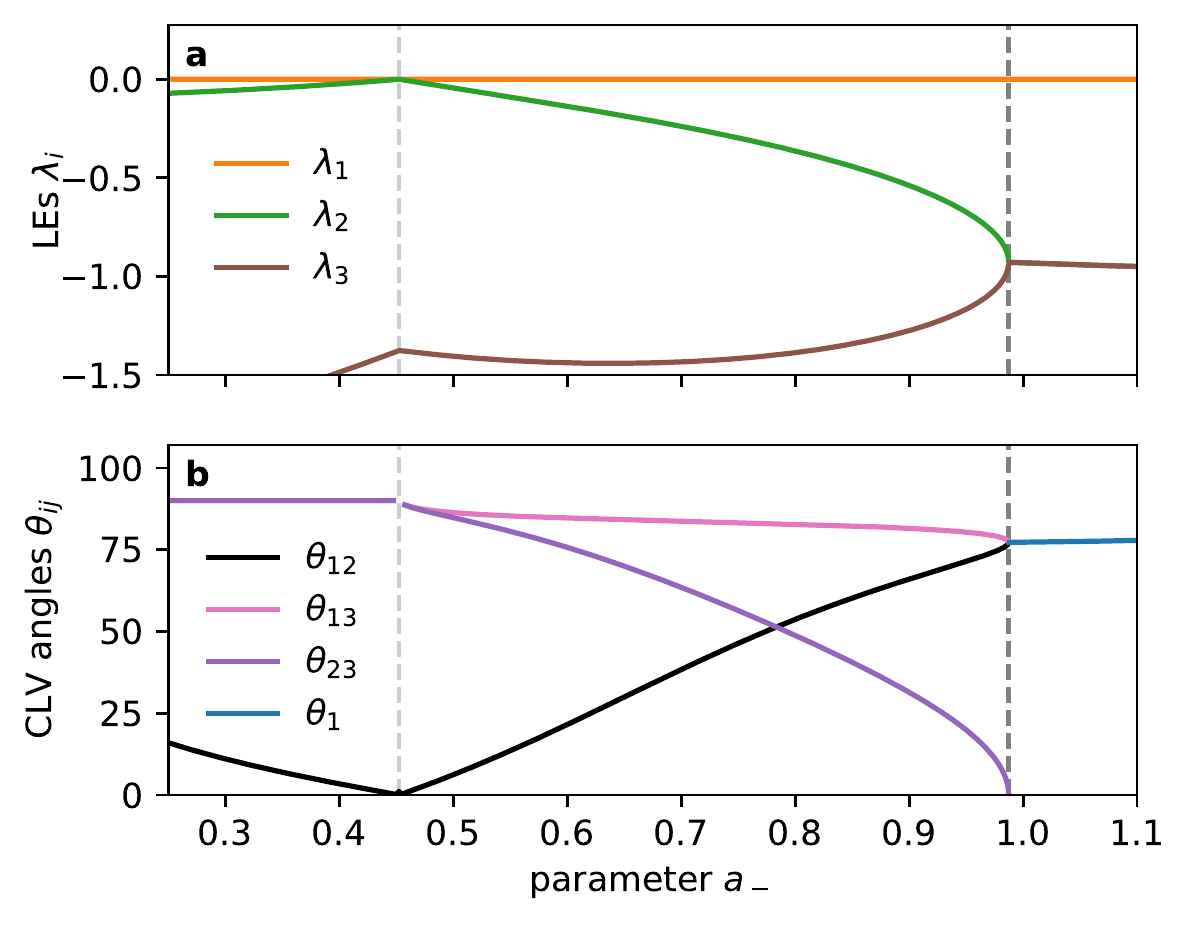}
     \caption{\label{fig:figure_finiteLE_EP}
    \strong{Finite Lyapunov Exponent Exceptional Point}
    We simulate the coupled nonlinear oscillators given in Eq. \eqref{eq:coupledhopf}. At the  point $a_{-} = a_{-}^{\text{f}} = 0.9872$ (dark gray dashed line), 
    a non-zero Lyapunov exponent exceptional point occurs. In panel a, the 3 smallest Lyapunov exponents  are plotted.  The angle $\theta_{23}$ between the CLVs ${c}_2$ and ${c}_3$ transitions from being aligned to being in a degenerate space. $\theta_1$ is the angle from the CLV corresponding to $\lambda_1$ to the plane that contains the CLVs of $\lambda_1$ and $\lambda_2$ in the oscillatory region.
    We have set $a=\omega=-b =1, a_{+}=0.1$. In panel a, $a_{-}$ is varied from \num{0.2} to \num{1.1}.
    }
\end{figure}

In the main text, we focus on generalized exceptional points (EP) with a vanishing Lyapunov exponent.
In this Appendix, we discuss generalized EP that have a nonzero Lyapunov exponent. 
As an example, Fig.~\ref{fig:figure_finiteLE_EP} shows the Lyapunov exponents (panel a) and the angles between the corresponding covariant Lyapunov vectors (panel b) for the system of coupled nonlinear oscillators described by Eq.~\eqref{eq:coupledhopf} and Fig.~\ref{fig:hopf}. We use the same parameters as in Fig.~\ref{fig:hopf}e-f of the main text, with an extended range for the parameter $a_{-}$. As shown in Fig.~\ref{fig:figure_finiteLE_EP}, there is a generalized exceptional point with Lyapunov exponent $\lambda \sim \num{-1}$ at $a_{-}^{\text{f}} \simeq \num{0.9872}$.

In the case of limit cycles, insights about the effect of these generalized exceptional points can be gained using Floquet theory, see App.~\ref{floquet_basics}. For $a_{-} > a_{-}^{\text{f}}$, the two relevant Floquet exponents form a complex conjugate pair. Therefore, the CLVs form a degenerate subspace corresponding to the same Lyapunov exponents. For $a_{-} < a_{-}^{\text{f}}$, the Floquet exponents have different magnitudes, so the Lyapunov exponents are different (the degeneracy is lifted). 
The generalized EP at $a_{-}^{\text{f}}$ marks a change in the transient behavior of the system similar to the transition from overdamped to underdamped oscillations towards a fixed point. In the case of a periodic orbit, Floquet theory allows us to decompose the evolution of a perturbation $\delta X(0)$ into two parts by writing $U(t) = V(t) \ee^{t F}$ where $V(t+T) = V(t)$, see App.~\ref{floquet_basics}. The periodic part corresponds to $V(t)$, while the non-periodic part corresponds to $F$ (the eigenvalues of which are the Floquet exponents), and can be measured by performing stroboscopic measurements of $\delta X(t)$ (i.e. measuring it every period). This non-periodic part of the evolution describes the overall relaxation towards the limit cycle, which can be monotonic or oscillatory.

\section{Estimation of the phase diffusion constant}
\label{app_noise}

In this section, we provide details about the effect of noise in the solvable model discussed in the main text.
We estimate the phase diffusion constant in two cases: for an Ornstein-Uhlenbeck noise (away from the generalized exceptional) in Sec.~\ref{ioup}, and for a quartic potential using results from Ref.~\cite{Coffey1996} in Sec.~\ref{iquartic}.

\subsection{Integrated Ornstein-Uhlenbeck process}
\label{ioup}

We consider the set of coupled stochastic differential equations
\begin{subequations}
\begin{align}
\dd x_t &= -a \, x_t \, \dd t + \sigma \, \dd W_t \\
\dd y_t &= \omega_1 \, x_t \, \dd t
\end{align}
\end{subequations}
in which $W_t$ is a standard Wiener process and $\sigma \equiv \sqrt{2T}$.
With $x \sim w$ and $y \sim \phi$, they correspond to a version of Eqs.~\eqref{noisy_parity_breaking} of the main text where the non-linear term $w^3$ has been removed and $\omega_0 = 0$. (This gives a rough idea of the behavior of the system when $r$ is large enough for the non-linear term to be mostly negligible, which is far from the generalized exceptional point.)
Our goal is to evaluate the variance of the integral $y_t$ of the Ornstein-Uhlenbeck process $x_t \equiv x(t)$.

The stationary distribution associated with the SDE above is
\begin{equation}
p(x) = \mathcal{N} \ee^{- V(x)/T}
\end{equation}
where $V(x) = a x^2/2$ and $\mathcal{N}$ is a normalization factor such that $p$ integrates to unity.
In addition, the mean and covariance of the stationary (unconditioned) process are
\begin{equation}
\mathbb{E}[x_t] = 0 
\qquad
\text{and}
\qquad
\label{cov_x_OUP}
\text{cov}(x_s, x_t) \equiv \mathbb{E}\big[(x_s - \mathbb{E}[x_s]) (x_t - \mathbb{E}[x_t])\big]  = \mathbb{E}[x_s x_t] = \frac{\sigma^2}{2 a} e^{-a |t-s|}
\end{equation}
In particuliar the variance of $x_t$ is $\frac{\sigma^2}{2 a} = \frac{T}{a}$. 

We now estimate the variance of $y_t$. Integrating the SDE, we find
\begin{equation}
    y_t = \omega_1 \int_{0}^{t} x(s) \dd s.
\end{equation}
Therefore, we have
\begin{equation}
    \mathbb{E}[y_t] = \omega_1 \int_{0}^{t} \mathbb{E}[x(s)] ds = 0
\end{equation}
and
\begin{equation}
    \label{varyt}
    \mathbb{E}[y_t^2] 
    = \omega_1^2
    \mathbb{E}\left[
    \left( \int_{0}^{t} x(r) dr \right)
    \left( \int_{0}^{t} x(s) ds \right)
    \right]
    = \omega_1^2
    \int_{0}^{t} dr 
    \int_{0}^{t} ds
    \mathbb{E}[x(r) x(s)]
    = \omega_1^2
    \,
    \frac{\sigma^2}{2 a}
    \,
    \frac{2}{a^2} \left( 
    a t + e^{- a t} - 1
    \right)
    \sim
    \omega_1^2
    \,
    \frac{\sigma^2}{a^2} 
    \,
    t
\end{equation}
in which we have used the expression of the covariance Eq.~\eqref{cov_x_OUP} and the property
\begin{equation}
    \label{annoying_exp_identity}
    \int_{0}^{t} dr 
    \int_{0}^{t} ds
    e^{- \alpha |r-s|}
    = 
    \frac{2}{\alpha^2} \left( 
    \alpha t + e^{- \alpha t} - 1
    \right)
    \underset{\alpha t \gg 1}{\sim}
    \frac{2}{\alpha} t
\end{equation}
with $\alpha = a$.
From Eq.~\eqref{varyt}, we obtain the effective diffusion coefficient quoted in the main text for a process $y_t$ submitted to an Ornstein-Uhlenbeck noise $x_t$.

\subsection{Integrated quartic-potential noise}
\label{iquartic}

We now consider the situation in which the quartic potential of the Ornstein-Uhlenbeck process is replaced with a quartic potential. This occurs in Eq.~\eqref{noisy_parity_breaking} as we approach the generalized exceptional point.
The stochastic differential equations are
\begin{subequations}    
\label{quartic_SDE}
\begin{align}
\dd x_t &= (-a x_t - b x_t^3) \dd t + \sigma \dd W_t = - V'(x_t) \, \dd t + \sigma \dd W_t \\
\dd y_t &= \omega_1 \, x_t \, \dd t
\end{align}
\end{subequations}
with $\sigma = \sqrt{2T}$. Again, our goal is to evaluate the variance of $y_t$.
These equations correspond to Eqs.~\eqref{noisy_parity_breaking} with $x \sim w$ and $y \sim \phi$ (and $\omega_0 = 0$). 
In contrast with the previous section, we did not remove the non-linear term ($w^3$ or $x^3$). 
The potential in Eq.~\eqref{quartic_SDE} is
\begin{equation}
V(x) = \frac{1}{2} a x^2 + \frac{1}{4} b x^4.
\end{equation}
Here, we consider the case where the quartic potential has a single minimum (i.e., it is not a double-well potential) that corresponds on one side of the bifurcation (the handle side of the pitchfork). Similar calculations can be performed on the other side of the bifurcation~\cite{Coffey1996}.
The stationary distribution is
\begin{equation}
p(x) = \mathcal{N} e^{- U(x)/T}
\qquad
\text{where}
\qquad
\mathcal{N} = \sqrt{\frac{a}{2 b}}
\,
e^{\frac{a^2}{8 b T}} \,
K_{\frac{1}{4}}\left(\frac{a^2}{8 b T}\right)
\end{equation}
in which $K_n(z)$ is the modified Bessel function of the second kind of order $n$ \cite[\href{https://dlmf.nist.gov/10.25}{\S 10.25}]{NIST_DLMF}.

Computing the moments yields the mean and the variance
\begin{equation}
\mathbb{E}[x_t] = 0 
\qquad
\text{and}
\qquad
\def\nfrac#1#2{#1/#2}
\text{Var}[x_t] = 
\frac{\pi  T}{\sqrt{2} a}
\,
\frac{\left(z^2 \left(I_{\nfrac{5}{4}}(z^2/4)-I_{\nfrac{3}{4}}(z^2/4)\right)+\left(z^2+2\right) I_{\nfrac{1}{4}}(z^2/4)-z^2 I_{-\nfrac{1}{4}}(z^2/4)\right)}{K_{\nfrac{1}{4}}(z^2/4)}
\end{equation}
in which $I_{n}(z)$ is the modified Bessel function of the first kind of order $n$ \cite[\href{https://dlmf.nist.gov/10.25}{\S 10.25}]{NIST_DLMF} and we have defined $z \equiv a/\sqrt{2 b T}$.

We are going to be interested in the limit $a \to 0$, and it is enough for our purposes to consider the case $a=0^{+}$. Computing the limit yields
\begin{equation}
\label{variance_dwp_atozero}
\def\nfrac#1#2{#1/#2}
\text{Var}[x_t ; a = 0^{+}] = \langle x_{t}^2 \rangle =
\sqrt{\frac{T}{b}} \, \frac{\sqrt{2} \, \Gamma^2(3/4)}{\pi}
\end{equation}
in which $\Gamma$ is the Euler gamma function \cite[\href{https://dlmf.nist.gov/5.2}{\S 5.2}]{NIST_DLMF}.

Chapter 6 of Ref.~\cite{Coffey1996} gives an analytic expression for the correlation function. Combined with the variance in Eq.~\eqref{variance_dwp_atozero}, this gives the covariance of $x_t$, from which we will obtain the variance of $y_t$. For small enough $a$ (i.e., small $z$) the covariance is reasonably approximated by the simple expression~\cite{Coffey1996}
\begin{equation}
\def\braket#1{\langle #1 \rangle}
\braket{x_{t} x_{s}} = \braket{x_{t}^2} e^{-t/T_{\text{c}}}
\end{equation}
in which $T_{\text{c}}$ is the correlation time (defined as the area under the curve of the normalized autocorrelation function), that can in turn be approximated as
\begin{equation}
T_{\text{c}}(q) = \tau_0 \, \frac{e^{q} - 1}{2 q} \left( \pi \sqrt{q} + 2^{1 - \sqrt{q}} \right)
\end{equation}
where $q = z^2/2$ and $\tau_0 = 1/{\sqrt{2 b T}}$.
When $a=0$ (i.e. $z=0$, i.e. $q=0$), one has
\begin{equation}
\label{Tcovertauzero}
\frac{T_{\text{c}}}{\tau_0} = \frac{\Gamma^2(1/4)}{4 \sqrt{2} \pi} \left[\pi -2 \log \left(\sqrt{2}+1\right)\right] \simeq 1.01989...
\end{equation}
Hence, we get in the limit $a = 0^{+}$ and when $t$ is large compared to $T_{\text{c}}$
\begin{align}
    \text{Var}[y_t] &= 
    \mathbb{E}[y_t^2] 
    = \omega_1^2
    \int_{0}^{t} dr 
    \int_{0}^{t} ds
    \braket{x(r) x(s)}
\underset{t \gg T_{\text{c}}}{\sim}
\left[ \frac{ 2 \, \Gamma^2(3/4)}{\pi} \right] \left[ \frac{T_{\text{c}}}{\tau_0} \right]
    \omega_1^2 \, \frac{1}{b} \, t \end{align}
using Eq.~\eqref{variance_dwp_atozero} and Eq.~\eqref{annoying_exp_identity} with $\alpha = 1/T_{\text{c}}$. 
In the limit $a \to 0$, the quantity $T_{\text{c}}/\tau_0$ is given in Eq.~\eqref{Tcovertauzero}. This gives the effective diffusion coefficient quoted in the main text (Sec.~\ref{noise_flat}).

\subsection{Analysis of the plankton food web model}
\label{app_plankton}

In this Appendix, we analyze a set of coupled Rosenzweig-MacArthur-like equations used in Ref.~\cite{Beninca2009} to model laboratory experiments on a plankton food web isolated from the Baltic Sea \cite{Beninca2008}. 
Following these references, we consider $N_\text{pred}$ consumers (predators) feeding on $N_\text{prey}$ resources (prey). The populations $u_i(t)$ of resources and $v_k(t)$ of consumers evolve according to
\begin{subequations}
\label{rma_pfw}
\begin{align}
    \frac{\dd u_i}{\dd t} &= r_i u_i \left(1-\sum _{j=1}^{N_\text{prey}} \frac{u_j \alpha_{ij}}{K_i}\right)-\sum_{k=1}^{N_\text{pred}} \frac{g_k u_i v_k \beta_{ik}}{\sum _{j=1}^{N_\text{prey}} u_j \beta_{jk}+H_k} 
    \\
    \frac{\dd v_k}{\dd t} &= \frac{g_k v_k \left(\sum _{j=1}^{N_\text{prey}} u_j \beta_{jk}\right)}{\sum_{j=1}^{N_\text{prey}} u_j \beta_{j,k}+H_k}-m_k v_k
\end{align}
\end{subequations}
Here, we take $N_\text{pred} = N_\text{prey} = 2$ (so the state of the system is described by $X\equiv(u_1,u_2,v_1,v_2)$), and 
$r_1 = r_2 = 1$, 
$K_1 = K_2 = 1$, 
$m_1 = m_2 =  m \simeq \num{0.12}$,
$g_1 = g_2 =  g \simeq \num{1.5}$,
$H_1 = H_2 =  H \simeq \num{1.1}$,
$\alpha_{11} = \alpha_{22} = 1$,
$\alpha_{12} = \alpha_{21} =  \alpha \simeq{1.2}$,
$\beta_{11} = \beta_{22} =  1$,
$\beta_{12} = \beta_{21} = \beta$. The parameter $\beta$ quantifies the coupling through predation, and is taken as a bifurcation parameter. For these values, we find that a bifurcation takes place at $\beta_{\text{c}} \simeq \num{0.02275}$.
We have slightly modified the numerical values used in Ref.~\cite{Beninca2009} without crossing any bifurcation in order to make the system less stiff, and the calculations of the LEs and CLVs easier.
We take as initial conditions $X(0) \equiv (u_1(0), u_2(0), v_1(0), v_2(0)) = (0.5, 0.1, 0.5, 0.5)$ and $(0.1,0.5,0.5,0.5)$. For the parameters considered, these are in the basin of attraction of the only limit cycle below the bifurcation (Fig.~\ref{figure_model_plankton}a-c), and in the respective basins of attractions of the two stable limit cycles above the bifurcation (Fig.~\ref{figure_model_plankton}d-i).

A direct integration of Eqs.~\eqref{rma_pfw} is performed using a LSODA approach with automatic switching between a non-stiff Adams method and a stiff BDF method as implemented in Mathematica \texttt{NDSolve}. The numerical integration is carried out for a long number of periods (of the order of \num{5000} periods) in order to obtain a good approximation $X(t)$ of the limit cycle. A precise estimation of the period $T$ is obtained by tracking the extrema of $u_i(t)$ using a Brent-Dekker root-finding method as implemented by Mathematica \texttt{WhenEvent}. We then solve
\begin{equation}
\delta \dot{U} = J(X(t)) \, U
\end{equation}
with initial condition $U(0) = \text{Id}$ (\text{Id} is the identity matrix) from $t=0$ to $t=T$, in which the Jacobian of the vector field defined by the right-hand side of Eqs.~\eqref{rma_pfw} is computed analytically. Evaluating the solution $U(t)$ at the final time gives the Floquet operator $U(T)$. 

Diagonalizing the Floquet operator $U(T)$ produces eigenvalues $\mu_i$ and eigenvectors $\vec{w_i}$. The LEs are identified as $\lambda_i = \log |\mu_i|$, while the CLVs at time $t=0$ (modulo a period) are $\vec{c_i}(0) = \vec{w_i}$, perhaps up to a normalization. We then apply the evolution operator $U(t)$ to find $\vec{c_i}(t) = U(t) \vec{c_i}(0)$.
Finally, we compute the angles $\theta_{i j}(t)$ between the CLVs, that satisfy $\cos \theta_{i j}(t) = \vec{c_i}(t)  \cdot \vec{c}_j(t) /(\lVert \vec{c_i}(t) \rVert \, \lVert \vec{c_j}(t) \rVert)$. We emphasize that these angles depend on time. In order to quantify how align are the CLVs, we compute the average of $[\sin \theta_{ij}(t)]^2$ over one period.

Typical trajectories are shown in Fig.~\ref{figure_model_plankton}a-i. We observe that below a critical value $\beta_{\text{c}} \simeq \num{0.02275}$ of the parameter $\beta$ describing coupling through predation, a single limit cycle exists, with the two predators in antiphase (as well as the prey), see Fig.~\ref{figure_model_plankton}a-c. As shown in Fig.~\ref{figure_model_plankton}c, the projections of the cycle in the $(u_1, v_1)$ and $(u_2, v_2)$ planes are identical. For $\beta > \beta_{\text{c}}$, two limit cycles are present, with the two predators oscillating with a dephasing $\delta \phi$ that is neither zero nor $\pi$ (Fig.~\ref{figure_model_plankton}d-i). Figure~\ref{figure_model_plankton}f and i show that the projections of each cycle on the $(u_1, v_1)$ and $(u_2, v_2)$ planes are now different, and the roles of both populations are exchanged between the two limit cycles. 
We plot as a function of $\beta$ the two Lyapunov exponent closest to zero ($\lambda_1 \equiv 0$ and $\lambda_2$) in Fig.~\ref{figure_model_plankton}j and the average angle between the corresponding covariant vectors in Fig.~\ref{figure_model_plankton}k (more precisely, we plot the average $\langle\sin^2(\theta_{12})\rangle$ over one period, in which $\theta_{12}$ is the angle between the CLV $\vec{c}_i$ ($i=1,2$), because only the directions corresponding to the CLVs are physically meaningful - so $\pm \vec{c}_i$ are equivalent). We find that the Lyapunov exponents vanish at the bifurcation. The angle $\theta_{12}$ between the CLVs also vanishes, and its behavior is discontinuous from the left through the bifurcation. Panel l shows the dephasing between the predator densities.
Finally, Fig.~\ref{figure_model_plankton}m shows the instantaneous quantity $\sin^2\theta_{12}(t)$ over one period for different values of $\beta$. We observe that the angle between the CLVs depends on the location on the periodic orbit. Even away from the generalized EP, small values of the angles may occur at some points of the limit cycle. For $\beta \to \beta_{\text{c}}^+$, the instantaneous angle $\theta_{12}(t)$ becomes arbitrarily small all along the limit cycle (black curve).

\twocolumngrid


\begin{thebibliography}{199}\makeatletter
\providecommand \@ifxundefined [1]{\@ifx{#1\undefined}
}\providecommand \@ifnum [1]{\ifnum #1\expandafter \@firstoftwo
 \else \expandafter \@secondoftwo
 \fi
}\providecommand \@ifx [1]{\ifx #1\expandafter \@firstoftwo
 \else \expandafter \@secondoftwo
 \fi
}\providecommand \natexlab [1]{#1}\providecommand \enquote  [1]{``#1''}\providecommand \bibnamefont  [1]{#1}\providecommand \bibfnamefont [1]{#1}\providecommand \citenamefont [1]{#1}\providecommand \href@noop [0]{\@secondoftwo}\providecommand \href [0]{\begingroup \@sanitize@url \@href}\providecommand \@href[1]{\@@startlink{#1}\@@href}\providecommand \@@href[1]{\endgroup#1\@@endlink}\providecommand \@sanitize@url [0]{\catcode `\\12\catcode `\$12\catcode
  `\&12\catcode `\#12\catcode `\^12\catcode `\_12\catcode `\%12\relax}\providecommand \@@startlink[1]{}\providecommand \@@endlink[0]{}\providecommand \url  [0]{\begingroup\@sanitize@url \@url }\providecommand \@url [1]{\endgroup\@href {#1}{\urlprefix }}\providecommand \urlprefix  [0]{URL }\providecommand \Eprint [0]{\href }\providecommand \doibase [0]{https://doi.org/}\providecommand \selectlanguage [0]{\@gobble}\providecommand \bibinfo  [0]{\@secondoftwo}\providecommand \bibfield  [0]{\@secondoftwo}\providecommand \translation [1]{[#1]}\providecommand \BibitemOpen [0]{}\providecommand \bibitemStop [0]{}\providecommand \bibitemNoStop [0]{.\EOS\space}\providecommand \EOS [0]{\spacefactor3000\relax}\providecommand \BibitemShut  [1]{\csname bibitem#1\endcsname}\let\auto@bib@innerbib\@empty
\bibitem [{\citenamefont {Shankar}\ \emph {et~al.}(2022)\citenamefont
  {Shankar}, \citenamefont {Souslov}, \citenamefont {Bowick}, \citenamefont
  {Marchetti},\ and\ \citenamefont {Vitelli}}]{Shankar2022}\BibitemOpen
  \bibfield  {author} {\bibinfo {author} {\bibfnamefont {S.}~\bibnamefont
  {Shankar}}, \bibinfo {author} {\bibfnamefont {A.}~\bibnamefont {Souslov}},
  \bibinfo {author} {\bibfnamefont {M.~J.}\ \bibnamefont {Bowick}}, \bibinfo
  {author} {\bibfnamefont {M.~C.}\ \bibnamefont {Marchetti}},\ and\ \bibinfo
  {author} {\bibfnamefont {V.}~\bibnamefont {Vitelli}},\ }\bibfield  {title}
  {\bibinfo {title} {Topological active matter},\ }\href
  {https://doi.org/10.1038/s42254-022-00445-3} {\bibfield  {journal} {\bibinfo
  {journal} {Nature Reviews Physics}\ }\textbf {\bibinfo {volume} {4}},\
  \bibinfo {pages} {380–398} (\bibinfo {year} {2022})}\BibitemShut {NoStop}\bibitem [{\citenamefont {Ashida}\ \emph {et~al.}(2020)\citenamefont {Ashida},
  \citenamefont {Gong},\ and\ \citenamefont {Ueda}}]{Ashida2020}\BibitemOpen
  \bibfield  {author} {\bibinfo {author} {\bibfnamefont {Y.}~\bibnamefont
  {Ashida}}, \bibinfo {author} {\bibfnamefont {Z.}~\bibnamefont {Gong}},\ and\
  \bibinfo {author} {\bibfnamefont {M.}~\bibnamefont {Ueda}},\ }\bibfield
  {title} {\bibinfo {title} {Non-{H}ermitian physics},\ }\href
  {https://doi.org/10.1080/00018732.2021.1876991} {\bibfield  {journal}
  {\bibinfo  {journal} {Advances in Physics}\ }\textbf {\bibinfo {volume}
  {69}},\ \bibinfo {pages} {249–435} (\bibinfo {year} {2020})}\BibitemShut
  {NoStop}\bibitem [{\citenamefont {Nassar}\ \emph {et~al.}(2020)\citenamefont {Nassar},
  \citenamefont {Yousefzadeh}, \citenamefont {Fleury}, \citenamefont {Ruzzene},
  \citenamefont {Alù}, \citenamefont {Daraio}, \citenamefont {Norris},
  \citenamefont {Huang},\ and\ \citenamefont {Haberman}}]{Nassar2020}\BibitemOpen
  \bibfield  {author} {\bibinfo {author} {\bibfnamefont {H.}~\bibnamefont
  {Nassar}}, \bibinfo {author} {\bibfnamefont {B.}~\bibnamefont {Yousefzadeh}},
  \bibinfo {author} {\bibfnamefont {R.}~\bibnamefont {Fleury}}, \bibinfo
  {author} {\bibfnamefont {M.}~\bibnamefont {Ruzzene}}, \bibinfo {author}
  {\bibfnamefont {A.}~\bibnamefont {Alù}}, \bibinfo {author} {\bibfnamefont
  {C.}~\bibnamefont {Daraio}}, \bibinfo {author} {\bibfnamefont {A.~N.}\
  \bibnamefont {Norris}}, \bibinfo {author} {\bibfnamefont {G.}~\bibnamefont
  {Huang}},\ and\ \bibinfo {author} {\bibfnamefont {M.~R.}\ \bibnamefont
  {Haberman}},\ }\bibfield  {title} {\bibinfo {title} {Nonreciprocity in
  acoustic and elastic materials},\ }\href
  {https://doi.org/10.1038/s41578-020-0206-0} {\bibfield  {journal} {\bibinfo
  {journal} {Nature Reviews Materials}\ }\textbf {\bibinfo {volume} {5}},\
  \bibinfo {pages} {667–685} (\bibinfo {year} {2020})}\BibitemShut {NoStop}\bibitem [{\citenamefont {Bergholtz}\ \emph {et~al.}(2021)\citenamefont
  {Bergholtz}, \citenamefont {Budich},\ and\ \citenamefont
  {Kunst}}]{Bergholtz2021}\BibitemOpen
  \bibfield  {author} {\bibinfo {author} {\bibfnamefont {E.~J.}\ \bibnamefont
  {Bergholtz}}, \bibinfo {author} {\bibfnamefont {J.~C.}\ \bibnamefont
  {Budich}},\ and\ \bibinfo {author} {\bibfnamefont {F.~K.}\ \bibnamefont
  {Kunst}},\ }\bibfield  {title} {\bibinfo {title} {Exceptional topology of
  non-{H}ermitian systems},\ }\href
  {https://doi.org/10.1103/revmodphys.93.015005} {\bibfield  {journal}
  {\bibinfo  {journal} {Reviews of Modern Physics}\ }\textbf {\bibinfo {volume}
  {93}},\ \bibinfo {pages} {015005} (\bibinfo {year} {2021})}\BibitemShut
  {NoStop}\bibitem [{\citenamefont {Clerk}(2022)}]{Clerk2022}\BibitemOpen
  \bibfield  {author} {\bibinfo {author} {\bibfnamefont {A.}~\bibnamefont
  {Clerk}},\ }\bibfield  {title} {\bibinfo {title} {Introduction to quantum
  non-reciprocal interactions: from non-{H}ermitian {H}amiltonians to quantum
  master equations and quantum feedforward schemes},\ }\bibfield  {journal}
  {\bibinfo  {journal} {SciPost Physics Lecture Notes}\ }\href
  {https://doi.org/10.21468/scipostphyslectnotes.44}
  {10.21468/scipostphyslectnotes.44} (\bibinfo {year} {2022})\BibitemShut
  {NoStop}\bibitem [{\citenamefont {Fruchart}\ \emph {et~al.}(2023)\citenamefont
  {Fruchart}, \citenamefont {Scheibner},\ and\ \citenamefont
  {Vitelli}}]{Fruchart2023}\BibitemOpen
  \bibfield  {author} {\bibinfo {author} {\bibfnamefont {M.}~\bibnamefont
  {Fruchart}}, \bibinfo {author} {\bibfnamefont {C.}~\bibnamefont
  {Scheibner}},\ and\ \bibinfo {author} {\bibfnamefont {V.}~\bibnamefont
  {Vitelli}},\ }\bibfield  {title} {\bibinfo {title} {Odd viscosity and odd
  elasticity},\ }\href
  {https://doi.org/10.1146/annurev-conmatphys-040821-125506} {\bibfield
  {journal} {\bibinfo  {journal} {Annual Review of Condensed Matter Physics}\
  }\textbf {\bibinfo {volume} {14}},\ \bibinfo {pages} {471–510} (\bibinfo
  {year} {2023})}\BibitemShut {NoStop}\bibitem [{\citenamefont {Kato}(1984)}]{Kato1984}\BibitemOpen
  \bibfield  {author} {\bibinfo {author} {\bibfnamefont {T.}~\bibnamefont
  {Kato}},\ }\href@noop {} {\emph {\bibinfo {title} {Perturbation theory for
  linear operators}}},\ \bibinfo {edition} {2nd}\ ed.\ (\bibinfo  {publisher}
  {Springer},\ \bibinfo {year} {1984})\BibitemShut {NoStop}\bibitem [{\citenamefont {Landau}\ and\ \citenamefont
  {Lifshitz}(1976)}]{landaumechanics}\BibitemOpen
  \bibfield  {author} {\bibinfo {author} {\bibfnamefont {L.}~\bibnamefont
  {Landau}}\ and\ \bibinfo {author} {\bibfnamefont {E.}~\bibnamefont
  {Lifshitz}},\ }\href@noop {} {\emph {\bibinfo {title} {Mechanics: Volume
  1}}}\ (\bibinfo  {publisher} {Butterworth Heinemann},\ \bibinfo {year}
  {1976})\BibitemShut {NoStop}\bibitem [{\citenamefont {Strogatz}(2019)}]{StrogatzNL}\BibitemOpen
  \bibfield  {author} {\bibinfo {author} {\bibfnamefont {S.}~\bibnamefont
  {Strogatz}},\ }\href@noop {} {\emph {\bibinfo {title} {Nonlinear Dynamics and
  Chaos: With Applications to Physics, Biology, Chemistry, and Engineering}}},\
  Chapman \& Hall book\ (\bibinfo  {publisher} {CRC Press},\ \bibinfo {year}
  {2019})\BibitemShut {NoStop}\bibitem [{\citenamefont {Manneville}(2010)}]{Manneville2010}\BibitemOpen
  \bibfield  {author} {\bibinfo {author} {\bibfnamefont {P.}~\bibnamefont
  {Manneville}},\ }\href@noop {} {\emph {\bibinfo {title} {Instabilities, Chaos
  and Turbulence}}},\ ICP fluid mechanics\ (\bibinfo  {publisher} {Imperial
  College Press},\ \bibinfo {year} {2010})\BibitemShut {NoStop}\bibitem [{\citenamefont {Murray}(2013)}]{Murray2013}\BibitemOpen
  \bibfield  {author} {\bibinfo {author} {\bibfnamefont {J.}~\bibnamefont
  {Murray}},\ }\href@noop {} {\emph {\bibinfo {title} {Mathematical Biology: I.
  An Introduction}}},\ Interdisciplinary Applied Mathematics\ (\bibinfo
  {publisher} {Springer New York},\ \bibinfo {year} {2013})\BibitemShut
  {NoStop}\bibitem [{\citenamefont {Golubitsky}\ and\ \citenamefont
  {Schaeffer}(1985)}]{Golubitsky1985b}\BibitemOpen
  \bibfield  {author} {\bibinfo {author} {\bibfnamefont {M.}~\bibnamefont
  {Golubitsky}}\ and\ \bibinfo {author} {\bibfnamefont {D.~G.}\ \bibnamefont
  {Schaeffer}},\ }\href {https://doi.org/10.1007/978-1-4612-5034-0} {\emph
  {\bibinfo {title} {{Singularities and Groups in Bifurcation Theory}}}},\
  Vol.~\bibinfo {volume} {I}\ (\bibinfo  {publisher} {Springer New York},\
  \bibinfo {year} {1985})\BibitemShut {NoStop}\bibitem [{\citenamefont {Golubitsky}\ \emph {et~al.}(1988)\citenamefont
  {Golubitsky}, \citenamefont {Stewart},\ and\ \citenamefont
  {Schaeffer}}]{Golubitsky1988}\BibitemOpen
  \bibfield  {author} {\bibinfo {author} {\bibfnamefont {M.}~\bibnamefont
  {Golubitsky}}, \bibinfo {author} {\bibfnamefont {I.}~\bibnamefont
  {Stewart}},\ and\ \bibinfo {author} {\bibfnamefont {D.~G.}\ \bibnamefont
  {Schaeffer}},\ }\href {https://doi.org/10.1007/978-1-4612-4574-2} {\emph
  {\bibinfo {title} {{Singularities and Groups in Bifurcation Theory}}}},\
  Vol.~\bibinfo {volume} {II}\ (\bibinfo  {publisher} {Springer New York},\
  \bibinfo {year} {1988})\BibitemShut {NoStop}\bibitem [{\citenamefont {Golubitsky}\ and\ \citenamefont
  {Stewart}(2002)}]{Golubitsky2002}\BibitemOpen
  \bibfield  {author} {\bibinfo {author} {\bibfnamefont {M.}~\bibnamefont
  {Golubitsky}}\ and\ \bibinfo {author} {\bibfnamefont {I.}~\bibnamefont
  {Stewart}},\ }\href {https://doi.org/10.1007/978-3-0348-8167-8} {\emph
  {\bibinfo {title} {{The Symmetry Perspective}}}}\ (\bibinfo  {publisher}
  {Birkhäuser Basel},\ \bibinfo {year} {2002})\BibitemShut {NoStop}\bibitem [{\citenamefont {Crawford}\ and\ \citenamefont
  {Knobloch}(1991)}]{Crawford1991}\BibitemOpen
  \bibfield  {author} {\bibinfo {author} {\bibfnamefont {J.~D.}\ \bibnamefont
  {Crawford}}\ and\ \bibinfo {author} {\bibfnamefont {E.}~\bibnamefont
  {Knobloch}},\ }\bibfield  {title} {\bibinfo {title} {Symmetry and
  symmetry-breaking bifurcations in fluid dynamics},\ }\href
  {https://doi.org/10.1146/annurev.fl.23.010191.002013} {\bibfield  {journal}
  {\bibinfo  {journal} {Annual Review of Fluid Mechanics}\ }\textbf {\bibinfo
  {volume} {23}},\ \bibinfo {pages} {341–387} (\bibinfo {year}
  {1991})}\BibitemShut {NoStop}\bibitem [{\citenamefont {Timoshenko}\ and\ \citenamefont
  {Gere}(1961)}]{Timoshenko1961}\BibitemOpen
  \bibfield  {author} {\bibinfo {author} {\bibfnamefont {S.}~\bibnamefont
  {Timoshenko}}\ and\ \bibinfo {author} {\bibfnamefont {J.}~\bibnamefont
  {Gere}},\ }\href@noop {} {\emph {\bibinfo {title} {Theory of Elastic
  Stability}}},\ Dover Civil and Mechanical Engineering\ (\bibinfo  {publisher}
  {Dover Publications},\ \bibinfo {year} {1961})\BibitemShut {NoStop}\bibitem [{\citenamefont {Fruchart}\ \emph {et~al.}(2021)\citenamefont
  {Fruchart}, \citenamefont {Hanai}, \citenamefont {Littlewood},\ and\
  \citenamefont {Vitelli}}]{Fruchart2021}\BibitemOpen
  \bibfield  {author} {\bibinfo {author} {\bibfnamefont {M.}~\bibnamefont
  {Fruchart}}, \bibinfo {author} {\bibfnamefont {R.}~\bibnamefont {Hanai}},
  \bibinfo {author} {\bibfnamefont {P.~B.}\ \bibnamefont {Littlewood}},\ and\
  \bibinfo {author} {\bibfnamefont {V.}~\bibnamefont {Vitelli}},\ }\bibfield
  {title} {\bibinfo {title} {Non-reciprocal phase transitions},\ }\href
  {https://doi.org/10.1038/s41586-021-03375-9} {\bibfield  {journal} {\bibinfo
  {journal} {Nature}\ }\textbf {\bibinfo {volume} {592}},\ \bibinfo {pages}
  {363} (\bibinfo {year} {2021})}\BibitemShut {NoStop}\bibitem [{\citenamefont {Pikovsky}\ and\ \citenamefont
  {Politi}(2016)}]{Pikovsky2016}\BibitemOpen
  \bibfield  {author} {\bibinfo {author} {\bibfnamefont {A.}~\bibnamefont
  {Pikovsky}}\ and\ \bibinfo {author} {\bibfnamefont {A.}~\bibnamefont
  {Politi}},\ }\href@noop {} {\emph {\bibinfo {title} {Lyapunov Exponents: A
  Tool to Explore Complex Dynamics}}}\ (\bibinfo  {publisher} {Cambridge
  University Press},\ \bibinfo {year} {2016})\BibitemShut {NoStop}\bibitem [{\citenamefont {Eckmann}\ and\ \citenamefont
  {Ruelle}(1985)}]{Eckmann1985}\BibitemOpen
  \bibfield  {author} {\bibinfo {author} {\bibfnamefont {J.~P.}\ \bibnamefont
  {Eckmann}}\ and\ \bibinfo {author} {\bibfnamefont {D.}~\bibnamefont
  {Ruelle}},\ }\bibfield  {title} {\bibinfo {title} {Ergodic theory of chaos
  and strange attractors},\ }\href {https://doi.org/10.1103/revmodphys.57.617}
  {\bibfield  {journal} {\bibinfo  {journal} {Reviews of Modern Physics}\
  }\textbf {\bibinfo {volume} {57}},\ \bibinfo {pages} {617} (\bibinfo {year}
  {1985})}\BibitemShut {NoStop}\bibitem [{\citenamefont {Benincà}\ \emph {et~al.}(2008)\citenamefont
  {Benincà}, \citenamefont {Huisman}, \citenamefont {Heerkloss}, \citenamefont
  {Jöhnk}, \citenamefont {Branco}, \citenamefont {Van~Nes}, \citenamefont
  {Scheffer},\ and\ \citenamefont {Ellner}}]{Beninca2008}\BibitemOpen
  \bibfield  {author} {\bibinfo {author} {\bibfnamefont {E.}~\bibnamefont
  {Benincà}}, \bibinfo {author} {\bibfnamefont {J.}~\bibnamefont {Huisman}},
  \bibinfo {author} {\bibfnamefont {R.}~\bibnamefont {Heerkloss}}, \bibinfo
  {author} {\bibfnamefont {K.~D.}\ \bibnamefont {Jöhnk}}, \bibinfo {author}
  {\bibfnamefont {P.}~\bibnamefont {Branco}}, \bibinfo {author} {\bibfnamefont
  {E.~H.}\ \bibnamefont {Van~Nes}}, \bibinfo {author} {\bibfnamefont
  {M.}~\bibnamefont {Scheffer}},\ and\ \bibinfo {author} {\bibfnamefont
  {S.~P.}\ \bibnamefont {Ellner}},\ }\bibfield  {title} {\bibinfo {title}
  {Chaos in a long-term experiment with a plankton community},\ }\href
  {https://doi.org/10.1038/nature06512} {\bibfield  {journal} {\bibinfo
  {journal} {Nature}\ }\textbf {\bibinfo {volume} {451}},\ \bibinfo {pages}
  {822–825} (\bibinfo {year} {2008})}\BibitemShut {NoStop}\bibitem [{\citenamefont {Wang}\ and\ \citenamefont {Zhang}(2023)}]{Wang2023}\BibitemOpen
  \bibfield  {author} {\bibinfo {author} {\bibfnamefont {K.}~\bibnamefont
  {Wang}}\ and\ \bibinfo {author} {\bibfnamefont {J.}~\bibnamefont {Zhang}},\
  }\bibfield  {title} {\bibinfo {title} {Persistent corotation of the
  large-scale flow of thermal convection and an immersed free body},\
  }\bibfield  {journal} {\bibinfo  {journal} {Proceedings of the National
  Academy of Sciences}\ }\textbf {\bibinfo {volume} {120}},\ \href
  {https://doi.org/10.1073/pnas.2217705120} {10.1073/pnas.2217705120} (\bibinfo
  {year} {2023})\BibitemShut {NoStop}\bibitem [{Note1()}]{Note1}\BibitemOpen
  \bibinfo {note} {Technically, $\delta X(t)$ is a tangent vector, i.e. an
  element of the tangent space to the unperturbed trajectory at point $X(t)$.
  For a fixed point, $X(t) = X_0$.}\BibitemShut {Stop}\bibitem [{\citenamefont {Seyranian}\ \emph {et~al.}(2005)\citenamefont
  {Seyranian}, \citenamefont {Kirillov},\ and\ \citenamefont
  {Mailybaev}}]{Seyranian2005}\BibitemOpen
  \bibfield  {author} {\bibinfo {author} {\bibfnamefont {A.~P.}\ \bibnamefont
  {Seyranian}}, \bibinfo {author} {\bibfnamefont {O.~N.}\ \bibnamefont
  {Kirillov}},\ and\ \bibinfo {author} {\bibfnamefont {A.~A.}\ \bibnamefont
  {Mailybaev}},\ }\bibfield  {title} {\bibinfo {title} {Coupling of eigenvalues
  of complex matrices at diabolic and exceptional points},\ }\href
  {https://doi.org/10.1088/0305-4470/38/8/009} {\bibfield  {journal} {\bibinfo
  {journal} {Journal of Physics A: Mathematical and General}\ }\textbf
  {\bibinfo {volume} {38}},\ \bibinfo {pages} {1723–1740} (\bibinfo {year}
  {2005})}\BibitemShut {NoStop}\bibitem [{\citenamefont {Lang}(2010)}]{Lang2010}\BibitemOpen
  \bibfield  {author} {\bibinfo {author} {\bibfnamefont {S.}~\bibnamefont
  {Lang}},\ }\href@noop {} {\emph {\bibinfo {title} {Linear Algebra}}},\
  Undergraduate Texts in Mathematics\ (\bibinfo  {publisher} {Springer New
  York},\ \bibinfo {year} {2010})\BibitemShut {NoStop}\bibitem [{\citenamefont {Trefethen}\ and\ \citenamefont
  {Embree}(2005)}]{Trefethen2005}\BibitemOpen
  \bibfield  {author} {\bibinfo {author} {\bibfnamefont {L.}~\bibnamefont
  {Trefethen}}\ and\ \bibinfo {author} {\bibfnamefont {M.}~\bibnamefont
  {Embree}},\ }\href@noop {} {\emph {\bibinfo {title} {Spectra and
  Pseudospectra: The Behavior of Nonnormal Matrices and Operators}}}\ (\bibinfo
   {publisher} {Princeton University Press},\ \bibinfo {year}
  {2005})\BibitemShut {NoStop}\bibitem [{\citenamefont {Kuznetsov}(2004)}]{Kuznetsov2004}\BibitemOpen
  \bibfield  {author} {\bibinfo {author} {\bibfnamefont {Y.~A.}\ \bibnamefont
  {Kuznetsov}},\ }\href {https://doi.org/10.1007/978-1-4757-3978-7} {\emph
  {\bibinfo {title} {Elements of Applied Bifurcation Theory}}}\ (\bibinfo
  {publisher} {Springer New York},\ \bibinfo {year} {2004})\BibitemShut
  {NoStop}\bibitem [{\citenamefont {Bazykin}(1998)}]{Bazykin1998}\BibitemOpen
  \bibfield  {author} {\bibinfo {author} {\bibfnamefont {A.~D.}\ \bibnamefont
  {Bazykin}},\ }\href@noop {} {\emph {\bibinfo {title} {Nonlinear dynamics of
  interacting populations}}},\ World Scientific series on nonlinear science,
  Series A, Monographs and treatises, v. 11\ (\bibinfo  {publisher} {World
  Scientific},\ \bibinfo {year} {1998})\BibitemShut {NoStop}\bibitem [{\citenamefont {Izhikevich}(2007)}]{Izhikevich2007}\BibitemOpen
  \bibfield  {author} {\bibinfo {author} {\bibfnamefont {E.~M.}\ \bibnamefont
  {Izhikevich}},\ }\href@noop {} {\emph {\bibinfo {title} {Dynamical Systems In
  Neuroscience}}}\ (\bibinfo  {publisher} {MIT Press},\ \bibinfo {year}
  {2007})\BibitemShut {NoStop}\bibitem [{\citenamefont {Renardy}\ \emph {et~al.}(1999)\citenamefont
  {Renardy}, \citenamefont {Renardy},\ and\ \citenamefont
  {Fujimura}}]{Renardy1999}\BibitemOpen
  \bibfield  {author} {\bibinfo {author} {\bibfnamefont {Y.~Y.}\ \bibnamefont
  {Renardy}}, \bibinfo {author} {\bibfnamefont {M.}~\bibnamefont {Renardy}},\
  and\ \bibinfo {author} {\bibfnamefont {K.}~\bibnamefont {Fujimura}},\
  }\bibfield  {title} {\bibinfo {title} {Takens–{B}ogdanov bifurcation on the
  hexagonal lattice for double-layer convection},\ }\href
  {https://doi.org/10.1016/s0167-2789(99)00007-x} {\bibfield  {journal}
  {\bibinfo  {journal} {Physica D: Nonlinear Phenomena}\ }\textbf {\bibinfo
  {volume} {129}},\ \bibinfo {pages} {171–202} (\bibinfo {year}
  {1999})}\BibitemShut {NoStop}\bibitem [{\citenamefont {Jaeger}\ and\ \citenamefont
  {Kantz}(1997)}]{Jaeger1997}\BibitemOpen
  \bibfield  {author} {\bibinfo {author} {\bibfnamefont {L.}~\bibnamefont
  {Jaeger}}\ and\ \bibinfo {author} {\bibfnamefont {H.}~\bibnamefont {Kantz}},\
  }\bibfield  {title} {\bibinfo {title} {Homoclinic tangencies and non-normal
  jacobians — effects of noise in nonhyperbolic chaotic systems},\ }\href
  {https://doi.org/10.1016/s0167-2789(97)00247-9} {\bibfield  {journal}
  {\bibinfo  {journal} {Physica D: Nonlinear Phenomena}\ }\textbf {\bibinfo
  {volume} {105}},\ \bibinfo {pages} {79–96} (\bibinfo {year}
  {1997})}\BibitemShut {NoStop}\bibitem [{\citenamefont {Biancalani}\ \emph {et~al.}(2017)\citenamefont
  {Biancalani}, \citenamefont {Jafarpour},\ and\ \citenamefont
  {Goldenfeld}}]{Biancalani2017}\BibitemOpen
  \bibfield  {author} {\bibinfo {author} {\bibfnamefont {T.}~\bibnamefont
  {Biancalani}}, \bibinfo {author} {\bibfnamefont {F.}~\bibnamefont
  {Jafarpour}},\ and\ \bibinfo {author} {\bibfnamefont {N.}~\bibnamefont
  {Goldenfeld}},\ }\bibfield  {title} {\bibinfo {title} {Giant amplification of
  noise in fluctuation-induced pattern formation},\ }\href
  {https://doi.org/10.1103/physrevlett.118.018101} {\bibfield  {journal}
  {\bibinfo  {journal} {Physical Review Letters}\ }\textbf {\bibinfo {volume}
  {118}},\ \bibinfo {pages} {018101} (\bibinfo {year} {2017})}\BibitemShut
  {NoStop}\bibitem [{\citenamefont {Chajwa}\ \emph {et~al.}(2020)\citenamefont {Chajwa},
  \citenamefont {Menon}, \citenamefont {Ramaswamy},\ and\ \citenamefont
  {Govindarajan}}]{Chajwa2020}\BibitemOpen
  \bibfield  {author} {\bibinfo {author} {\bibfnamefont {R.}~\bibnamefont
  {Chajwa}}, \bibinfo {author} {\bibfnamefont {N.}~\bibnamefont {Menon}},
  \bibinfo {author} {\bibfnamefont {S.}~\bibnamefont {Ramaswamy}},\ and\
  \bibinfo {author} {\bibfnamefont {R.}~\bibnamefont {Govindarajan}},\
  }\bibfield  {title} {\bibinfo {title} {Waves, algebraic growth, and clumping
  in sedimenting disk arrays},\ }\href
  {https://doi.org/10.1103/physrevx.10.041016} {\bibfield  {journal} {\bibinfo
  {journal} {Physical Review X}\ }\textbf {\bibinfo {volume} {10}},\ \bibinfo
  {pages} {041016} (\bibinfo {year} {2020})}\BibitemShut {NoStop}\bibitem [{Note2()}]{Note2}\BibitemOpen
  \bibinfo {note} {Formally, the evolution operator $U(t,t_0)$ is defined as
  the unique solution of the Cauchy problem \protect \textup {{\protect
  \normalfont (\ref {eom_perturbation}}\protect \normalfont )} with $\delta
  X(t_0) = \protect \text {Id}$, with $\protect \text {Id}$ the identity
  matrix.}\BibitemShut {Stop}\bibitem [{\citenamefont {Sakurai}(1993)}]{SakuraiModernQM}\BibitemOpen
  \bibfield  {author} {\bibinfo {author} {\bibfnamefont {J.~J.}\ \bibnamefont
  {Sakurai}},\ }\href@noop {} {\emph {\bibinfo {title} {Modern Quantum
  Mechanics}}},\ \bibinfo {edition} {1st}\ ed.\ (\bibinfo  {publisher} {Addison
  Wesley},\ \bibinfo {year} {1993})\BibitemShut {NoStop}\bibitem [{\citenamefont {Le~Bellac}(2012)}]{LeBellac}\BibitemOpen
  \bibfield  {author} {\bibinfo {author} {\bibfnamefont {M.}~\bibnamefont
  {Le~Bellac}},\ }\href@noop {} {\emph {\bibinfo {title} {Quantum Physics}}}\
  (\bibinfo  {publisher} {Cambridge University Press},\ \bibinfo {year}
  {2012})\BibitemShut {NoStop}\bibitem [{\citenamefont {Kuptsov}\ and\ \citenamefont
  {Parlitz}(2012)}]{Kuptsov2012}\BibitemOpen
  \bibfield  {author} {\bibinfo {author} {\bibfnamefont {P.~V.}\ \bibnamefont
  {Kuptsov}}\ and\ \bibinfo {author} {\bibfnamefont {U.}~\bibnamefont
  {Parlitz}},\ }\bibfield  {title} {\bibinfo {title} {Theory and computation of
  covariant {L}yapunov vectors},\ }\href
  {https://doi.org/10.1007/s00332-012-9126-5} {\bibfield  {journal} {\bibinfo
  {journal} {Journal of Nonlinear Science}\ }\textbf {\bibinfo {volume} {22}},\
  \bibinfo {pages} {727} (\bibinfo {year} {2012})}\BibitemShut {NoStop}\bibitem [{\citenamefont {Schubert}\ and\ \citenamefont
  {Lucarini}(2015)}]{Schubert2015}\BibitemOpen
  \bibfield  {author} {\bibinfo {author} {\bibfnamefont {S.}~\bibnamefont
  {Schubert}}\ and\ \bibinfo {author} {\bibfnamefont {V.}~\bibnamefont
  {Lucarini}},\ }\bibfield  {title} {\bibinfo {title} {Covariant {L}yapunov
  vectors of a quasi‐geostrophic baroclinic model: analysis of instabilities
  and feedbacks},\ }\href {https://doi.org/10.1002/qj.2588} {\bibfield
  {journal} {\bibinfo  {journal} {Quarterly Journal of the Royal Meteorological
  Society}\ }\textbf {\bibinfo {volume} {141}},\ \bibinfo {pages} {3040–3055}
  (\bibinfo {year} {2015})}\BibitemShut {NoStop}\bibitem [{\citenamefont {Gaspard}\ and\ \citenamefont
  {Dorfman}(1995)}]{Gaspard1995}\BibitemOpen
  \bibfield  {author} {\bibinfo {author} {\bibfnamefont {P.}~\bibnamefont
  {Gaspard}}\ and\ \bibinfo {author} {\bibfnamefont {J.~R.}\ \bibnamefont
  {Dorfman}},\ }\bibfield  {title} {\bibinfo {title} {Chaotic scattering
  theory, thermodynamic formalism, and transport coefficients},\ }\href
  {https://doi.org/10.1103/physreve.52.3525} {\bibfield  {journal} {\bibinfo
  {journal} {Physical Review E}\ }\textbf {\bibinfo {volume} {52}},\ \bibinfo
  {pages} {3525–3552} (\bibinfo {year} {1995})}\BibitemShut {NoStop}\bibitem [{Note3()}]{Note3}\BibitemOpen
  \bibinfo {note} {We refer the reader to \cite {Trevisan1998,Huhn2019}, that
  includes a discussion of degenerate cases.}\BibitemShut {Stop}\bibitem [{\citenamefont {Ginelli}\ \emph {et~al.}(2007)\citenamefont
  {Ginelli}, \citenamefont {Poggi}, \citenamefont {Turchi}, \citenamefont
  {Chat{\'{e}}}, \citenamefont {Livi},\ and\ \citenamefont
  {Politi}}]{Ginelli2007}\BibitemOpen
  \bibfield  {author} {\bibinfo {author} {\bibfnamefont {F.}~\bibnamefont
  {Ginelli}}, \bibinfo {author} {\bibfnamefont {P.}~\bibnamefont {Poggi}},
  \bibinfo {author} {\bibfnamefont {A.}~\bibnamefont {Turchi}}, \bibinfo
  {author} {\bibfnamefont {H.}~\bibnamefont {Chat{\'{e}}}}, \bibinfo {author}
  {\bibfnamefont {R.}~\bibnamefont {Livi}},\ and\ \bibinfo {author}
  {\bibfnamefont {A.}~\bibnamefont {Politi}},\ }\bibfield  {title} {\bibinfo
  {title} {Characterizing dynamics with covariant {L}yapunov vectors},\ }\href
  {https://doi.org/10.1103/physrevlett.99.130601} {\bibfield  {journal}
  {\bibinfo  {journal} {Physical Review Letters}\ }\textbf {\bibinfo {volume}
  {99}},\ \bibinfo {pages} {130601} (\bibinfo {year} {2007})}\BibitemShut
  {NoStop}\bibitem [{\citenamefont {Wolfe}\ and\ \citenamefont
  {Samelson}(2007)}]{Wolfe2007}\BibitemOpen
  \bibfield  {author} {\bibinfo {author} {\bibfnamefont {C.~L.}\ \bibnamefont
  {Wolfe}}\ and\ \bibinfo {author} {\bibfnamefont {R.~M.}\ \bibnamefont
  {Samelson}},\ }\bibfield  {title} {\bibinfo {title} {An efficient method for
  recovering {L}yapunov vectors from singular vectors},\ }\href
  {https://doi.org/10.1111/j.1600-0870.2007.00234.x} {\bibfield  {journal}
  {\bibinfo  {journal} {Tellus A: Dynamic Meteorology and Oceanography}\
  }\textbf {\bibinfo {volume} {59}},\ \bibinfo {pages} {355} (\bibinfo {year}
  {2007})}\BibitemShut {NoStop}\bibitem [{\citenamefont {Ginelli}\ \emph {et~al.}(2013)\citenamefont
  {Ginelli}, \citenamefont {Chat{\'{e}}}, \citenamefont {Livi},\ and\
  \citenamefont {Politi}}]{Ginelli2013}\BibitemOpen
  \bibfield  {author} {\bibinfo {author} {\bibfnamefont {F.}~\bibnamefont
  {Ginelli}}, \bibinfo {author} {\bibfnamefont {H.}~\bibnamefont
  {Chat{\'{e}}}}, \bibinfo {author} {\bibfnamefont {R.}~\bibnamefont {Livi}},\
  and\ \bibinfo {author} {\bibfnamefont {A.}~\bibnamefont {Politi}},\
  }\bibfield  {title} {\bibinfo {title} {Covariant {L}yapunov vectors},\ }\href
  {https://doi.org/10.1088/1751-8113/46/25/254005} {\bibfield  {journal}
  {\bibinfo  {journal} {Journal of Physics A: Mathematical and Theoretical}\
  }\textbf {\bibinfo {volume} {46}},\ \bibinfo {pages} {254005} (\bibinfo
  {year} {2013})}\BibitemShut {NoStop}\bibitem [{\citenamefont {Takeuchi}\ and\ \citenamefont
  {Chat{\'{e}}}(2013)}]{Takeuchi2013}\BibitemOpen
  \bibfield  {author} {\bibinfo {author} {\bibfnamefont {K.~A.}\ \bibnamefont
  {Takeuchi}}\ and\ \bibinfo {author} {\bibfnamefont {H.}~\bibnamefont
  {Chat{\'{e}}}},\ }\bibfield  {title} {\bibinfo {title} {Collective {L}yapunov
  modes},\ }\href {https://doi.org/10.1088/1751-8113/46/25/254007} {\bibfield
  {journal} {\bibinfo  {journal} {Journal of Physics A: Mathematical and
  Theoretical}\ }\textbf {\bibinfo {volume} {46}},\ \bibinfo {pages} {254007}
  (\bibinfo {year} {2013})}\BibitemShut {NoStop}\bibitem [{\citenamefont {liu Yang}\ and\ \citenamefont
  {Radons}(2010)}]{Yang2010}\BibitemOpen
  \bibfield  {author} {\bibinfo {author} {\bibfnamefont {H.}~\bibnamefont {liu
  Yang}}\ and\ \bibinfo {author} {\bibfnamefont {G.}~\bibnamefont {Radons}},\
  }\bibfield  {title} {\bibinfo {title} {Comparison between covariant and
  orthogonal {L}yapunov vectors},\ }\href
  {https://doi.org/10.1103/physreve.82.046204} {\bibfield  {journal} {\bibinfo
  {journal} {Physical Review E}\ }\textbf {\bibinfo {volume} {82}},\ \bibinfo
  {pages} {046204} (\bibinfo {year} {2010})}\BibitemShut {NoStop}\bibitem [{\citenamefont {Froyland}\ \emph {et~al.}(2013)\citenamefont
  {Froyland}, \citenamefont {Hüls}, \citenamefont {Morriss},\ and\
  \citenamefont {Watson}}]{Froyland2013}\BibitemOpen
  \bibfield  {author} {\bibinfo {author} {\bibfnamefont {G.}~\bibnamefont
  {Froyland}}, \bibinfo {author} {\bibfnamefont {T.}~\bibnamefont {Hüls}},
  \bibinfo {author} {\bibfnamefont {G.~P.}\ \bibnamefont {Morriss}},\ and\
  \bibinfo {author} {\bibfnamefont {T.~M.}\ \bibnamefont {Watson}},\ }\bibfield
   {title} {\bibinfo {title} {Computing covariant {L}yapunov vectors,
  {O}seledets vectors, and dichotomy projectors: A comparative numerical
  study},\ }\href {https://doi.org/10.1016/j.physd.2012.12.005} {\bibfield
  {journal} {\bibinfo  {journal} {Physica D: Nonlinear Phenomena}\ }\textbf
  {\bibinfo {volume} {247}},\ \bibinfo {pages} {18} (\bibinfo {year}
  {2013})}\BibitemShut {NoStop}\bibitem [{\citenamefont {Trevisan}\ and\ \citenamefont
  {Pancotti}(1998)}]{Trevisan1998}\BibitemOpen
  \bibfield  {author} {\bibinfo {author} {\bibfnamefont {A.}~\bibnamefont
  {Trevisan}}\ and\ \bibinfo {author} {\bibfnamefont {F.}~\bibnamefont
  {Pancotti}},\ }\bibfield  {title} {\bibinfo {title} {Periodic orbits,
  {L}yapunov vectors, and singular vectors in the {L}orenz system},\ }\href
  {https://doi.org/10.1175/1520-0469(1998)055<0390:polvas>2.0.co;2} {\bibfield
  {journal} {\bibinfo  {journal} {Journal of the Atmospheric Sciences}\
  }\textbf {\bibinfo {volume} {55}},\ \bibinfo {pages} {390} (\bibinfo {year}
  {1998})}\BibitemShut {NoStop}\bibitem [{\citenamefont {Noethen}(2019)}]{Noethen2019}\BibitemOpen
  \bibfield  {author} {\bibinfo {author} {\bibfnamefont {F.}~\bibnamefont
  {Noethen}},\ }\bibfield  {title} {\bibinfo {title} {A projector-based
  convergence proof of the {G}inelli algorithm for covariant {L}yapunov
  vectors},\ }\href {https://doi.org/10.1016/j.physd.2019.02.012} {\bibfield
  {journal} {\bibinfo  {journal} {Physica D: Nonlinear Phenomena}\ }\textbf
  {\bibinfo {volume} {396}},\ \bibinfo {pages} {18} (\bibinfo {year}
  {2019})}\BibitemShut {NoStop}\bibitem [{\citenamefont {Wiesel}(1993{\natexlab{a}})}]{Wiesel1993}\BibitemOpen
  \bibfield  {author} {\bibinfo {author} {\bibfnamefont {W.~E.}\ \bibnamefont
  {Wiesel}},\ }\bibfield  {title} {\bibinfo {title} {Continuous time algorithm
  for {L}yapunov exponents. i},\ }\href
  {https://doi.org/10.1103/physreve.47.3686} {\bibfield  {journal} {\bibinfo
  {journal} {Physical Review E}\ }\textbf {\bibinfo {volume} {47}},\ \bibinfo
  {pages} {3686–3691} (\bibinfo {year} {1993}{\natexlab{a}})}\BibitemShut
  {NoStop}\bibitem [{\citenamefont {Wiesel}(1993{\natexlab{b}})}]{Wiesel1993b}\BibitemOpen
  \bibfield  {author} {\bibinfo {author} {\bibfnamefont {W.~E.}\ \bibnamefont
  {Wiesel}},\ }\bibfield  {title} {\bibinfo {title} {Continuous time algorithm
  for {L}yapunov exponents. ii},\ }\href
  {https://doi.org/10.1103/physreve.47.3692} {\bibfield  {journal} {\bibinfo
  {journal} {Physical Review E}\ }\textbf {\bibinfo {volume} {47}},\ \bibinfo
  {pages} {3692–3697} (\bibinfo {year} {1993}{\natexlab{b}})}\BibitemShut
  {NoStop}\bibitem [{\citenamefont {Hejazi}\ \emph {et~al.}(2017)\citenamefont {Hejazi},
  \citenamefont {Mehlig},\ and\ \citenamefont {Voth}}]{Hejazi2017}\BibitemOpen
  \bibfield  {author} {\bibinfo {author} {\bibfnamefont {B.}~\bibnamefont
  {Hejazi}}, \bibinfo {author} {\bibfnamefont {B.}~\bibnamefont {Mehlig}},\
  and\ \bibinfo {author} {\bibfnamefont {G.~A.}\ \bibnamefont {Voth}},\
  }\bibfield  {title} {\bibinfo {title} {Emergent scar lines in chaotic
  advection of passive directors},\ }\href
  {https://doi.org/10.1103/physrevfluids.2.124501} {\bibfield  {journal}
  {\bibinfo  {journal} {Physical Review Fluids}\ }\textbf {\bibinfo {volume}
  {2}},\ \bibinfo {pages} {124501} (\bibinfo {year} {2017})}\BibitemShut
  {NoStop}\bibitem [{\citenamefont {Voth}\ \emph {et~al.}(2002)\citenamefont {Voth},
  \citenamefont {Haller},\ and\ \citenamefont {Gollub}}]{Voth2002}\BibitemOpen
  \bibfield  {author} {\bibinfo {author} {\bibfnamefont {G.~A.}\ \bibnamefont
  {Voth}}, \bibinfo {author} {\bibfnamefont {G.}~\bibnamefont {Haller}},\ and\
  \bibinfo {author} {\bibfnamefont {J.~P.}\ \bibnamefont {Gollub}},\ }\bibfield
   {title} {\bibinfo {title} {Experimental measurements of stretching fields in
  fluid mixing},\ }\href {https://doi.org/10.1103/physrevlett.88.254501}
  {\bibfield  {journal} {\bibinfo  {journal} {Physical Review Letters}\
  }\textbf {\bibinfo {volume} {88}},\ \bibinfo {pages} {254501} (\bibinfo
  {year} {2002})}\BibitemShut {NoStop}\bibitem [{\citenamefont {Arnold}\ \emph {et~al.}(1999)\citenamefont {Arnold},
  \citenamefont {Nguyen},\ and\ \citenamefont {Oseledets}}]{Arnold1999}\BibitemOpen
  \bibfield  {author} {\bibinfo {author} {\bibfnamefont {L.}~\bibnamefont
  {Arnold}}, \bibinfo {author} {\bibfnamefont {D.~C.}\ \bibnamefont {Nguyen}},\
  and\ \bibinfo {author} {\bibfnamefont {V.}~\bibnamefont {Oseledets}},\
  }\bibfield  {title} {\bibinfo {title} {Jordan normal form for linear
  cocycles},\ }\bibfield  {journal} {\bibinfo  {journal} {Random Operators and
  Stochastic Equations}\ }\textbf {\bibinfo {volume} {7}},\ \href
  {https://doi.org/10.1515/rose.1999.7.4.303} {10.1515/rose.1999.7.4.303}
  (\bibinfo {year} {1999})\BibitemShut {NoStop}\bibitem [{\citenamefont {González-Tokman}\ and\ \citenamefont
  {Quas}(2013)}]{GonzalezTokman2013}\BibitemOpen
  \bibfield  {author} {\bibinfo {author} {\bibfnamefont {C.}~\bibnamefont
  {González-Tokman}}\ and\ \bibinfo {author} {\bibfnamefont {A.}~\bibnamefont
  {Quas}},\ }\bibfield  {title} {\bibinfo {title} {Multiplicative ergodic
  theorems and applications},\ }\href
  {https://notes.math.ca/archives/Notesv45n5.pdf} {\bibfield  {journal}
  {\bibinfo  {journal} {CMS Notes}\ }\textbf {\bibinfo {volume} {45}} (\bibinfo
  {year} {2013})}\BibitemShut {NoStop}\bibitem [{Note4()}]{Note4}\BibitemOpen
  \bibinfo {note} {Two remarks are in order. First, it is possible to have
  equal Lyapunov exponents with different CLVs (a trivial example consists in
  two identical decoupled dynamical systems, considered as a single dynamical
  system). Conversely, two identical CLVs must have the same Lyapunov exponent
  (because they are the same vector; this is a tautology).}\BibitemShut {Stop}\bibitem [{Note5()}]{Note5}\BibitemOpen
  \bibinfo {note} {In the main text, all the examples we consider correspond to
  a Jordan block of size two. The case of higher-order singularities is
  discussed in appendix section \ref {appendix_tori_higher}.}\BibitemShut
  {Stop}\bibitem [{Note6()}]{Note6}\BibitemOpen
  \bibinfo {note} {In general, the angles between CLVs depend on time
  (equivalently, on the base point on the trajectory). However, two CLVs that
  are exactly parallel at one point are also exactly parallel along the whole
  trajectory. In order to illustrate the presence of a tangency in the
  examples, we either average the angles $\theta _{i j}$ between the CLVs
  $\protect \mathaccentV {vec}17E{c}_i$ and $\protect \mathaccentV
  {vec}17E{c}_j$ over the trajectory, or, in the case of (quasi)periodic
  dynamics we evaluate them at a distinguished point (e.g. where an arbitrary
  coordinate is maximal). See Appendix \ref {app_plankton} (in particular
  Fig.~\ref {figure_model_plankton}m) for an example.}\BibitemShut {Stop}\bibitem [{\citenamefont {liu Yang}\ \emph {et~al.}(2009)\citenamefont {liu
  Yang}, \citenamefont {Takeuchi}, \citenamefont {Ginelli}, \citenamefont
  {Chat{\'{e}}},\ and\ \citenamefont {Radons}}]{Yang2009}\BibitemOpen
  \bibfield  {author} {\bibinfo {author} {\bibfnamefont {H.}~\bibnamefont {liu
  Yang}}, \bibinfo {author} {\bibfnamefont {K.~A.}\ \bibnamefont {Takeuchi}},
  \bibinfo {author} {\bibfnamefont {F.}~\bibnamefont {Ginelli}}, \bibinfo
  {author} {\bibfnamefont {H.}~\bibnamefont {Chat{\'{e}}}},\ and\ \bibinfo
  {author} {\bibfnamefont {G.}~\bibnamefont {Radons}},\ }\bibfield  {title}
  {\bibinfo {title} {Hyperbolicity and the effective dimension of spatially
  extended dissipative systems},\ }\href
  {https://doi.org/10.1103/physrevlett.102.074102} {\bibfield  {journal}
  {\bibinfo  {journal} {Physical Review Letters}\ }\textbf {\bibinfo {volume}
  {102}},\ \bibinfo {pages} {074102} (\bibinfo {year} {2009})}\BibitemShut
  {NoStop}\bibitem [{\citenamefont {Takeuchi}\ \emph {et~al.}(2011)\citenamefont
  {Takeuchi}, \citenamefont {liu Yang}, \citenamefont {Ginelli}, \citenamefont
  {Radons},\ and\ \citenamefont {Chat{\'{e}}}}]{Takeuchi2011}\BibitemOpen
  \bibfield  {author} {\bibinfo {author} {\bibfnamefont {K.~A.}\ \bibnamefont
  {Takeuchi}}, \bibinfo {author} {\bibfnamefont {H.}~\bibnamefont {liu Yang}},
  \bibinfo {author} {\bibfnamefont {F.}~\bibnamefont {Ginelli}}, \bibinfo
  {author} {\bibfnamefont {G.}~\bibnamefont {Radons}},\ and\ \bibinfo {author}
  {\bibfnamefont {H.}~\bibnamefont {Chat{\'{e}}}},\ }\bibfield  {title}
  {\bibinfo {title} {Hyperbolic decoupling of tangent space and effective
  dimension of dissipative systems},\ }\href
  {https://doi.org/10.1103/physreve.84.046214} {\bibfield  {journal} {\bibinfo
  {journal} {Physical Review E}\ }\textbf {\bibinfo {volume} {84}},\ \bibinfo
  {pages} {046214} (\bibinfo {year} {2011})}\BibitemShut {NoStop}\bibitem [{\citenamefont {Sharafi}\ \emph {et~al.}(2017)\citenamefont
  {Sharafi}, \citenamefont {Timme},\ and\ \citenamefont
  {Hallerberg}}]{Sharafi2017}\BibitemOpen
  \bibfield  {author} {\bibinfo {author} {\bibfnamefont {N.}~\bibnamefont
  {Sharafi}}, \bibinfo {author} {\bibfnamefont {M.}~\bibnamefont {Timme}},\
  and\ \bibinfo {author} {\bibfnamefont {S.}~\bibnamefont {Hallerberg}},\
  }\bibfield  {title} {\bibinfo {title} {Critical transitions and perturbation
  growth directions},\ }\href {https://doi.org/10.1103/physreve.96.032220}
  {\bibfield  {journal} {\bibinfo  {journal} {Physical Review E}\ }\textbf
  {\bibinfo {volume} {96}},\ \bibinfo {pages} {032220} (\bibinfo {year}
  {2017})}\BibitemShut {NoStop}\bibitem [{\citenamefont {Xu}\ and\ \citenamefont {Paul}(2016)}]{Xu2016}\BibitemOpen
  \bibfield  {author} {\bibinfo {author} {\bibfnamefont {M.}~\bibnamefont
  {Xu}}\ and\ \bibinfo {author} {\bibfnamefont {M.~R.}\ \bibnamefont {Paul}},\
  }\bibfield  {title} {\bibinfo {title} {Covariant {L}yapunov vectors of
  chaotic {R}ayleigh-{B}énard convection},\ }\href
  {https://doi.org/10.1103/physreve.93.062208} {\bibfield  {journal} {\bibinfo
  {journal} {Physical Review E}\ }\textbf {\bibinfo {volume} {93}},\ \bibinfo
  {pages} {062208} (\bibinfo {year} {2016})}\BibitemShut {NoStop}\bibitem [{\citenamefont {Bosetti}\ and\ \citenamefont
  {Posch}(2010)}]{Bosetti2010}\BibitemOpen
  \bibfield  {author} {\bibinfo {author} {\bibfnamefont {H.}~\bibnamefont
  {Bosetti}}\ and\ \bibinfo {author} {\bibfnamefont {H.~A.}\ \bibnamefont
  {Posch}},\ }\bibfield  {title} {\bibinfo {title} {Covariant {L}yapunov
  vectors for rigid disk systems},\ }\href
  {https://doi.org/10.1016/j.chemphys.2010.06.010} {\bibfield  {journal}
  {\bibinfo  {journal} {Chemical Physics}\ }\textbf {\bibinfo {volume} {375}},\
  \bibinfo {pages} {296–308} (\bibinfo {year} {2010})}\BibitemShut {NoStop}\bibitem [{\citenamefont {Vannitsem}\ and\ \citenamefont
  {Lucarini}(2016)}]{Vannitsem2016}\BibitemOpen
  \bibfield  {author} {\bibinfo {author} {\bibfnamefont {S.}~\bibnamefont
  {Vannitsem}}\ and\ \bibinfo {author} {\bibfnamefont {V.}~\bibnamefont
  {Lucarini}},\ }\bibfield  {title} {\bibinfo {title} {Statistical and
  dynamical properties of covariant {L}yapunov vectors in a coupled
  atmosphere-ocean model—multiscale effects, geometric degeneracy, and error
  dynamics},\ }\href {https://doi.org/10.1088/1751-8113/49/22/224001}
  {\bibfield  {journal} {\bibinfo  {journal} {Journal of Physics A:
  Mathematical and Theoretical}\ }\textbf {\bibinfo {volume} {49}},\ \bibinfo
  {pages} {224001} (\bibinfo {year} {2016})}\BibitemShut {NoStop}\bibitem [{Note7()}]{Note7}\BibitemOpen
  \bibinfo {note} {Consider for instance the matrix \begin {equation}
  J(\epsilon ) = \begin {pmatrix} \sigma & \epsilon \\ 1 - \epsilon & \sigma
  \end {pmatrix}. \end {equation} This matrix has exceptional points $\epsilon
  = 0$ and $\epsilon = 1$. Consider now the linear dynamical system $\protect
  \mathaccentV {dot}05F{X} = J(\epsilon ) X \equiv f_{\epsilon }(X)$ for $X \in
  \protect \mathbb {R}^2$. It has a unique fixed point at $X = 0$. In the case
  of a fixed points, the CLVs are the eigenvectors of the Jacobian (here
  $J(\epsilon )$), so there is indeed a tangency of the CLVs at $\epsilon = 0$
  (and at $\epsilon = 1$). The fixed point $X=0$ does change nature at
  $\epsilon = 0$ from a node (for $\epsilon = 0^-$, where the streamlines of
  the vector field are essentially straight), to a focus (for $\epsilon = 0^+$,
  where the streamlines are spiraling). Nevertheless, the dynamical systems at
  $\epsilon = 0^{\pm }$ are topologically equivalent (see for instance Example
  2.1 in Ref.~\cite {Kuznetsov2004}), so there is no bifurcation.}\BibitemShut
  {Stop}\bibitem [{\citenamefont {Knobloch}\ \emph {et~al.}(1995)\citenamefont
  {Knobloch}, \citenamefont {Hettel},\ and\ \citenamefont
  {Dangelmayr}}]{Knobloch1995}\BibitemOpen
  \bibfield  {author} {\bibinfo {author} {\bibfnamefont {E.}~\bibnamefont
  {Knobloch}}, \bibinfo {author} {\bibfnamefont {J.}~\bibnamefont {Hettel}},\
  and\ \bibinfo {author} {\bibfnamefont {G.}~\bibnamefont {Dangelmayr}},\
  }\bibfield  {title} {\bibinfo {title} {Parity breaking bifurcation in
  inhomogeneous systems},\ }\href {https://doi.org/10.1103/physrevlett.74.4839}
  {\bibfield  {journal} {\bibinfo  {journal} {Physical Review Letters}\
  }\textbf {\bibinfo {volume} {74}},\ \bibinfo {pages} {4839–4842} (\bibinfo
  {year} {1995})}\BibitemShut {NoStop}\bibitem [{Note8()}]{Note8}\BibitemOpen
  \bibinfo {note} {Up to a non-linear change of variable, the circle and line
  can be mapped to arbitrary closed and open curves.}\BibitemShut {Stop}\bibitem [{\citenamefont {Coullet}\ and\ \citenamefont
  {Fauve}(1985)}]{Coullet1985b}\BibitemOpen
  \bibfield  {author} {\bibinfo {author} {\bibfnamefont {P.}~\bibnamefont
  {Coullet}}\ and\ \bibinfo {author} {\bibfnamefont {S.}~\bibnamefont
  {Fauve}},\ }\bibfield  {title} {\bibinfo {title} {Propagative phase dynamics
  for systems with {G}alilean invariance},\ }\href
  {https://doi.org/10.1103/physrevlett.55.2857} {\bibfield  {journal} {\bibinfo
   {journal} {Physical Review Letters}\ }\textbf {\bibinfo {volume} {55}},\
  \bibinfo {pages} {2857} (\bibinfo {year} {1985})}\BibitemShut {NoStop}\bibitem [{\citenamefont {Coullet}\ \emph {et~al.}(1989)\citenamefont
  {Coullet}, \citenamefont {Goldstein},\ and\ \citenamefont
  {Gunaratne}}]{Coullet1989}\BibitemOpen
  \bibfield  {author} {\bibinfo {author} {\bibfnamefont {P.}~\bibnamefont
  {Coullet}}, \bibinfo {author} {\bibfnamefont {R.~E.}\ \bibnamefont
  {Goldstein}},\ and\ \bibinfo {author} {\bibfnamefont {G.~H.}\ \bibnamefont
  {Gunaratne}},\ }\bibfield  {title} {\bibinfo {title} {Parity-breaking
  transitions of modulated patterns in hydrodynamic systems},\ }\href
  {https://doi.org/10.1103/physrevlett.63.1954} {\bibfield  {journal} {\bibinfo
   {journal} {Physical Review Letters}\ }\textbf {\bibinfo {volume} {63}},\
  \bibinfo {pages} {1954} (\bibinfo {year} {1989})}\BibitemShut {NoStop}\bibitem [{\citenamefont {Malomed}\ and\ \citenamefont
  {Tribelsky}(1984)}]{Malomed1984}\BibitemOpen
  \bibfield  {author} {\bibinfo {author} {\bibfnamefont {B.}~\bibnamefont
  {Malomed}}\ and\ \bibinfo {author} {\bibfnamefont {M.}~\bibnamefont
  {Tribelsky}},\ }\bibfield  {title} {\bibinfo {title} {Bifurcations in
  distributed kinetic systems with aperiodic instability},\ }\href
  {https://doi.org/10.1016/0167-2789(84)90005-8} {\bibfield  {journal}
  {\bibinfo  {journal} {Physica D: Nonlinear Phenomena}\ }\textbf {\bibinfo
  {volume} {14}},\ \bibinfo {pages} {67} (\bibinfo {year} {1984})}\BibitemShut
  {NoStop}\bibitem [{\citenamefont {Douady}\ \emph {et~al.}(1989)\citenamefont {Douady},
  \citenamefont {Fauve},\ and\ \citenamefont {Thual}}]{Douady1989}\BibitemOpen
  \bibfield  {author} {\bibinfo {author} {\bibfnamefont {S.}~\bibnamefont
  {Douady}}, \bibinfo {author} {\bibfnamefont {S.}~\bibnamefont {Fauve}},\ and\
  \bibinfo {author} {\bibfnamefont {O.}~\bibnamefont {Thual}},\ }\bibfield
  {title} {\bibinfo {title} {Oscillatory phase modulation of parametrically
  forced surface waves},\ }\href {https://doi.org/10.1209/0295-5075/10/4/005}
  {\bibfield  {journal} {\bibinfo  {journal} {Europhysics Letters (EPL)}\
  }\textbf {\bibinfo {volume} {10}},\ \bibinfo {pages} {309–315} (\bibinfo
  {year} {1989})}\BibitemShut {NoStop}\bibitem [{\citenamefont {Brachet}\ \emph {et~al.}(1987)\citenamefont
  {Brachet}, \citenamefont {Coullet},\ and\ \citenamefont
  {Fauve}}]{Brachet1987}\BibitemOpen
  \bibfield  {author} {\bibinfo {author} {\bibfnamefont {M.~E.}\ \bibnamefont
  {Brachet}}, \bibinfo {author} {\bibfnamefont {P.}~\bibnamefont {Coullet}},\
  and\ \bibinfo {author} {\bibfnamefont {S.}~\bibnamefont {Fauve}},\ }\bibfield
   {title} {\bibinfo {title} {Propagative phase dynamics in temporally
  intermittent systems},\ }\href {https://doi.org/10.1209/0295-5075/4/9/011}
  {\bibfield  {journal} {\bibinfo  {journal} {Europhysics Letters ({EPL})}\
  }\textbf {\bibinfo {volume} {4}},\ \bibinfo {pages} {1017} (\bibinfo {year}
  {1987})}\BibitemShut {NoStop}\bibitem [{\citenamefont {Pan}\ and\ \citenamefont {de~Bruyn}(1994)}]{Pan1994}\BibitemOpen
  \bibfield  {author} {\bibinfo {author} {\bibfnamefont {L.}~\bibnamefont
  {Pan}}\ and\ \bibinfo {author} {\bibfnamefont {J.~R.}\ \bibnamefont
  {de~Bruyn}},\ }\bibfield  {title} {\bibinfo {title} {Spatially uniform
  traveling cellular patterns at a driven interface},\ }\href
  {https://doi.org/10.1103/physreve.49.483} {\bibfield  {journal} {\bibinfo
  {journal} {Physical Review E}\ }\textbf {\bibinfo {volume} {49}},\ \bibinfo
  {pages} {483} (\bibinfo {year} {1994})}\BibitemShut {NoStop}\bibitem [{\citenamefont {Bensimon}\ \emph {et~al.}(1989)\citenamefont
  {Bensimon}, \citenamefont {Pumir},\ and\ \citenamefont
  {Shraiman}}]{Bensimon1989}\BibitemOpen
  \bibfield  {author} {\bibinfo {author} {\bibfnamefont {D.}~\bibnamefont
  {Bensimon}}, \bibinfo {author} {\bibfnamefont {A.}~\bibnamefont {Pumir}},\
  and\ \bibinfo {author} {\bibfnamefont {B.}~\bibnamefont {Shraiman}},\
  }\bibfield  {title} {\bibinfo {title} {Nonlinear theory of traveling wave
  convection in binary mixtures},\ }\href
  {https://doi.org/10.1051/jphys:0198900500200308900} {\bibfield  {journal}
  {\bibinfo  {journal} {Journal de Physique}\ }\textbf {\bibinfo {volume}
  {50}},\ \bibinfo {pages} {3089–3108} (\bibinfo {year} {1989})}\BibitemShut
  {NoStop}\bibitem [{\citenamefont {Xin}\ \emph {et~al.}(1998)\citenamefont {Xin},
  \citenamefont {Le~Quéré},\ and\ \citenamefont {Tuckerman}}]{Xin1998}\BibitemOpen
  \bibfield  {author} {\bibinfo {author} {\bibfnamefont {S.}~\bibnamefont
  {Xin}}, \bibinfo {author} {\bibfnamefont {P.}~\bibnamefont {Le~Quéré}},\
  and\ \bibinfo {author} {\bibfnamefont {L.~S.}\ \bibnamefont {Tuckerman}},\
  }\bibfield  {title} {\bibinfo {title} {Bifurcation analysis of
  double-diffusive convection with opposing horizontal thermal and solutal
  gradients},\ }\href {https://doi.org/10.1063/1.869608} {\bibfield  {journal}
  {\bibinfo  {journal} {Physics of Fluids}\ }\textbf {\bibinfo {volume} {10}},\
  \bibinfo {pages} {850–858} (\bibinfo {year} {1998})}\BibitemShut {NoStop}\bibitem [{\citenamefont {Thiele}\ \emph {et~al.}(2004)\citenamefont {Thiele},
  \citenamefont {John},\ and\ \citenamefont {Bär}}]{Thiele2004}\BibitemOpen
  \bibfield  {author} {\bibinfo {author} {\bibfnamefont {U.}~\bibnamefont
  {Thiele}}, \bibinfo {author} {\bibfnamefont {K.}~\bibnamefont {John}},\ and\
  \bibinfo {author} {\bibfnamefont {M.}~\bibnamefont {Bär}},\ }\bibfield
  {title} {\bibinfo {title} {Dynamical model for chemically driven running
  droplets},\ }\href {https://doi.org/10.1103/physrevlett.93.027802} {\bibfield
   {journal} {\bibinfo  {journal} {Physical Review Letters}\ }\textbf {\bibinfo
  {volume} {93}},\ \bibinfo {pages} {027802} (\bibinfo {year}
  {2004})}\BibitemShut {NoStop}\bibitem [{\citenamefont {Kness}\ \emph {et~al.}(1992)\citenamefont {Kness},
  \citenamefont {Tuckerman},\ and\ \citenamefont {Barkley}}]{Kness1992}\BibitemOpen
  \bibfield  {author} {\bibinfo {author} {\bibfnamefont {M.}~\bibnamefont
  {Kness}}, \bibinfo {author} {\bibfnamefont {L.~S.}\ \bibnamefont
  {Tuckerman}},\ and\ \bibinfo {author} {\bibfnamefont {D.}~\bibnamefont
  {Barkley}},\ }\bibfield  {title} {\bibinfo {title} {Symmetry-breaking
  bifurcations in one-dimensional excitable media},\ }\href
  {https://doi.org/10.1103/physreva.46.5054} {\bibfield  {journal} {\bibinfo
  {journal} {Physical Review A}\ }\textbf {\bibinfo {volume} {46}},\ \bibinfo
  {pages} {5054–5062} (\bibinfo {year} {1992})}\BibitemShut {NoStop}\bibitem [{\citenamefont {Krischer}\ and\ \citenamefont
  {Mikhailov}(1994)}]{Krischer1994}\BibitemOpen
  \bibfield  {author} {\bibinfo {author} {\bibfnamefont {K.}~\bibnamefont
  {Krischer}}\ and\ \bibinfo {author} {\bibfnamefont {A.}~\bibnamefont
  {Mikhailov}},\ }\bibfield  {title} {\bibinfo {title} {Bifurcation to
  traveling spots in reaction-diffusion systems},\ }\href
  {https://doi.org/10.1103/physrevlett.73.3165} {\bibfield  {journal} {\bibinfo
   {journal} {Physical Review Letters}\ }\textbf {\bibinfo {volume} {73}},\
  \bibinfo {pages} {3165–3168} (\bibinfo {year} {1994})}\BibitemShut
  {NoStop}\bibitem [{\citenamefont {Hassan}\ \emph {et~al.}(2015)\citenamefont {Hassan},
  \citenamefont {Hodaei}, \citenamefont {Miri}, \citenamefont {Khajavikhan},\
  and\ \citenamefont {Christodoulides}}]{Hassan2015}\BibitemOpen
  \bibfield  {author} {\bibinfo {author} {\bibfnamefont {A.~U.}\ \bibnamefont
  {Hassan}}, \bibinfo {author} {\bibfnamefont {H.}~\bibnamefont {Hodaei}},
  \bibinfo {author} {\bibfnamefont {M.-A.}\ \bibnamefont {Miri}}, \bibinfo
  {author} {\bibfnamefont {M.}~\bibnamefont {Khajavikhan}},\ and\ \bibinfo
  {author} {\bibfnamefont {D.~N.}\ \bibnamefont {Christodoulides}},\ }\bibfield
   {title} {\bibinfo {title} {{Nonlinear reversal of the PT-symmetric phase
  transition in a system of coupled semiconductor microring resonators}},\
  }\href {https://doi.org/10.1103/physreva.92.063807} {\bibfield  {journal}
  {\bibinfo  {journal} {Physical Review A}\ }\textbf {\bibinfo {volume} {92}},\
  \bibinfo {pages} {063807} (\bibinfo {year} {2015})}\BibitemShut {NoStop}\bibitem [{\citenamefont {Clerkin}\ \emph {et~al.}(2014)\citenamefont
  {Clerkin}, \citenamefont {O’Brien},\ and\ \citenamefont
  {Amann}}]{Clerkin2014}\BibitemOpen
  \bibfield  {author} {\bibinfo {author} {\bibfnamefont {E.}~\bibnamefont
  {Clerkin}}, \bibinfo {author} {\bibfnamefont {S.}~\bibnamefont {O’Brien}},\
  and\ \bibinfo {author} {\bibfnamefont {A.}~\bibnamefont {Amann}},\ }\bibfield
   {title} {\bibinfo {title} {Multistabilities and symmetry-broken one-color
  and two-color states in closely coupled single-mode lasers},\ }\href
  {https://doi.org/10.1103/physreve.89.032919} {\bibfield  {journal} {\bibinfo
  {journal} {Physical Review E}\ }\textbf {\bibinfo {volume} {89}},\ \bibinfo
  {pages} {032919} (\bibinfo {year} {2014})}\BibitemShut {NoStop}\bibitem [{\citenamefont {Soriano}\ \emph {et~al.}(2013)\citenamefont
  {Soriano}, \citenamefont {García-Ojalvo}, \citenamefont {Mirasso},\ and\
  \citenamefont {Fischer}}]{Soriano2013}\BibitemOpen
  \bibfield  {author} {\bibinfo {author} {\bibfnamefont {M.~C.}\ \bibnamefont
  {Soriano}}, \bibinfo {author} {\bibfnamefont {J.}~\bibnamefont
  {García-Ojalvo}}, \bibinfo {author} {\bibfnamefont {C.~R.}\ \bibnamefont
  {Mirasso}},\ and\ \bibinfo {author} {\bibfnamefont {I.}~\bibnamefont
  {Fischer}},\ }\bibfield  {title} {\bibinfo {title} {Complex photonics:
  Dynamics and applications of delay-coupled semiconductors lasers},\ }\href
  {https://doi.org/10.1103/revmodphys.85.421} {\bibfield  {journal} {\bibinfo
  {journal} {Reviews of Modern Physics}\ }\textbf {\bibinfo {volume} {85}},\
  \bibinfo {pages} {421–470} (\bibinfo {year} {2013})}\BibitemShut {NoStop}\bibitem [{\citenamefont {Hong}\ and\ \citenamefont
  {Strogatz}(2011{\natexlab{a}})}]{Hong2011}\BibitemOpen
  \bibfield  {author} {\bibinfo {author} {\bibfnamefont {H.}~\bibnamefont
  {Hong}}\ and\ \bibinfo {author} {\bibfnamefont {S.~H.}\ \bibnamefont
  {Strogatz}},\ }\bibfield  {title} {\bibinfo {title} {Kuramoto model of
  coupled oscillators with positive and negative coupling parameters: An
  example of conformist and contrarian oscillators},\ }\href
  {https://doi.org/10.1103/physrevlett.106.054102} {\bibfield  {journal}
  {\bibinfo  {journal} {Physical Review Letters}\ }\textbf {\bibinfo {volume}
  {106}},\ \bibinfo {pages} {054102} (\bibinfo {year}
  {2011}{\natexlab{a}})}\BibitemShut {NoStop}\bibitem [{\citenamefont {Hong}\ and\ \citenamefont
  {Strogatz}(2011{\natexlab{b}})}]{Hong2011b}\BibitemOpen
  \bibfield  {author} {\bibinfo {author} {\bibfnamefont {H.}~\bibnamefont
  {Hong}}\ and\ \bibinfo {author} {\bibfnamefont {S.~H.}\ \bibnamefont
  {Strogatz}},\ }\bibfield  {title} {\bibinfo {title} {{Conformists and
  contrarians in a Kuramoto model with identical natural frequencies}},\ }\href
  {https://doi.org/10.1103/physreve.84.046202} {\bibfield  {journal} {\bibinfo
  {journal} {Physical Review E}\ }\textbf {\bibinfo {volume} {84}},\ \bibinfo
  {pages} {046202} (\bibinfo {year} {2011}{\natexlab{b}})}\BibitemShut
  {NoStop}\bibitem [{\citenamefont {Hong}(2014)}]{Hong2014}\BibitemOpen
  \bibfield  {author} {\bibinfo {author} {\bibfnamefont {H.}~\bibnamefont
  {Hong}},\ }\bibfield  {title} {\bibinfo {title} {{Periodic synchronization
  and chimera in conformist and contrarian oscillators}},\ }\href
  {https://doi.org/10.1103/physreve.89.062924} {\bibfield  {journal} {\bibinfo
  {journal} {Physical Review E}\ }\textbf {\bibinfo {volume} {89}},\ \bibinfo
  {pages} {062924} (\bibinfo {year} {2014})}\BibitemShut {NoStop}\bibitem [{\citenamefont {Leonetti}\ \emph {et~al.}(2006)\citenamefont
  {Leonetti}, \citenamefont {Nuebler},\ and\ \citenamefont
  {Homble}}]{Leonetti2006}\BibitemOpen
  \bibfield  {author} {\bibinfo {author} {\bibfnamefont {M.}~\bibnamefont
  {Leonetti}}, \bibinfo {author} {\bibfnamefont {J.}~\bibnamefont {Nuebler}},\
  and\ \bibinfo {author} {\bibfnamefont {F.}~\bibnamefont {Homble}},\
  }\bibfield  {title} {\bibinfo {title} {Parity-breaking bifurcation and global
  oscillation in patterns of ion channels},\ }\href
  {https://doi.org/10.1103/physrevlett.96.218101} {\bibfield  {journal}
  {\bibinfo  {journal} {Physical Review Letters}\ }\textbf {\bibinfo {volume}
  {96}},\ \bibinfo {pages} {218101} (\bibinfo {year} {2006})}\BibitemShut
  {NoStop}\bibitem [{\citenamefont {Hanai}\ \emph {et~al.}(2019)\citenamefont {Hanai},
  \citenamefont {Edelman}, \citenamefont {Ohashi},\ and\ \citenamefont
  {Littlewood}}]{Hanai2019}\BibitemOpen
  \bibfield  {author} {\bibinfo {author} {\bibfnamefont {R.}~\bibnamefont
  {Hanai}}, \bibinfo {author} {\bibfnamefont {A.}~\bibnamefont {Edelman}},
  \bibinfo {author} {\bibfnamefont {Y.}~\bibnamefont {Ohashi}},\ and\ \bibinfo
  {author} {\bibfnamefont {P.~B.}\ \bibnamefont {Littlewood}},\ }\bibfield
  {title} {\bibinfo {title} {Non-{H}ermitian phase transition from a polariton
  {B}ose-{E}instein condensate to a photon laser},\ }\href
  {https://doi.org/10.1103/physrevlett.122.185301} {\bibfield  {journal}
  {\bibinfo  {journal} {Physical Review Letters}\ }\textbf {\bibinfo {volume}
  {122}},\ \bibinfo {pages} {185301} (\bibinfo {year} {2019})}\BibitemShut
  {NoStop}\bibitem [{\citenamefont {Hanai}\ and\ \citenamefont
  {Littlewood}(2020)}]{Hanai2020}\BibitemOpen
  \bibfield  {author} {\bibinfo {author} {\bibfnamefont {R.}~\bibnamefont
  {Hanai}}\ and\ \bibinfo {author} {\bibfnamefont {P.~B.}\ \bibnamefont
  {Littlewood}},\ }\bibfield  {title} {\bibinfo {title} {Critical fluctuations
  at a many-body exceptional point},\ }\href
  {https://doi.org/10.1103/physrevresearch.2.033018} {\bibfield  {journal}
  {\bibinfo  {journal} {Physical Review Research}\ }\textbf {\bibinfo {volume}
  {2}},\ \bibinfo {pages} {033018} (\bibinfo {year} {2020})}\BibitemShut
  {NoStop}\bibitem [{\citenamefont {Saha}\ \emph {et~al.}(2020)\citenamefont {Saha},
  \citenamefont {Agudo-Canalejo},\ and\ \citenamefont
  {Golestanian}}]{Saha2020}\BibitemOpen
  \bibfield  {author} {\bibinfo {author} {\bibfnamefont {S.}~\bibnamefont
  {Saha}}, \bibinfo {author} {\bibfnamefont {J.}~\bibnamefont
  {Agudo-Canalejo}},\ and\ \bibinfo {author} {\bibfnamefont {R.}~\bibnamefont
  {Golestanian}},\ }\bibfield  {title} {\bibinfo {title} {Scalar active
  mixtures: The nonreciprocal {C}ahn-{H}illiard model},\ }\href
  {https://doi.org/10.1103/physrevx.10.041009} {\bibfield  {journal} {\bibinfo
  {journal} {Physical Review X}\ }\textbf {\bibinfo {volume} {10}},\ \bibinfo
  {pages} {041009} (\bibinfo {year} {2020})}\BibitemShut {NoStop}\bibitem [{\citenamefont {You}\ \emph {et~al.}(2020)\citenamefont {You},
  \citenamefont {Baskaran},\ and\ \citenamefont {Marchetti}}]{You2020}\BibitemOpen
  \bibfield  {author} {\bibinfo {author} {\bibfnamefont {Z.}~\bibnamefont
  {You}}, \bibinfo {author} {\bibfnamefont {A.}~\bibnamefont {Baskaran}},\ and\
  \bibinfo {author} {\bibfnamefont {M.~C.}\ \bibnamefont {Marchetti}},\
  }\bibfield  {title} {\bibinfo {title} {Nonreciprocity as a generic route to
  traveling states},\ }\href {https://doi.org/10.1073/pnas.2010318117}
  {\bibfield  {journal} {\bibinfo  {journal} {Proceedings of the National
  Academy of Sciences}\ }\textbf {\bibinfo {volume} {117}},\ \bibinfo {pages}
  {19767} (\bibinfo {year} {2020})}\BibitemShut {NoStop}\bibitem [{\citenamefont {Frohoff-Hülsmann}\ \emph {et~al.}(2023)\citenamefont
  {Frohoff-Hülsmann}, \citenamefont {Holl}, \citenamefont {Knobloch},
  \citenamefont {Gurevich},\ and\ \citenamefont
  {Thiele}}]{FrohoffHulsmann2023}\BibitemOpen
  \bibfield  {author} {\bibinfo {author} {\bibfnamefont {T.}~\bibnamefont
  {Frohoff-Hülsmann}}, \bibinfo {author} {\bibfnamefont {M.~P.}\ \bibnamefont
  {Holl}}, \bibinfo {author} {\bibfnamefont {E.}~\bibnamefont {Knobloch}},
  \bibinfo {author} {\bibfnamefont {S.~V.}\ \bibnamefont {Gurevich}},\ and\
  \bibinfo {author} {\bibfnamefont {U.}~\bibnamefont {Thiele}},\ }\bibfield
  {title} {\bibinfo {title} {Stationary broken parity states in active matter
  models},\ }\href {https://doi.org/10.1103/physreve.107.064210} {\bibfield
  {journal} {\bibinfo  {journal} {Physical Review E}\ }\textbf {\bibinfo
  {volume} {107}},\ \bibinfo {pages} {064210} (\bibinfo {year}
  {2023})}\BibitemShut {NoStop}\bibitem [{\citenamefont {Ophaus}\ \emph {et~al.}(2018)\citenamefont {Ophaus},
  \citenamefont {Gurevich},\ and\ \citenamefont {Thiele}}]{Ophaus2018}\BibitemOpen
  \bibfield  {author} {\bibinfo {author} {\bibfnamefont {L.}~\bibnamefont
  {Ophaus}}, \bibinfo {author} {\bibfnamefont {S.~V.}\ \bibnamefont
  {Gurevich}},\ and\ \bibinfo {author} {\bibfnamefont {U.}~\bibnamefont
  {Thiele}},\ }\bibfield  {title} {\bibinfo {title} {Resting and traveling
  localized states in an active phase-field-crystal model},\ }\href
  {https://doi.org/10.1103/physreve.98.022608} {\bibfield  {journal} {\bibinfo
  {journal} {Physical Review E}\ }\textbf {\bibinfo {volume} {98}},\ \bibinfo
  {pages} {022608} (\bibinfo {year} {2018})}\BibitemShut {NoStop}\bibitem [{Note9()}]{Note9}\BibitemOpen
  \bibinfo {note} {It is interesting to compare the parity-breaking bifurcation
  to the so-called Bogdanov-Takens bifurcation~\cite {Kuznetsov2004}, typically
  associated with a non-diagonalizable Jacobian. Despite having the same linear
  part (corresponding to the non-diagonalizable Jacobian), their non-linear
  parts do not match. This is because the dynamical system \protect \textup
  {{\protect \normalfont (\ref {parity_breaking}}\protect \normalfont )} does
  not satisfy the genericity conditions that are usually assumed to obtain the
  normal form of the Bogdanov-Takens bifurcation (equations (BT.0-3) in Theorem
  8.4 of Ref.~\cite {Kuznetsov2004}, for instance). The parity-breaking can
  then be thought as a degenerate version of a Bogdanov-Takens bifurcation (see
  Ref.~\cite {Kuznetsov2005} and references therein for other similar
  situations).}\BibitemShut {Stop}\bibitem [{Note10()}]{Note10}\BibitemOpen
  \bibinfo {note} {Let $p$ be the vector of parameters needed to parameterize
  $f(x; p)$. We can choose a function $\pi (w)$ and set $p=\pi (w)$. By
  choosing a reference parameter $p_0$ and setting $g(x,w) = f(x; \pi (w)) -
  f(x; p_0)$, and $f(x) = f(x; p_0)$, we go back to the case discussed in the
  main text.}\BibitemShut {Stop}\bibitem [{\citenamefont {Frohoff-Hülsmann}\ \emph {et~al.}(2021)\citenamefont
  {Frohoff-Hülsmann}, \citenamefont {Wrembel},\ and\ \citenamefont
  {Thiele}}]{FrohoffHulsmann2021}\BibitemOpen
  \bibfield  {author} {\bibinfo {author} {\bibfnamefont {T.}~\bibnamefont
  {Frohoff-Hülsmann}}, \bibinfo {author} {\bibfnamefont {J.}~\bibnamefont
  {Wrembel}},\ and\ \bibinfo {author} {\bibfnamefont {U.}~\bibnamefont
  {Thiele}},\ }\bibfield  {title} {\bibinfo {title} {Suppression of coarsening
  and emergence of oscillatory behavior in a cahn-hilliard model with
  nonvariational coupling},\ }\href
  {https://doi.org/10.1103/physreve.103.042602} {\bibfield  {journal} {\bibinfo
   {journal} {Physical Review E}\ }\textbf {\bibinfo {volume} {103}},\ \bibinfo
  {pages} {042602} (\bibinfo {year} {2021})}\BibitemShut {NoStop}\bibitem [{\citenamefont {Ophaus}\ \emph {et~al.}(2021)\citenamefont {Ophaus},
  \citenamefont {Knobloch}, \citenamefont {Gurevich},\ and\ \citenamefont
  {Thiele}}]{Ophaus2021}\BibitemOpen
  \bibfield  {author} {\bibinfo {author} {\bibfnamefont {L.}~\bibnamefont
  {Ophaus}}, \bibinfo {author} {\bibfnamefont {E.}~\bibnamefont {Knobloch}},
  \bibinfo {author} {\bibfnamefont {S.~V.}\ \bibnamefont {Gurevich}},\ and\
  \bibinfo {author} {\bibfnamefont {U.}~\bibnamefont {Thiele}},\ }\bibfield
  {title} {\bibinfo {title} {Two-dimensional localized states in an active
  phase-field-crystal model},\ }\href
  {https://doi.org/10.1103/physreve.103.032601} {\bibfield  {journal} {\bibinfo
   {journal} {Physical Review E}\ }\textbf {\bibinfo {volume} {103}},\ \bibinfo
  {pages} {032601} (\bibinfo {year} {2021})}\BibitemShut {NoStop}\bibitem [{\citenamefont {Knobloch}(2015)}]{Knobloch2015}\BibitemOpen
  \bibfield  {author} {\bibinfo {author} {\bibfnamefont {E.}~\bibnamefont
  {Knobloch}},\ }\bibfield  {title} {\bibinfo {title} {Spatial localization in
  dissipative systems},\ }\href
  {https://doi.org/10.1146/annurev-conmatphys-031214-014514} {\bibfield
  {journal} {\bibinfo  {journal} {Annual Review of Condensed Matter Physics}\
  }\textbf {\bibinfo {volume} {6}},\ \bibinfo {pages} {325–359} (\bibinfo
  {year} {2015})}\BibitemShut {NoStop}\bibitem [{\citenamefont {Brusch}\ \emph {et~al.}(2000)\citenamefont {Brusch},
  \citenamefont {Zimmermann}, \citenamefont {van Hecke}, \citenamefont {Bär},\
  and\ \citenamefont {Torcini}}]{Brusch2000}\BibitemOpen
  \bibfield  {author} {\bibinfo {author} {\bibfnamefont {L.}~\bibnamefont
  {Brusch}}, \bibinfo {author} {\bibfnamefont {M.~G.}\ \bibnamefont
  {Zimmermann}}, \bibinfo {author} {\bibfnamefont {M.}~\bibnamefont {van
  Hecke}}, \bibinfo {author} {\bibfnamefont {M.}~\bibnamefont {Bär}},\ and\
  \bibinfo {author} {\bibfnamefont {A.}~\bibnamefont {Torcini}},\ }\bibfield
  {title} {\bibinfo {title} {Modulated amplitude waves and the transition from
  phase to defect chaos},\ }\href {https://doi.org/10.1103/physrevlett.85.86}
  {\bibfield  {journal} {\bibinfo  {journal} {Physical Review Letters}\
  }\textbf {\bibinfo {volume} {85}},\ \bibinfo {pages} {86–89} (\bibinfo
  {year} {2000})}\BibitemShut {NoStop}\bibitem [{Note11()}]{Note11}\BibitemOpen
  \bibinfo {note} {Here, we assume that a limit cycle occurs on both sides of
  the bifurcation. This is not necessarily the case. Setting $\alpha = \beta =
  0$ and $h(w) = w$, we find that Eq.~\protect \textup {{\protect \normalfont
  (\ref {pitchfork_lc_polar}}\protect \normalfont )} reduces to the normal form
  of the parity-breaking bifurcation \protect \textup {{\protect \normalfont
  (\ref {parity_breaking}}\protect \normalfont )}, in which a fixed point
  exists on one side, and two limit cycles on the other side.}\BibitemShut
  {Stop}\bibitem [{\citenamefont {Nikolaev}\ and\ \citenamefont
  {Shnol}(1998{\natexlab{a}})}]{Nikolaev1998a}\BibitemOpen
  \bibfield  {author} {\bibinfo {author} {\bibfnamefont {E.~V.}\ \bibnamefont
  {Nikolaev}}\ and\ \bibinfo {author} {\bibfnamefont {E.~E.}\ \bibnamefont
  {Shnol}},\ }\bibfield  {title} {\bibinfo {title} {Bifurcations of cycles in
  systems of differential equations with a finite symmetry group – i},\
  }\href {https://doi.org/10.1023/a:1022832331959} {\bibfield  {journal}
  {\bibinfo  {journal} {Journal of Dynamical and Control Systems}\ }\textbf
  {\bibinfo {volume} {4}},\ \bibinfo {pages} {315–341} (\bibinfo {year}
  {1998}{\natexlab{a}})}\BibitemShut {NoStop}\bibitem [{\citenamefont {Nikolaev}\ and\ \citenamefont
  {Shnol}(1998{\natexlab{b}})}]{Nikolaev1998b}\BibitemOpen
  \bibfield  {author} {\bibinfo {author} {\bibfnamefont {E.~V.}\ \bibnamefont
  {Nikolaev}}\ and\ \bibinfo {author} {\bibfnamefont {E.~E.}\ \bibnamefont
  {Shnol}},\ }\bibfield  {title} {\bibinfo {title} {Bifurcations of cycles in
  systems of differential equations with a finite symmetry group – ii},\
  }\href {https://doi.org/10.1023/a:1022884316030} {\bibfield  {journal}
  {\bibinfo  {journal} {Journal of Dynamical and Control Systems}\ }\textbf
  {\bibinfo {volume} {4}},\ \bibinfo {pages} {343–363} (\bibinfo {year}
  {1998}{\natexlab{b}})}\BibitemShut {NoStop}\bibitem [{\citenamefont {Nikolaev}(1995)}]{Nikolaev1995}\BibitemOpen
  \bibfield  {author} {\bibinfo {author} {\bibfnamefont {E.~V.}\ \bibnamefont
  {Nikolaev}},\ }\bibfield  {title} {\bibinfo {title} {Bifurcations of limit
  cycles of differential equations admitting an involutive symmetry},\ }\href
  {https://doi.org/10.1070/sm1995v186n04abeh000033} {\bibfield  {journal}
  {\bibinfo  {journal} {Sbornik: Mathematics}\ }\textbf {\bibinfo {volume}
  {186}},\ \bibinfo {pages} {611–627} (\bibinfo {year} {1995})}\BibitemShut
  {NoStop}\bibitem [{\citenamefont {Abshagen}\ \emph {et~al.}(2005)\citenamefont
  {Abshagen}, \citenamefont {Lopez}, \citenamefont {Marques},\ and\
  \citenamefont {Pfister}}]{Abshagen2005}\BibitemOpen
  \bibfield  {author} {\bibinfo {author} {\bibfnamefont {J.}~\bibnamefont
  {Abshagen}}, \bibinfo {author} {\bibfnamefont {J.~M.}\ \bibnamefont {Lopez}},
  \bibinfo {author} {\bibfnamefont {F.}~\bibnamefont {Marques}},\ and\ \bibinfo
  {author} {\bibfnamefont {G.}~\bibnamefont {Pfister}},\ }\bibfield  {title}
  {\bibinfo {title} {Mode competition of rotating waves in reflection-symmetric
  taylor–couette flow},\ }\href {https://doi.org/10.1017/s0022112005005811}
  {\bibfield  {journal} {\bibinfo  {journal} {Journal of Fluid Mechanics}\
  }\textbf {\bibinfo {volume} {540}},\ \bibinfo {pages} {269} (\bibinfo {year}
  {2005})}\BibitemShut {NoStop}\bibitem [{\citenamefont {Pedersen}\ \emph {et~al.}(2022)\citenamefont
  {Pedersen}, \citenamefont {Brøns},\ and\ \citenamefont
  {Sørensen}}]{Pedersen2022}\BibitemOpen
  \bibfield  {author} {\bibinfo {author} {\bibfnamefont {M.~G.}\ \bibnamefont
  {Pedersen}}, \bibinfo {author} {\bibfnamefont {M.}~\bibnamefont {Brøns}},\
  and\ \bibinfo {author} {\bibfnamefont {M.~P.}\ \bibnamefont {Sørensen}},\
  }\bibfield  {title} {\bibinfo {title} {Amplitude-modulated spiking as a novel
  route to bursting: Coupling-induced mixed-mode oscillations by symmetry
  breaking},\ }\href {https://doi.org/10.1063/5.0072497} {\bibfield  {journal}
  {\bibinfo  {journal} {Chaos: An Interdisciplinary Journal of Nonlinear
  Science}\ }\textbf {\bibinfo {volume} {32}},\ \bibinfo {pages} {013121}
  (\bibinfo {year} {2022})}\BibitemShut {NoStop}\bibitem [{\citenamefont {Sherman}(1994)}]{Sherman1994}\BibitemOpen
  \bibfield  {author} {\bibinfo {author} {\bibfnamefont {A.}~\bibnamefont
  {Sherman}},\ }\bibfield  {title} {\bibinfo {title} {Anti-phase, asymmetric
  and aperiodic oscillations in excitable cells—i. coupled bursters},\ }\href
  {https://doi.org/10.1007/bf02458269} {\bibfield  {journal} {\bibinfo
  {journal} {Bulletin of Mathematical Biology}\ }\textbf {\bibinfo {volume}
  {56}},\ \bibinfo {pages} {811–835} (\bibinfo {year} {1994})}\BibitemShut
  {NoStop}\bibitem [{\citenamefont {Röhm}\ \emph {et~al.}(2018)\citenamefont {Röhm},
  \citenamefont {Lüdge},\ and\ \citenamefont {Schneider}}]{Rohm2018}\BibitemOpen
  \bibfield  {author} {\bibinfo {author} {\bibfnamefont {A.}~\bibnamefont
  {Röhm}}, \bibinfo {author} {\bibfnamefont {K.}~\bibnamefont {Lüdge}},\ and\
  \bibinfo {author} {\bibfnamefont {I.}~\bibnamefont {Schneider}},\ }\bibfield
  {title} {\bibinfo {title} {Bistability in two simple symmetrically coupled
  oscillators with symmetry-broken amplitude- and phase-locking},\ }\href
  {https://doi.org/10.1063/1.5018262} {\bibfield  {journal} {\bibinfo
  {journal} {Chaos: An Interdisciplinary Journal of Nonlinear Science}\
  }\textbf {\bibinfo {volume} {28}},\ \bibinfo {pages} {063114} (\bibinfo
  {year} {2018})}\BibitemShut {NoStop}\bibitem [{\citenamefont {Marques}\ \emph {et~al.}(2013)\citenamefont
  {Marques}, \citenamefont {Mellibovsky},\ and\ \citenamefont
  {Meseguer}}]{Marques2013}\BibitemOpen
  \bibfield  {author} {\bibinfo {author} {\bibfnamefont {F.}~\bibnamefont
  {Marques}}, \bibinfo {author} {\bibfnamefont {F.}~\bibnamefont
  {Mellibovsky}},\ and\ \bibinfo {author} {\bibfnamefont {A.}~\bibnamefont
  {Meseguer}},\ }\bibfield  {title} {\bibinfo {title} {Fold-pitchfork
  bifurcation for maps with z2 symmetry in pipe flow},\ }\href
  {https://doi.org/10.1103/physreve.88.013006} {\bibfield  {journal} {\bibinfo
  {journal} {Physical Review E}\ }\textbf {\bibinfo {volume} {88}},\ \bibinfo
  {pages} {013006} (\bibinfo {year} {2013})}\BibitemShut {NoStop}\bibitem [{\citenamefont {Bačić}\ and\ \citenamefont
  {Franović}(2020)}]{Bacic2020}\BibitemOpen
  \bibfield  {author} {\bibinfo {author} {\bibfnamefont {I.}~\bibnamefont
  {Bačić}}\ and\ \bibinfo {author} {\bibfnamefont {I.}~\bibnamefont
  {Franović}},\ }\bibfield  {title} {\bibinfo {title} {Two paradigmatic
  scenarios for inverse stochastic resonance},\ }\bibfield  {journal} {\bibinfo
   {journal} {Chaos: An Interdisciplinary Journal of Nonlinear Science}\
  }\textbf {\bibinfo {volume} {30}},\ \href {https://doi.org/10.1063/1.5139628}
  {10.1063/1.5139628} (\bibinfo {year} {2020})\BibitemShut {NoStop}\bibitem [{\citenamefont {Willms}\ \emph {et~al.}(2017)\citenamefont {Willms},
  \citenamefont {Kitanov},\ and\ \citenamefont {Langford}}]{Willms2017}\BibitemOpen
  \bibfield  {author} {\bibinfo {author} {\bibfnamefont {A.~R.}\ \bibnamefont
  {Willms}}, \bibinfo {author} {\bibfnamefont {P.~M.}\ \bibnamefont
  {Kitanov}},\ and\ \bibinfo {author} {\bibfnamefont {W.~F.}\ \bibnamefont
  {Langford}},\ }\bibfield  {title} {\bibinfo {title} {Huygens’ clocks
  revisited},\ }\href {https://doi.org/10.1098/rsos.170777} {\bibfield
  {journal} {\bibinfo  {journal} {Royal Society Open Science}\ }\textbf
  {\bibinfo {volume} {4}},\ \bibinfo {pages} {170777} (\bibinfo {year}
  {2017})}\BibitemShut {NoStop}\bibitem [{\citenamefont {Huhn}\ and\ \citenamefont {Magri}(2019)}]{Huhn2019}\BibitemOpen
  \bibfield  {author} {\bibinfo {author} {\bibfnamefont {F.}~\bibnamefont
  {Huhn}}\ and\ \bibinfo {author} {\bibfnamefont {L.}~\bibnamefont {Magri}},\
  }\bibfield  {title} {\bibinfo {title} {Stability, sensitivity and
  optimisation of chaotic acoustic oscillations},\ }\bibfield  {journal}
  {\bibinfo  {journal} {Journal of Fluid Mechanics}\ }\textbf {\bibinfo
  {volume} {882}},\ \href {https://doi.org/10.1017/jfm.2019.828}
  {10.1017/jfm.2019.828} (\bibinfo {year} {2019})\BibitemShut {NoStop}\bibitem [{\citenamefont {Haken}(1983)}]{Haken1983}\BibitemOpen
  \bibfield  {author} {\bibinfo {author} {\bibfnamefont {H.}~\bibnamefont
  {Haken}},\ }\href@noop {} {\emph {\bibinfo {title} {Synergetics: An
  Introduction}}},\ Springer Series in Synergetics\ (\bibinfo  {publisher}
  {Springer Berlin Heidelberg},\ \bibinfo {year} {1983})\BibitemShut {NoStop}\bibitem [{\citenamefont {Wilson}\ and\ \citenamefont
  {Cowan}(1972)}]{Wilson1972}\BibitemOpen
  \bibfield  {author} {\bibinfo {author} {\bibfnamefont {H.~R.}\ \bibnamefont
  {Wilson}}\ and\ \bibinfo {author} {\bibfnamefont {J.~D.}\ \bibnamefont
  {Cowan}},\ }\bibfield  {title} {\bibinfo {title} {Excitatory and inhibitory
  interactions in localized populations of model neurons},\ }\href
  {https://doi.org/10.1016/S0006-3495(72)86068-5} {\bibfield  {journal}
  {\bibinfo  {journal} {Biophysical Journal}\ }\textbf {\bibinfo {volume}
  {12}},\ \bibinfo {pages} {1} (\bibinfo {year} {1972})}\BibitemShut {NoStop}\bibitem [{\citenamefont {P{\'{e}}rez-Cervera}\ \emph
  {et~al.}(2019)\citenamefont {P{\'{e}}rez-Cervera}, \citenamefont {Ashwin},
  \citenamefont {Huguet}, \citenamefont {Seara},\ and\ \citenamefont
  {Rankin}}]{PrezCervera2019}\BibitemOpen
  \bibfield  {author} {\bibinfo {author} {\bibfnamefont {A.}~\bibnamefont
  {P{\'{e}}rez-Cervera}}, \bibinfo {author} {\bibfnamefont {P.}~\bibnamefont
  {Ashwin}}, \bibinfo {author} {\bibfnamefont {G.}~\bibnamefont {Huguet}},
  \bibinfo {author} {\bibfnamefont {T.~M.}\ \bibnamefont {Seara}},\ and\
  \bibinfo {author} {\bibfnamefont {J.}~\bibnamefont {Rankin}},\ }\bibfield
  {title} {\bibinfo {title} {The uncoupling limit of identical {H}opf
  bifurcations with an application to perceptual bistability},\ }\bibfield
  {journal} {\bibinfo  {journal} {The Journal of Mathematical Neuroscience}\
  }\textbf {\bibinfo {volume} {9}},\ \href
  {https://doi.org/10.1186/s13408-019-0075-2} {10.1186/s13408-019-0075-2}
  (\bibinfo {year} {2019})\BibitemShut {NoStop}\bibitem [{\citenamefont {Rankin}\ \emph {et~al.}(2015)\citenamefont {Rankin},
  \citenamefont {Sussman},\ and\ \citenamefont {Rinzel}}]{Rankin2015}\BibitemOpen
  \bibfield  {author} {\bibinfo {author} {\bibfnamefont {J.}~\bibnamefont
  {Rankin}}, \bibinfo {author} {\bibfnamefont {E.}~\bibnamefont {Sussman}},\
  and\ \bibinfo {author} {\bibfnamefont {J.}~\bibnamefont {Rinzel}},\
  }\bibfield  {title} {\bibinfo {title} {Neuromechanistic model of auditory
  bistability},\ }\href {https://doi.org/10.1371/journal.pcbi.1004555}
  {\bibfield  {journal} {\bibinfo  {journal} {PLOS Computational Biology}\
  }\textbf {\bibinfo {volume} {11}},\ \bibinfo {pages} {e1004555} (\bibinfo
  {year} {2015})}\BibitemShut {NoStop}\bibitem [{\citenamefont {Pressnitzer}\ and\ \citenamefont
  {Hupé}(2006)}]{Pressnitzer2006}\BibitemOpen
  \bibfield  {author} {\bibinfo {author} {\bibfnamefont {D.}~\bibnamefont
  {Pressnitzer}}\ and\ \bibinfo {author} {\bibfnamefont {J.-M.}\ \bibnamefont
  {Hupé}},\ }\bibfield  {title} {\bibinfo {title} {Temporal dynamics of
  auditory and visual bistability reveal common principles of perceptual
  organization},\ }\href {https://doi.org/10.1016/j.cub.2006.05.054} {\bibfield
   {journal} {\bibinfo  {journal} {Current Biology}\ }\textbf {\bibinfo
  {volume} {16}},\ \bibinfo {pages} {1351–1357} (\bibinfo {year}
  {2006})}\BibitemShut {NoStop}\bibitem [{per(2022)}]{perceptual}\BibitemOpen
  \href {https://www.illusionsindex.org/i/bistable-audio} {\bibinfo {title}
  {Bistable auditory stimulus}} (\bibinfo {year} {2022})\BibitemShut {NoStop}\bibitem [{\citenamefont {Vandermeer}(2006)}]{Vandermeer2006}\BibitemOpen
  \bibfield  {author} {\bibinfo {author} {\bibfnamefont {J.}~\bibnamefont
  {Vandermeer}},\ }\bibfield  {title} {\bibinfo {title} {Oscillating
  populations and biodiversity maintenance},\ }\href
  {https://doi.org/10.1641/0006-3568(2006)56[967:opabm]2.0.co;2} {\bibfield
  {journal} {\bibinfo  {journal} {BioScience}\ }\textbf {\bibinfo {volume}
  {56}},\ \bibinfo {pages} {967} (\bibinfo {year} {2006})}\BibitemShut
  {NoStop}\bibitem [{\citenamefont {Chesson}(2000)}]{Chesson2000}\BibitemOpen
  \bibfield  {author} {\bibinfo {author} {\bibfnamefont {P.}~\bibnamefont
  {Chesson}},\ }\bibfield  {title} {\bibinfo {title} {Mechanisms of maintenance
  of species diversity},\ }\href
  {https://doi.org/10.1146/annurev.ecolsys.31.1.343} {\bibfield  {journal}
  {\bibinfo  {journal} {Annual Review of Ecology and Systematics}\ }\textbf
  {\bibinfo {volume} {31}},\ \bibinfo {pages} {343–366} (\bibinfo {year}
  {2000})}\BibitemShut {NoStop}\bibitem [{\citenamefont {Armstrong}\ and\ \citenamefont
  {McGehee}(1980)}]{Armstrong1980}\BibitemOpen
  \bibfield  {author} {\bibinfo {author} {\bibfnamefont {R.~A.}\ \bibnamefont
  {Armstrong}}\ and\ \bibinfo {author} {\bibfnamefont {R.}~\bibnamefont
  {McGehee}},\ }\bibfield  {title} {\bibinfo {title} {Competitive exclusion},\
  }\href {https://doi.org/10.1086/283553} {\bibfield  {journal} {\bibinfo
  {journal} {The American Naturalist}\ }\textbf {\bibinfo {volume} {115}},\
  \bibinfo {pages} {151–170} (\bibinfo {year} {1980})}\BibitemShut {NoStop}\bibitem [{\citenamefont {Benincà}\ \emph {et~al.}(2009)\citenamefont
  {Benincà}, \citenamefont {Jöhnk}, \citenamefont {Heerkloss},\ and\
  \citenamefont {Huisman}}]{Beninca2009}\BibitemOpen
  \bibfield  {author} {\bibinfo {author} {\bibfnamefont {E.}~\bibnamefont
  {Benincà}}, \bibinfo {author} {\bibfnamefont {K.~D.}\ \bibnamefont
  {Jöhnk}}, \bibinfo {author} {\bibfnamefont {R.}~\bibnamefont {Heerkloss}},\
  and\ \bibinfo {author} {\bibfnamefont {J.}~\bibnamefont {Huisman}},\
  }\bibfield  {title} {\bibinfo {title} {Coupled predator-prey oscillations in
  a chaotic food web},\ }\href
  {https://doi.org/10.1111/j.1461-0248.2009.01391.x} {\bibfield  {journal}
  {\bibinfo  {journal} {Ecology Letters}\ }\textbf {\bibinfo {volume} {12}},\
  \bibinfo {pages} {1367–1378} (\bibinfo {year} {2009})}\BibitemShut
  {NoStop}\bibitem [{\citenamefont {Turchin}(2013)}]{Turchin2013}\BibitemOpen
  \bibfield  {author} {\bibinfo {author} {\bibfnamefont {P.}~\bibnamefont
  {Turchin}},\ }\href@noop {} {\emph {\bibinfo {title} {Complex Population
  Dynamics: A Theoretical/Empirical Synthesis (MPB-35)}}},\ Monographs in
  Population Biology\ (\bibinfo  {publisher} {Princeton University Press},\
  \bibinfo {year} {2013})\BibitemShut {NoStop}\bibitem [{\citenamefont {Maynard-Smith}(1978)}]{Maynard1978}\BibitemOpen
  \bibfield  {author} {\bibinfo {author} {\bibfnamefont {J.}~\bibnamefont
  {Maynard-Smith}},\ }\href@noop {} {\emph {\bibinfo {title} {Models in
  Ecology}}}\ (\bibinfo  {publisher} {Cambridge University Press},\ \bibinfo
  {year} {1978})\BibitemShut {NoStop}\bibitem [{\citenamefont {Rosenzweig}\ and\ \citenamefont
  {MacArthur}(1963)}]{Rosenzweig1963}\BibitemOpen
  \bibfield  {author} {\bibinfo {author} {\bibfnamefont {M.~L.}\ \bibnamefont
  {Rosenzweig}}\ and\ \bibinfo {author} {\bibfnamefont {R.~H.}\ \bibnamefont
  {MacArthur}},\ }\bibfield  {title} {\bibinfo {title} {Graphical
  representation and stability conditions of predator-prey interactions},\
  }\href {https://doi.org/10.1086/282272} {\bibfield  {journal} {\bibinfo
  {journal} {The American Naturalist}\ }\textbf {\bibinfo {volume} {97}},\
  \bibinfo {pages} {209–223} (\bibinfo {year} {1963})}\BibitemShut {NoStop}\bibitem [{\citenamefont {Vandermeer}(2004)}]{Vandermeer2004}\BibitemOpen
  \bibfield  {author} {\bibinfo {author} {\bibfnamefont {J.}~\bibnamefont
  {Vandermeer}},\ }\bibfield  {title} {\bibinfo {title} {Coupled oscillations
  in food webs: Balancing competition and mutualism in simple ecological
  models},\ }\href {https://doi.org/10.1086/420776} {\bibfield  {journal}
  {\bibinfo  {journal} {The American Naturalist}\ }\textbf {\bibinfo {volume}
  {163}},\ \bibinfo {pages} {857–867} (\bibinfo {year} {2004})}\BibitemShut
  {NoStop}\bibitem [{\citenamefont {Vandermeer}(1993)}]{Vandermeer1993}\BibitemOpen
  \bibfield  {author} {\bibinfo {author} {\bibfnamefont {J.}~\bibnamefont
  {Vandermeer}},\ }\bibfield  {title} {\bibinfo {title} {Loose coupling of
  predator-prey cycles: Entrainment, chaos, and intermittency in the classic
  macarthur consumer-resource equations},\ }\href
  {https://doi.org/10.1086/285500} {\bibfield  {journal} {\bibinfo  {journal}
  {The American Naturalist}\ }\textbf {\bibinfo {volume} {141}},\ \bibinfo
  {pages} {687–716} (\bibinfo {year} {1993})}\BibitemShut {NoStop}\bibitem [{\citenamefont {Xue}\ and\ \citenamefont
  {Goldenfeld}(2017)}]{Xue2017}\BibitemOpen
  \bibfield  {author} {\bibinfo {author} {\bibfnamefont {C.}~\bibnamefont
  {Xue}}\ and\ \bibinfo {author} {\bibfnamefont {N.}~\bibnamefont
  {Goldenfeld}},\ }\bibfield  {title} {\bibinfo {title} {Coevolution maintains
  diversity in the stochastic “kill the winner” model},\ }\href
  {https://doi.org/10.1103/physrevlett.119.268101} {\bibfield  {journal}
  {\bibinfo  {journal} {Physical Review Letters}\ }\textbf {\bibinfo {volume}
  {119}},\ \bibinfo {pages} {268101} (\bibinfo {year} {2017})}\BibitemShut
  {NoStop}\bibitem [{\citenamefont {Lorenz}(1963)}]{Lorenz1963}\BibitemOpen
  \bibfield  {author} {\bibinfo {author} {\bibfnamefont {E.~N.}\ \bibnamefont
  {Lorenz}},\ }\bibfield  {title} {\bibinfo {title} {Deterministic nonperiodic
  flow},\ }\href {https://doi.org/10.1175/1520-0469(1963)020<0130:dnf>2.0.co;2}
  {\bibfield  {journal} {\bibinfo  {journal} {Journal of the Atmospheric
  Sciences}\ }\textbf {\bibinfo {volume} {20}},\ \bibinfo {pages} {130–141}
  (\bibinfo {year} {1963})}\BibitemShut {NoStop}\bibitem [{\citenamefont {Ott}(2002)}]{Ott2002}\BibitemOpen
  \bibfield  {author} {\bibinfo {author} {\bibfnamefont {E.}~\bibnamefont
  {Ott}},\ }\href@noop {} {\emph {\bibinfo {title} {Chaos in Dynamical
  Systems}}}\ (\bibinfo  {publisher} {Cambridge University Press},\ \bibinfo
  {year} {2002})\BibitemShut {NoStop}\bibitem [{Note12()}]{Note12}\BibitemOpen
  \bibinfo {note} {The model \protect \textup {{\protect \normalfont (\ref
  {eq:chaos}}\protect \normalfont )} does not directly describes coupled
  chaotic attractors. The parallels between this situation and the case of
  limit cycles suggest that a bifurcation accompanied by a coalescence of
  covariant Lyapunov attractors could occur in coupled chaotic attractors,
  maybe in relation with their synchronization~\cite
  {Boccaletti2002,Pecora1990}. However, this is not analyzed in this work.
  Relations between Lyapunov vectors and attractor merging crises~\cite
  {Grebogi1982,Grebogi1983,Grebogi1987}, that occur when attractors collide
  with each other have also been put forward~\cite
  {Beims2016,Tantet2017}.}\BibitemShut {Stop}\bibitem [{\citenamefont {Rackauckas}\ and\ \citenamefont
  {Nie}(2017)}]{Rackauckas2017}\BibitemOpen
  \bibfield  {author} {\bibinfo {author} {\bibfnamefont {C.}~\bibnamefont
  {Rackauckas}}\ and\ \bibinfo {author} {\bibfnamefont {Q.}~\bibnamefont
  {Nie}},\ }\bibfield  {title} {\bibinfo {title} {Differential{E}quations.jl--a
  performant and feature-rich ecosystem for solving differential equations in
  {J}ulia},\ }\href@noop {} {\bibfield  {journal} {\bibinfo  {journal} {Journal
  of Open Research Software}\ }\textbf {\bibinfo {volume} {5}} (\bibinfo {year}
  {2017})}\BibitemShut {NoStop}\bibitem [{\citenamefont {Benettin}\ \emph
  {et~al.}(1980{\natexlab{a}})\citenamefont {Benettin}, \citenamefont
  {Galgani}, \citenamefont {Giorgilli},\ and\ \citenamefont
  {Strelcyn}}]{Benettin1980}\BibitemOpen
  \bibfield  {author} {\bibinfo {author} {\bibfnamefont {G.}~\bibnamefont
  {Benettin}}, \bibinfo {author} {\bibfnamefont {L.}~\bibnamefont {Galgani}},
  \bibinfo {author} {\bibfnamefont {A.}~\bibnamefont {Giorgilli}},\ and\
  \bibinfo {author} {\bibfnamefont {J.-M.}\ \bibnamefont {Strelcyn}},\
  }\bibfield  {title} {\bibinfo {title} {Lyapunov characteristic exponents for
  smooth dynamical systems and for hamiltonian systems: a method for computing
  all of them. part 1: Theory},\ }\href {https://doi.org/10.1007/bf02128236}
  {\bibfield  {journal} {\bibinfo  {journal} {Meccanica}\ }\textbf {\bibinfo
  {volume} {15}},\ \bibinfo {pages} {9} (\bibinfo {year}
  {1980}{\natexlab{a}})}\BibitemShut {NoStop}\bibitem [{\citenamefont {Benettin}\ \emph
  {et~al.}(1980{\natexlab{b}})\citenamefont {Benettin}, \citenamefont
  {Galgani}, \citenamefont {Giorgilli},\ and\ \citenamefont
  {Strelcyn}}]{Benettin1980b}\BibitemOpen
  \bibfield  {author} {\bibinfo {author} {\bibfnamefont {G.}~\bibnamefont
  {Benettin}}, \bibinfo {author} {\bibfnamefont {L.}~\bibnamefont {Galgani}},
  \bibinfo {author} {\bibfnamefont {A.}~\bibnamefont {Giorgilli}},\ and\
  \bibinfo {author} {\bibfnamefont {J.-M.}\ \bibnamefont {Strelcyn}},\
  }\bibfield  {title} {\bibinfo {title} {Lyapunov characteristic exponents for
  smooth dynamical systems and for hamiltonian systems: A method for computing
  all of them. part 2: Numerical application},\ }\href
  {https://doi.org/10.1007/bf02128237} {\bibfield  {journal} {\bibinfo
  {journal} {Meccanica}\ }\textbf {\bibinfo {volume} {15}},\ \bibinfo {pages}
  {21} (\bibinfo {year} {1980}{\natexlab{b}})}\BibitemShut {NoStop}\bibitem [{\citenamefont {Muñoz}(2018)}]{Munoz2018}\BibitemOpen
  \bibfield  {author} {\bibinfo {author} {\bibfnamefont {M.~A.}\ \bibnamefont
  {Muñoz}},\ }\bibfield  {title} {\bibinfo {title} {Colloquium: Criticality
  and dynamical scaling in living systems},\ }\href
  {https://doi.org/10.1103/revmodphys.90.031001} {\bibfield  {journal}
  {\bibinfo  {journal} {Reviews of Modern Physics}\ }\textbf {\bibinfo {volume}
  {90}},\ \bibinfo {pages} {031001} (\bibinfo {year} {2018})}\BibitemShut
  {NoStop}\bibitem [{Note13()}]{Note13}\BibitemOpen
  \bibinfo {note} {Lyapunov exponent are asymptotic quantities describing the
  growth rate at large times. By definition, they do not pick up the transient
  effects we discuss here. These can be captured more easily by so-called
  finite-time Lyapunov exponents and similar quantities. It is remarkable that
  the long-time CLVs are still a partial proxy for some transient
  effects.}\BibitemShut {Stop}\bibitem [{NoteCaptionNRP()}]{NoteCaptionNRP}\BibitemOpen
  \bibinfo {note} {In general, there can be several missing directions
  $\vec{c}_n^\bot$. At an exceptional point, these would correspond to
  so-called generalized eigenvectors in the Jordan decomposition of a matrix.
  Note also that in the figure, the tangent space is shown as light-gray
  squares orthogonal to the black trajectory for graphical reasons; in
  practice, CLVs can be tangent to the trajectory.}\BibitemShut {Stop}\bibitem [{\citenamefont {Kuramoto}(1984)}]{Kuramoto1984}\BibitemOpen
  \bibfield  {author} {\bibinfo {author} {\bibfnamefont {Y.}~\bibnamefont
  {Kuramoto}},\ }\href {https://doi.org/10.1007/978-3-642-69689-3} {\emph
  {\bibinfo {title} {Chemical Oscillations, Waves, And Turbulence}}}\ (\bibinfo
   {publisher} {Springer},\ \bibinfo {year} {1984})\BibitemShut {NoStop}\bibitem [{\citenamefont {Winfree}(2001)}]{Winfree2001}\BibitemOpen
  \bibfield  {author} {\bibinfo {author} {\bibfnamefont {A.~T.}\ \bibnamefont
  {Winfree}},\ }\href {https://doi.org/10.1007/978-1-4757-3484-3} {\emph
  {\bibinfo {title} {{The Geometry of Biological Time}}}}\ (\bibinfo
  {publisher} {Springer New York},\ \bibinfo {year} {2001})\BibitemShut
  {NoStop}\bibitem [{\citenamefont {Guckenheimer}(1975)}]{Guckenheimer1975}\BibitemOpen
  \bibfield  {author} {\bibinfo {author} {\bibfnamefont {J.}~\bibnamefont
  {Guckenheimer}},\ }\bibfield  {title} {\bibinfo {title} {Isochrons and
  phaseless sets},\ }\href {https://doi.org/10.1007/bf01273747} {\bibfield
  {journal} {\bibinfo  {journal} {Journal of Mathematical Biology}\ }\textbf
  {\bibinfo {volume} {1}},\ \bibinfo {pages} {259–273} (\bibinfo {year}
  {1975})}\BibitemShut {NoStop}\bibitem [{\citenamefont {Winfree}(1974)}]{Winfree1974}\BibitemOpen
  \bibfield  {author} {\bibinfo {author} {\bibfnamefont {A.~T.}\ \bibnamefont
  {Winfree}},\ }\bibfield  {title} {\bibinfo {title} {Patterns of phase
  compromise in biological cycles},\ }\href
  {https://doi.org/10.1007/bf02339491} {\bibfield  {journal} {\bibinfo
  {journal} {Journal of Mathematical Biology}\ }\textbf {\bibinfo {volume}
  {1}},\ \bibinfo {pages} {73–93} (\bibinfo {year} {1974})}\BibitemShut
  {NoStop}\bibitem [{\citenamefont {Mauroy}\ and\ \citenamefont
  {Mezić}(2012)}]{Mauroy2012}\BibitemOpen
  \bibfield  {author} {\bibinfo {author} {\bibfnamefont {A.}~\bibnamefont
  {Mauroy}}\ and\ \bibinfo {author} {\bibfnamefont {I.}~\bibnamefont
  {Mezić}},\ }\bibfield  {title} {\bibinfo {title} {On the use of fourier
  averages to compute the global isochrons of (quasi)periodic dynamics},\
  }\href {https://doi.org/10.1063/1.4736859} {\bibfield  {journal} {\bibinfo
  {journal} {Chaos: An Interdisciplinary Journal of Nonlinear Science}\
  }\textbf {\bibinfo {volume} {22}},\ \bibinfo {pages} {033112} (\bibinfo
  {year} {2012})}\BibitemShut {NoStop}\bibitem [{\citenamefont {Mauroy}\ \emph {et~al.}(2013)\citenamefont {Mauroy},
  \citenamefont {Mezić},\ and\ \citenamefont {Moehlis}}]{Mauroy2013}\BibitemOpen
  \bibfield  {author} {\bibinfo {author} {\bibfnamefont {A.}~\bibnamefont
  {Mauroy}}, \bibinfo {author} {\bibfnamefont {I.}~\bibnamefont {Mezić}},\
  and\ \bibinfo {author} {\bibfnamefont {J.}~\bibnamefont {Moehlis}},\
  }\bibfield  {title} {\bibinfo {title} {Isostables, isochrons, and koopman
  spectrum for the action–angle representation of stable fixed point
  dynamics},\ }\href {https://doi.org/10.1016/j.physd.2013.06.004} {\bibfield
  {journal} {\bibinfo  {journal} {Physica D: Nonlinear Phenomena}\ }\textbf
  {\bibinfo {volume} {261}},\ \bibinfo {pages} {19–30} (\bibinfo {year}
  {2013})}\BibitemShut {NoStop}\bibitem [{\citenamefont {Mauroy}\ and\ \citenamefont
  {Mezic}(2016)}]{Mauroy2016}\BibitemOpen
  \bibfield  {author} {\bibinfo {author} {\bibfnamefont {A.}~\bibnamefont
  {Mauroy}}\ and\ \bibinfo {author} {\bibfnamefont {I.}~\bibnamefont {Mezic}},\
  }\bibfield  {title} {\bibinfo {title} {Global stability analysis using the
  eigenfunctions of the koopman operator},\ }\href
  {https://doi.org/10.1109/tac.2016.2518918} {\bibfield  {journal} {\bibinfo
  {journal} {IEEE Transactions on Automatic Control}\ }\textbf {\bibinfo
  {volume} {61}},\ \bibinfo {pages} {3356–3369} (\bibinfo {year}
  {2016})}\BibitemShut {NoStop}\bibitem [{\citenamefont {Ichinose}\ \emph {et~al.}(1998)\citenamefont
  {Ichinose}, \citenamefont {Aihara},\ and\ \citenamefont
  {Judd}}]{Ichinose1998}\BibitemOpen
  \bibfield  {author} {\bibinfo {author} {\bibfnamefont {N.}~\bibnamefont
  {Ichinose}}, \bibinfo {author} {\bibfnamefont {K.}~\bibnamefont {Aihara}},\
  and\ \bibinfo {author} {\bibfnamefont {K.}~\bibnamefont {Judd}},\ }\bibfield
  {title} {\bibinfo {title} {Extending the concept of isochrons from
  oscillatory to excitable systems for modeling an excitable neuron},\ }\href
  {https://doi.org/10.1142/s021812749800190x} {\bibfield  {journal} {\bibinfo
  {journal} {International Journal of Bifurcation and Chaos}\ }\textbf
  {\bibinfo {volume} {08}},\ \bibinfo {pages} {2375–2385} (\bibinfo {year}
  {1998})}\BibitemShut {NoStop}\bibitem [{\citenamefont {Shirasaka}\ \emph {et~al.}(2017)\citenamefont
  {Shirasaka}, \citenamefont {Kurebayashi},\ and\ \citenamefont
  {Nakao}}]{Shirasaka2017}\BibitemOpen
  \bibfield  {author} {\bibinfo {author} {\bibfnamefont {S.}~\bibnamefont
  {Shirasaka}}, \bibinfo {author} {\bibfnamefont {W.}~\bibnamefont
  {Kurebayashi}},\ and\ \bibinfo {author} {\bibfnamefont {H.}~\bibnamefont
  {Nakao}},\ }\bibfield  {title} {\bibinfo {title} {Phase-amplitude reduction
  of transient dynamics far from attractors for limit-cycling systems},\ }\href
  {https://doi.org/10.1063/1.4977195} {\bibfield  {journal} {\bibinfo
  {journal} {Chaos: An Interdisciplinary Journal of Nonlinear Science}\
  }\textbf {\bibinfo {volume} {27}},\ \bibinfo {pages} {023119} (\bibinfo
  {year} {2017})}\BibitemShut {NoStop}\bibitem [{\citenamefont {Wilson}\ and\ \citenamefont
  {Moehlis}(2015)}]{Wilson2015}\BibitemOpen
  \bibfield  {author} {\bibinfo {author} {\bibfnamefont {D.}~\bibnamefont
  {Wilson}}\ and\ \bibinfo {author} {\bibfnamefont {J.}~\bibnamefont
  {Moehlis}},\ }\bibfield  {title} {\bibinfo {title} {Extending phase reduction
  to excitable media: Theory and applications},\ }\href
  {https://doi.org/10.1137/140952478} {\bibfield  {journal} {\bibinfo
  {journal} {SIAM Review}\ }\textbf {\bibinfo {volume} {57}},\ \bibinfo {pages}
  {201–222} (\bibinfo {year} {2015})}\BibitemShut {NoStop}\bibitem [{\citenamefont {Shaw}\ \emph {et~al.}(2012)\citenamefont {Shaw},
  \citenamefont {Park}, \citenamefont {Chiel},\ and\ \citenamefont
  {Thomas}}]{Shaw2012}\BibitemOpen
  \bibfield  {author} {\bibinfo {author} {\bibfnamefont {K.~M.}\ \bibnamefont
  {Shaw}}, \bibinfo {author} {\bibfnamefont {Y.-M.}\ \bibnamefont {Park}},
  \bibinfo {author} {\bibfnamefont {H.~J.}\ \bibnamefont {Chiel}},\ and\
  \bibinfo {author} {\bibfnamefont {P.~J.}\ \bibnamefont {Thomas}},\ }\bibfield
   {title} {\bibinfo {title} {Phase resetting in an asymptotically phaseless
  system: On the phase response of limit cycles verging on a heteroclinic
  orbit},\ }\href {https://doi.org/10.1137/110828976} {\bibfield  {journal}
  {\bibinfo  {journal} {SIAM Journal on Applied Dynamical Systems}\ }\textbf
  {\bibinfo {volume} {11}},\ \bibinfo {pages} {350–391} (\bibinfo {year}
  {2012})}\BibitemShut {NoStop}\bibitem [{\citenamefont {Rosenblum}\ \emph {et~al.}(1996)\citenamefont
  {Rosenblum}, \citenamefont {Pikovsky},\ and\ \citenamefont
  {Kurths}}]{Rosenblum1996}\BibitemOpen
  \bibfield  {author} {\bibinfo {author} {\bibfnamefont {M.~G.}\ \bibnamefont
  {Rosenblum}}, \bibinfo {author} {\bibfnamefont {A.~S.}\ \bibnamefont
  {Pikovsky}},\ and\ \bibinfo {author} {\bibfnamefont {J.}~\bibnamefont
  {Kurths}},\ }\bibfield  {title} {\bibinfo {title} {Phase synchronization of
  chaotic oscillators},\ }\href {https://doi.org/10.1103/physrevlett.76.1804}
  {\bibfield  {journal} {\bibinfo  {journal} {Physical Review Letters}\
  }\textbf {\bibinfo {volume} {76}},\ \bibinfo {pages} {1804–1807} (\bibinfo
  {year} {1996})}\BibitemShut {NoStop}\bibitem [{\citenamefont {Schwabedal}\ \emph {et~al.}(2012)\citenamefont
  {Schwabedal}, \citenamefont {Pikovsky}, \citenamefont {Kralemann},\ and\
  \citenamefont {Rosenblum}}]{Schwabedal2012}\BibitemOpen
  \bibfield  {author} {\bibinfo {author} {\bibfnamefont {J.~T.~C.}\
  \bibnamefont {Schwabedal}}, \bibinfo {author} {\bibfnamefont
  {A.}~\bibnamefont {Pikovsky}}, \bibinfo {author} {\bibfnamefont
  {B.}~\bibnamefont {Kralemann}},\ and\ \bibinfo {author} {\bibfnamefont
  {M.}~\bibnamefont {Rosenblum}},\ }\bibfield  {title} {\bibinfo {title}
  {Optimal phase description of chaotic oscillators},\ }\href
  {https://doi.org/10.1103/physreve.85.026216} {\bibfield  {journal} {\bibinfo
  {journal} {Physical Review E}\ }\textbf {\bibinfo {volume} {85}},\ \bibinfo
  {pages} {026216} (\bibinfo {year} {2012})}\BibitemShut {NoStop}\bibitem [{\citenamefont {Tönjes}\ and\ \citenamefont
  {Kori}(2022)}]{Tonjes2022}\BibitemOpen
  \bibfield  {author} {\bibinfo {author} {\bibfnamefont {R.}~\bibnamefont
  {Tönjes}}\ and\ \bibinfo {author} {\bibfnamefont {H.}~\bibnamefont {Kori}},\
  }\bibfield  {title} {\bibinfo {title} {Phase and frequency linear response
  theory for hyperbolic chaotic oscillators},\ }\href
  {https://doi.org/10.1063/5.0064519} {\bibfield  {journal} {\bibinfo
  {journal} {Chaos: An Interdisciplinary Journal of Nonlinear Science}\
  }\textbf {\bibinfo {volume} {32}},\ \bibinfo {pages} {043124} (\bibinfo
  {year} {2022})}\BibitemShut {NoStop}\bibitem [{\citenamefont {Himona}\ \emph {et~al.}(2022)\citenamefont {Himona},
  \citenamefont {Kovanis},\ and\ \citenamefont {Kominis}}]{Himona2022}\BibitemOpen
  \bibfield  {author} {\bibinfo {author} {\bibfnamefont {G.}~\bibnamefont
  {Himona}}, \bibinfo {author} {\bibfnamefont {V.}~\bibnamefont {Kovanis}},\
  and\ \bibinfo {author} {\bibfnamefont {Y.}~\bibnamefont {Kominis}},\
  }\bibfield  {title} {\bibinfo {title} {Isochrons, phase response and
  synchronization dynamics of tunable photonic oscillators},\ }\href
  {https://doi.org/10.1103/physrevresearch.4.l012039} {\bibfield  {journal}
  {\bibinfo  {journal} {Physical Review Research}\ }\textbf {\bibinfo {volume}
  {4}},\ \bibinfo {pages} {l012039} (\bibinfo {year} {2022})}\BibitemShut
  {NoStop}\bibitem [{\citenamefont {Nakao}\ \emph {et~al.}(2014)\citenamefont {Nakao},
  \citenamefont {Yanagita},\ and\ \citenamefont {Kawamura}}]{Nakao2014}\BibitemOpen
  \bibfield  {author} {\bibinfo {author} {\bibfnamefont {H.}~\bibnamefont
  {Nakao}}, \bibinfo {author} {\bibfnamefont {T.}~\bibnamefont {Yanagita}},\
  and\ \bibinfo {author} {\bibfnamefont {Y.}~\bibnamefont {Kawamura}},\
  }\bibfield  {title} {\bibinfo {title} {Phase-reduction approach to
  synchronization of spatiotemporal rhythms in reaction-diffusion systems},\
  }\href {https://doi.org/10.1103/physrevx.4.021032} {\bibfield  {journal}
  {\bibinfo  {journal} {Physical Review X}\ }\textbf {\bibinfo {volume} {4}},\
  \bibinfo {pages} {021032} (\bibinfo {year} {2014})}\BibitemShut {NoStop}\bibitem [{Note14()}]{Note14}\BibitemOpen
  \bibinfo {note} {In Ref.~\cite {Guckenheimer1975}, it is proven that
  isochrons exist under the hypothesis that the limit cycle is hyperbolic (this
  means that the only Floquet multiplier on the unit circle corresponds to the
  Floquet eigenvector $\protect \mathaccentV {vec}17E{c}_*$ with multiplier
  $\mu = 1$, see Ref.~\cite {Kuznetsov2004}). Non-hyperbolic limit cycles
  typically have no well-defined isochrons, but this is not necessarily
  true~\cite {Castejon2013,Freire2007}. Conversely, limit cycles without
  well-defined isochrons are necessarily non-hyperbolic. This is because the
  hyperbolicity is defined in terms of Floquet multipliers (or Lyapunov
  exponents), while the existence or absence of isochrons is related to the
  Floquet eigenvectors (or CLVs). Namely, the isochrons cannot be well-defined
  when another CLV aligns with $\protect \mathaccentV {vec}17E{c}_*$. For
  instance, an exactly solvable planar system without isochrons can be found in
  Ref.~\cite {Demir2007}, in which the monodromy matrix [Eq~. (13) in the
  reference] is a Jordan block of size two. We also refer to Refs.~\cite
  {Tantet2020,Tantet2019,Chekroun2020} for a discussion of the effect of noise
  on an oscillator with twisted isochrons from the point of view of so-called
  Ruelle-Pollicott resonances.}\BibitemShut {Stop}\bibitem [{\citenamefont {Demir}(2007)}]{Demir2007}\BibitemOpen
  \bibfield  {author} {\bibinfo {author} {\bibfnamefont {A.}~\bibnamefont
  {Demir}},\ }\bibfield  {title} {\bibinfo {title} {Fully nonlinear oscillator
  noise analysis: an oscillator with no asymptotic phase},\ }\href
  {https://doi.org/10.1002/cta.387} {\bibfield  {journal} {\bibinfo  {journal}
  {International Journal of Circuit Theory and Applications}\ }\textbf
  {\bibinfo {volume} {35}},\ \bibinfo {pages} {175–203} (\bibinfo {year}
  {2007})}\BibitemShut {NoStop}\bibitem [{Note15()}]{Note15}\BibitemOpen
  \bibinfo {note} {If we assume that the phase correlations are eventually
  diffusive (namely, $\mathinner {\delimiter "426830A {\phi (t_0+t)\phi
  (t_0)}\delimiter "526930B } \sim D t$), these results can be guessed by
  dimensional analysis. To do so, let us consider the SDEs $\protect \text {d}w
  = r w \protect \text {d}t - b w^3 + \sigma \protect \text {d}W$ and $\protect
  \text {d}\phi = \omega _1 w$ in which $W$ is a Wiener process. We attribute
  independent dimensions $[w] = \protect \mathsf {W}$ and $[\phi ] = \Phi $ to
  the variables $w$ and $\phi $ (here, square brackets label the dimension of a
  quantity). In addition, the dimension symbol of time is $\protect \mathsf
  {T}$. The parameters are $[r] = \protect \mathsf {T}^{-1}$, $[b] = \protect
  \mathsf {T}^{-1} \protect \mathsf {W}^{-2}$, $[\omega _1] = \Phi \protect
  \mathsf {W}^{-1} \protect \mathsf {T}^{-1}$, and $[\sigma ] = \protect
  \mathsf {W} \protect \mathsf {T}^{-1/2}$ (recall that $[\protect \text {d}W]
  = \protect \sqrt {[\protect \text {d}t]}$). In addition, the diffusion
  coefficient we are looking for has dimension $[D] = \Phi ^2 \protect \mathsf
  {T}^{-1}$. First, assume $b=0$ (and $r \not =0$); dimensional analysis yields
  that the only combination of parameters with the appropriate dimension is
  $[D] = [\omega _1^2 \sigma ^2/r^2]$, so $D \sim \sigma ^2 \sim T_{w}$.
  Second, assume $r=0$ (and $b \not =0$); in the same way, dimensional analysis
  yields $[D] = [\omega _1^2/b]$ which does not depend on the noise strength
  $\sigma ^2$ (i.e. on the temperature $T_w$). A similar analysis can be
  performed when $\protect \text {d}\phi = \omega _1^{(2)} w^2$ (this
  corresponds to the case of Sec.~\ref {pflc} with discontinuous CLVs); in this
  case, we find that dimensionally, $D \sim [\omega _1^{(2)}]^2 \sigma b^{-3/2}
  \sim \protect \sqrt {T_w}$. We can therefore expect a significant increase of
  fluctuations near the bifurcation compared to the case far from the
  exceptional point where $r \gg b$ at low $T_w$. The general case in which
  both $r$ and $b$ are finite is not fully determined by dimensional analysis.
  We refer to Ref.~\cite {Shmakov2023} for a more complete scaling
  theory.}\BibitemShut {Stop}\bibitem [{Note16()}]{Note16}\BibitemOpen
  \bibinfo {note} {More precisely, the time evolution of $w$ can be
  approximated as a discrete-state continuous-time Markov process with two
  states $w_{\pm }$ (corresponding to the two minima of $U(w)$) with transition
  rates $1/\tau ^*$ between the states following the master equation \begin
  {equation} \partial _t \begin {pmatrix} p_+ \\ p_- \end {pmatrix} = \begin
  {pmatrix} - \lambda _{-+} & \lambda _{+-} \\ \lambda _{-+} & -\lambda _{+-}
  \end {pmatrix} \begin {pmatrix} p_+ \\ p_- \end {pmatrix} \end {equation} in
  which $\lambda _{i j} = \lambda _{j\to i}$ is the rate of transition from $i$
  to $j$, given by $\lambda _{+-} = \lambda _{-+} = 1/\tau ^*$ as the
  double-well is symmetric. As the transition rates are constant, the
  distribution of waiting times is exponential, reproducing Eq.~\protect
  \textup {{\protect \normalfont (\ref {expdisttimes}}\protect \normalfont )}.
  The telegraph noise is also known as a burst noise, a Markovian dichotomous
  noise, or a Kac process. It is related to a Poisson process in the sense that
  the number of changes between the two states in an interval of duration $\tau
  $ is Poissonian. See Refs.~\cite {vanKampen1992,Horsthemke2006,Bena2006} for
  details. We also note that a superposition of a large number of random
  telegraph processes can produce a $1/f$ noise \cite
  {Weissman1988,Kogan1996}.}\BibitemShut {Stop}\bibitem [{\citenamefont {Hänggi}\ \emph {et~al.}(1990)\citenamefont
  {Hänggi}, \citenamefont {Talkner},\ and\ \citenamefont
  {Borkovec}}]{Hanggi1990}\BibitemOpen
  \bibfield  {author} {\bibinfo {author} {\bibfnamefont {P.}~\bibnamefont
  {Hänggi}}, \bibinfo {author} {\bibfnamefont {P.}~\bibnamefont {Talkner}},\
  and\ \bibinfo {author} {\bibfnamefont {M.}~\bibnamefont {Borkovec}},\
  }\bibfield  {title} {\bibinfo {title} {Reaction-rate theory: fifty years
  after kramers},\ }\href {https://doi.org/10.1103/revmodphys.62.251}
  {\bibfield  {journal} {\bibinfo  {journal} {Reviews of Modern Physics}\
  }\textbf {\bibinfo {volume} {62}},\ \bibinfo {pages} {251–341} (\bibinfo
  {year} {1990})}\BibitemShut {NoStop}\bibitem [{\citenamefont {Berglund}(2011)}]{Berglund2011}\BibitemOpen
  \bibfield  {author} {\bibinfo {author} {\bibfnamefont {N.}~\bibnamefont
  {Berglund}},\ }\href@noop {} {\bibinfo {title} {Kramers' law: Validity,
  derivations and generalisations}} (\bibinfo {year} {2011}),\ \Eprint
  {https://arxiv.org/abs/1106.5799} {arXiv:1106.5799} \BibitemShut {NoStop}\bibitem [{\citenamefont {Bovier}\ \emph {et~al.}(2004)\citenamefont {Bovier},
  \citenamefont {Eckhoff}, \citenamefont {Gayrard},\ and\ \citenamefont
  {Klein}}]{Bovier2004}\BibitemOpen
  \bibfield  {author} {\bibinfo {author} {\bibfnamefont {A.}~\bibnamefont
  {Bovier}}, \bibinfo {author} {\bibfnamefont {M.}~\bibnamefont {Eckhoff}},
  \bibinfo {author} {\bibfnamefont {V.}~\bibnamefont {Gayrard}},\ and\ \bibinfo
  {author} {\bibfnamefont {M.}~\bibnamefont {Klein}},\ }\bibfield  {title}
  {\bibinfo {title} {Metastability in reversible diffusion processes i: Sharp
  asymptotics for capacities and exit times},\ }\href
  {https://doi.org/10.4171/jems/14} {\bibfield  {journal} {\bibinfo  {journal}
  {Journal of the European Mathematical Society}\ ,\ \bibinfo {pages} {399}}
  (\bibinfo {year} {2004})}\BibitemShut {NoStop}\bibitem [{\citenamefont {Bovier}\ \emph {et~al.}(2005)\citenamefont {Bovier},
  \citenamefont {Gayrard},\ and\ \citenamefont {Klein}}]{Bovier2005}\BibitemOpen
  \bibfield  {author} {\bibinfo {author} {\bibfnamefont {A.}~\bibnamefont
  {Bovier}}, \bibinfo {author} {\bibfnamefont {V.}~\bibnamefont {Gayrard}},\
  and\ \bibinfo {author} {\bibfnamefont {M.}~\bibnamefont {Klein}},\ }\bibfield
   {title} {\bibinfo {title} {Metastability in reversible diffusion processes
  {II}: precise asymptotics for small eigenvalues},\ }\href
  {https://doi.org/10.4171/jems/22} {\bibfield  {journal} {\bibinfo  {journal}
  {Journal of the European Mathematical Society}\ ,\ \bibinfo {pages} {69}}
  (\bibinfo {year} {2005})}\BibitemShut {NoStop}\bibitem [{\citenamefont {Bouchet}\ and\ \citenamefont
  {Reygner}(2016)}]{Bouchet2016}\BibitemOpen
  \bibfield  {author} {\bibinfo {author} {\bibfnamefont {F.}~\bibnamefont
  {Bouchet}}\ and\ \bibinfo {author} {\bibfnamefont {J.}~\bibnamefont
  {Reygner}},\ }\bibfield  {title} {\bibinfo {title} {Generalisation of the
  eyring–kramers transition rate formula to irreversible diffusion
  processes},\ }\href {https://doi.org/10.1007/s00023-016-0507-4} {\bibfield
  {journal} {\bibinfo  {journal} {Annales Henri Poincaré}\ }\textbf {\bibinfo
  {volume} {17}},\ \bibinfo {pages} {3499–3532} (\bibinfo {year}
  {2016})}\BibitemShut {NoStop}\bibitem [{\citenamefont {Kampen}(1992)}]{vanKampen1992}\BibitemOpen
  \bibfield  {author} {\bibinfo {author} {\bibfnamefont {N.~V.}\ \bibnamefont
  {Kampen}},\ }\href@noop {} {\emph {\bibinfo {title} {Stochastic Processes In
  Physics And Chemistry}}}\ (\bibinfo  {publisher} {Elsevier},\ \bibinfo {year}
  {1992})\BibitemShut {NoStop}\bibitem [{\citenamefont {Horsthemke}\ and\ \citenamefont
  {Lefever}(2006)}]{Horsthemke2006}\BibitemOpen
  \bibfield  {author} {\bibinfo {author} {\bibfnamefont {W.}~\bibnamefont
  {Horsthemke}}\ and\ \bibinfo {author} {\bibfnamefont {R.}~\bibnamefont
  {Lefever}},\ }\href@noop {} {\emph {\bibinfo {title} {Noise-Induced
  Transitions}}}\ (\bibinfo  {publisher} {Springer Science and Business
  Media},\ \bibinfo {year} {2006})\BibitemShut {NoStop}\bibitem [{\citenamefont {Bena}(2006)}]{Bena2006}\BibitemOpen
  \bibfield  {author} {\bibinfo {author} {\bibfnamefont {I.}~\bibnamefont
  {Bena}},\ }\bibfield  {title} {\bibinfo {title} {Dichotomous markov noise:
  Exact results for out-of-equilibrium systems},\ }\href
  {https://doi.org/10.1142/s0217979206034881} {\bibfield  {journal} {\bibinfo
  {journal} {International Journal of Modern Physics B}\ }\textbf {\bibinfo
  {volume} {20}},\ \bibinfo {pages} {2825–2888} (\bibinfo {year}
  {2006})}\BibitemShut {NoStop}\bibitem [{\citenamefont {H{\"a}nggi}\ and\ \citenamefont
  {Jung}(1995)}]{Hanggi1995}\BibitemOpen
  \bibfield  {author} {\bibinfo {author} {\bibfnamefont {P.}~\bibnamefont
  {H{\"a}nggi}}\ and\ \bibinfo {author} {\bibfnamefont {P.}~\bibnamefont
  {Jung}},\ }\bibfield  {title} {\bibinfo {title} {Colored noise in dynamical
  systems},\ }\href@noop {} {\bibfield  {journal} {\bibinfo  {journal}
  {Advances in chemical physics}\ }\textbf {\bibinfo {volume} {89}},\ \bibinfo
  {pages} {239} (\bibinfo {year} {1995})}\BibitemShut {NoStop}\bibitem [{\citenamefont {Weiss}(2002)}]{Weiss2002}\BibitemOpen
  \bibfield  {author} {\bibinfo {author} {\bibfnamefont {G.~H.}\ \bibnamefont
  {Weiss}},\ }\bibfield  {title} {\bibinfo {title} {Some applications of
  persistent random walks and the telegrapher’s equation},\ }\href
  {https://doi.org/10.1016/s0378-4371(02)00805-1} {\bibfield  {journal}
  {\bibinfo  {journal} {Physica A: Statistical Mechanics and its Applications}\
  }\textbf {\bibinfo {volume} {311}},\ \bibinfo {pages} {381–410} (\bibinfo
  {year} {2002})}\BibitemShut {NoStop}\bibitem [{\citenamefont {Masoliver}\ and\ \citenamefont
  {Lindenberg}(2017)}]{Masoliver2017}\BibitemOpen
  \bibfield  {author} {\bibinfo {author} {\bibfnamefont {J.}~\bibnamefont
  {Masoliver}}\ and\ \bibinfo {author} {\bibfnamefont {K.}~\bibnamefont
  {Lindenberg}},\ }\bibfield  {title} {\bibinfo {title} {Continuous time
  persistent random walk: a review and some generalizations},\ }\bibfield
  {journal} {\bibinfo  {journal} {The European Physical Journal B}\ }\textbf
  {\bibinfo {volume} {90}},\ \href {https://doi.org/10.1140/epjb/e2017-80123-7}
  {10.1140/epjb/e2017-80123-7} (\bibinfo {year} {2017})\BibitemShut {NoStop}\bibitem [{\citenamefont {Kutner}\ and\ \citenamefont
  {Masoliver}(2017)}]{Kutner2017}\BibitemOpen
  \bibfield  {author} {\bibinfo {author} {\bibfnamefont {R.}~\bibnamefont
  {Kutner}}\ and\ \bibinfo {author} {\bibfnamefont {J.}~\bibnamefont
  {Masoliver}},\ }\bibfield  {title} {\bibinfo {title} {The continuous time
  random walk, still trendy: fifty-year history, state of art and outlook},\
  }\bibfield  {journal} {\bibinfo  {journal} {The European Physical Journal B}\
  }\textbf {\bibinfo {volume} {90}},\ \href
  {https://doi.org/10.1140/epjb/e2016-70578-3} {10.1140/epjb/e2016-70578-3}
  (\bibinfo {year} {2017})\BibitemShut {NoStop}\bibitem [{\citenamefont {Tailleur}\ and\ \citenamefont
  {Cates}(2008)}]{Tailleur2008}\BibitemOpen
  \bibfield  {author} {\bibinfo {author} {\bibfnamefont {J.}~\bibnamefont
  {Tailleur}}\ and\ \bibinfo {author} {\bibfnamefont {M.~E.}\ \bibnamefont
  {Cates}},\ }\bibfield  {title} {\bibinfo {title} {Statistical mechanics of
  interacting run-and-tumble bacteria},\ }\href
  {https://doi.org/10.1103/physrevlett.100.218103} {\bibfield  {journal}
  {\bibinfo  {journal} {Physical Review Letters}\ }\textbf {\bibinfo {volume}
  {100}},\ \bibinfo {pages} {218103} (\bibinfo {year} {2008})}\BibitemShut
  {NoStop}\bibitem [{\citenamefont {Touzo}\ \emph {et~al.}(2023)\citenamefont {Touzo},
  \citenamefont {Doussal},\ and\ \citenamefont {Schehr}}]{Touzo2023}\BibitemOpen
  \bibfield  {author} {\bibinfo {author} {\bibfnamefont {L.}~\bibnamefont
  {Touzo}}, \bibinfo {author} {\bibfnamefont {P.~L.}\ \bibnamefont {Doussal}},\
  and\ \bibinfo {author} {\bibfnamefont {G.}~\bibnamefont {Schehr}},\
  }\href@noop {} {\bibinfo {title} {Interacting, running and tumbling: the
  active dyson brownian motion}} (\bibinfo {year} {2023}),\ \Eprint
  {https://arxiv.org/abs/2302.02937} {arXiv:2302.02937} \BibitemShut {NoStop}\bibitem [{\citenamefont {Luitz}\ and\ \citenamefont
  {Piazza}(2019)}]{Luitz2019}\BibitemOpen
  \bibfield  {author} {\bibinfo {author} {\bibfnamefont {D.~J.}\ \bibnamefont
  {Luitz}}\ and\ \bibinfo {author} {\bibfnamefont {F.}~\bibnamefont {Piazza}},\
  }\bibfield  {title} {\bibinfo {title} {Exceptional points and the topology of
  quantum many-body spectra},\ }\href
  {https://doi.org/10.1103/physrevresearch.1.033051} {\bibfield  {journal}
  {\bibinfo  {journal} {Physical Review Research}\ }\textbf {\bibinfo {volume}
  {1}},\ \bibinfo {pages} {033051} (\bibinfo {year} {2019})}\BibitemShut
  {NoStop}\bibitem [{\citenamefont {Miri}\ and\ \citenamefont {Alù}(2019)}]{Miri2019}\BibitemOpen
  \bibfield  {author} {\bibinfo {author} {\bibfnamefont {M.-A.}\ \bibnamefont
  {Miri}}\ and\ \bibinfo {author} {\bibfnamefont {A.}~\bibnamefont {Alù}},\
  }\bibfield  {title} {\bibinfo {title} {Exceptional points in optics and
  photonics},\ }\bibfield  {journal} {\bibinfo  {journal} {Science}\ }\textbf
  {\bibinfo {volume} {363}},\ \href {https://doi.org/10.1126/science.aar7709}
  {10.1126/science.aar7709} (\bibinfo {year} {2019})\BibitemShut {NoStop}\bibitem [{\citenamefont {Ghatak}\ \emph {et~al.}(2020)\citenamefont {Ghatak},
  \citenamefont {Brandenbourger}, \citenamefont {van Wezel},\ and\
  \citenamefont {Coulais}}]{Ghatak2020}\BibitemOpen
  \bibfield  {author} {\bibinfo {author} {\bibfnamefont {A.}~\bibnamefont
  {Ghatak}}, \bibinfo {author} {\bibfnamefont {M.}~\bibnamefont
  {Brandenbourger}}, \bibinfo {author} {\bibfnamefont {J.}~\bibnamefont {van
  Wezel}},\ and\ \bibinfo {author} {\bibfnamefont {C.}~\bibnamefont
  {Coulais}},\ }\bibfield  {title} {\bibinfo {title} {Observation of
  non-{H}ermitian topology and its bulk–edge correspondence in an active
  mechanical metamaterial},\ }\href {https://doi.org/10.1073/pnas.2010580117}
  {\bibfield  {journal} {\bibinfo  {journal} {Proceedings of the National
  Academy of Sciences}\ }\textbf {\bibinfo {volume} {117}},\ \bibinfo {pages}
  {29561–29568} (\bibinfo {year} {2020})}\BibitemShut {NoStop}\bibitem [{\citenamefont {Baconnier}\ \emph {et~al.}(2022)\citenamefont
  {Baconnier}, \citenamefont {Shohat}, \citenamefont {López}, \citenamefont
  {Coulais}, \citenamefont {Démery}, \citenamefont {Düring},\ and\
  \citenamefont {Dauchot}}]{Baconnier2022}\BibitemOpen
  \bibfield  {author} {\bibinfo {author} {\bibfnamefont {P.}~\bibnamefont
  {Baconnier}}, \bibinfo {author} {\bibfnamefont {D.}~\bibnamefont {Shohat}},
  \bibinfo {author} {\bibfnamefont {C.~H.}\ \bibnamefont {López}}, \bibinfo
  {author} {\bibfnamefont {C.}~\bibnamefont {Coulais}}, \bibinfo {author}
  {\bibfnamefont {V.}~\bibnamefont {Démery}}, \bibinfo {author} {\bibfnamefont
  {G.}~\bibnamefont {Düring}},\ and\ \bibinfo {author} {\bibfnamefont
  {O.}~\bibnamefont {Dauchot}},\ }\bibfield  {title} {\bibinfo {title}
  {Selective and collective actuation in active solids},\ }\href
  {https://doi.org/10.1038/s41567-022-01704-x} {\bibfield  {journal} {\bibinfo
  {journal} {Nature Physics}\ }\textbf {\bibinfo {volume} {18}},\ \bibinfo
  {pages} {1234–1239} (\bibinfo {year} {2022})}\BibitemShut {NoStop}\bibitem [{\citenamefont {Tlusty}(2021)}]{Tlusty2021}\BibitemOpen
  \bibfield  {author} {\bibinfo {author} {\bibfnamefont {T.}~\bibnamefont
  {Tlusty}},\ }\bibfield  {title} {\bibinfo {title} {Exceptional topology in
  ordinary soft matter},\ }\href {https://doi.org/10.1103/physreve.104.025002}
  {\bibfield  {journal} {\bibinfo  {journal} {Physical Review E}\ }\textbf
  {\bibinfo {volume} {104}},\ \bibinfo {pages} {025002} (\bibinfo {year}
  {2021})}\BibitemShut {NoStop}\bibitem [{\citenamefont {Wiersig}(2020)}]{Wiersig2020}\BibitemOpen
  \bibfield  {author} {\bibinfo {author} {\bibfnamefont {J.}~\bibnamefont
  {Wiersig}},\ }\bibfield  {title} {\bibinfo {title} {Review of exceptional
  point-based sensors},\ }\href {https://doi.org/10.1364/prj.396115} {\bibfield
   {journal} {\bibinfo  {journal} {Photonics Research}\ }\textbf {\bibinfo
  {volume} {8}},\ \bibinfo {pages} {1457} (\bibinfo {year} {2020})}\BibitemShut
  {NoStop}\bibitem [{\citenamefont {Doppler}\ \emph {et~al.}(2016)\citenamefont
  {Doppler}, \citenamefont {Mailybaev}, \citenamefont {Böhm}, \citenamefont
  {Kuhl}, \citenamefont {Girschik}, \citenamefont {Libisch}, \citenamefont
  {Milburn}, \citenamefont {Rabl}, \citenamefont {Moiseyev},\ and\
  \citenamefont {Rotter}}]{Doppler2016}\BibitemOpen
  \bibfield  {author} {\bibinfo {author} {\bibfnamefont {J.}~\bibnamefont
  {Doppler}}, \bibinfo {author} {\bibfnamefont {A.~A.}\ \bibnamefont
  {Mailybaev}}, \bibinfo {author} {\bibfnamefont {J.}~\bibnamefont {Böhm}},
  \bibinfo {author} {\bibfnamefont {U.}~\bibnamefont {Kuhl}}, \bibinfo {author}
  {\bibfnamefont {A.}~\bibnamefont {Girschik}}, \bibinfo {author}
  {\bibfnamefont {F.}~\bibnamefont {Libisch}}, \bibinfo {author} {\bibfnamefont
  {T.~J.}\ \bibnamefont {Milburn}}, \bibinfo {author} {\bibfnamefont
  {P.}~\bibnamefont {Rabl}}, \bibinfo {author} {\bibfnamefont {N.}~\bibnamefont
  {Moiseyev}},\ and\ \bibinfo {author} {\bibfnamefont {S.}~\bibnamefont
  {Rotter}},\ }\bibfield  {title} {\bibinfo {title} {Dynamically encircling an
  exceptional point for asymmetric mode switching},\ }\href
  {https://doi.org/10.1038/nature18605} {\bibfield  {journal} {\bibinfo
  {journal} {Nature}\ }\textbf {\bibinfo {volume} {537}},\ \bibinfo {pages}
  {76–79} (\bibinfo {year} {2016})}\BibitemShut {NoStop}\bibitem [{\citenamefont {Hassan}\ \emph {et~al.}(2017)\citenamefont {Hassan},
  \citenamefont {Zhen}, \citenamefont {Soljačić}, \citenamefont
  {Khajavikhan},\ and\ \citenamefont {Christodoulides}}]{Hassan2017}\BibitemOpen
  \bibfield  {author} {\bibinfo {author} {\bibfnamefont {A.~U.}\ \bibnamefont
  {Hassan}}, \bibinfo {author} {\bibfnamefont {B.}~\bibnamefont {Zhen}},
  \bibinfo {author} {\bibfnamefont {M.}~\bibnamefont {Soljačić}}, \bibinfo
  {author} {\bibfnamefont {M.}~\bibnamefont {Khajavikhan}},\ and\ \bibinfo
  {author} {\bibfnamefont {D.~N.}\ \bibnamefont {Christodoulides}},\ }\bibfield
   {title} {\bibinfo {title} {Dynamically encircling exceptional points: Exact
  evolution and polarization state conversion},\ }\href
  {https://doi.org/10.1103/physrevlett.118.093002} {\bibfield  {journal}
  {\bibinfo  {journal} {Physical Review Letters}\ }\textbf {\bibinfo {volume}
  {118}},\ \bibinfo {pages} {093002} (\bibinfo {year} {2017})}\BibitemShut
  {NoStop}\bibitem [{\citenamefont {Dembowski}\ \emph {et~al.}(2004)\citenamefont
  {Dembowski}, \citenamefont {Dietz}, \citenamefont {Gräf}, \citenamefont
  {Harney}, \citenamefont {Heine}, \citenamefont {Heiss},\ and\ \citenamefont
  {Richter}}]{Dembowski2004}\BibitemOpen
  \bibfield  {author} {\bibinfo {author} {\bibfnamefont {C.}~\bibnamefont
  {Dembowski}}, \bibinfo {author} {\bibfnamefont {B.}~\bibnamefont {Dietz}},
  \bibinfo {author} {\bibfnamefont {H.-D.}\ \bibnamefont {Gräf}}, \bibinfo
  {author} {\bibfnamefont {H.~L.}\ \bibnamefont {Harney}}, \bibinfo {author}
  {\bibfnamefont {A.}~\bibnamefont {Heine}}, \bibinfo {author} {\bibfnamefont
  {W.~D.}\ \bibnamefont {Heiss}},\ and\ \bibinfo {author} {\bibfnamefont
  {A.}~\bibnamefont {Richter}},\ }\bibfield  {title} {\bibinfo {title}
  {Encircling an exceptional point},\ }\href
  {https://doi.org/10.1103/physreve.69.056216} {\bibfield  {journal} {\bibinfo
  {journal} {Physical Review E}\ }\textbf {\bibinfo {volume} {69}},\ \bibinfo
  {pages} {056216} (\bibinfo {year} {2004})}\BibitemShut {NoStop}\bibitem [{\citenamefont {Floquet}(1883)}]{Floquet1883}\BibitemOpen
  \bibfield  {author} {\bibinfo {author} {\bibfnamefont {G.}~\bibnamefont
  {Floquet}},\ }\bibfield  {title} {\bibinfo {title} {Sur les équations
  différentielles linéaires à coefficients périodiques},\ }\href
  {https://doi.org/10.24033/asens.220} {\bibfield  {journal} {\bibinfo
  {journal} {Annales scientifiques de l'École normale supérieure}\ }\textbf
  {\bibinfo {volume} {12}},\ \bibinfo {pages} {47} (\bibinfo {year}
  {1883})}\BibitemShut {NoStop}\bibitem [{\citenamefont {Lyapunov}(1907)}]{Lyapunov1907}\BibitemOpen
  \bibfield  {author} {\bibinfo {author} {\bibfnamefont {A.~M.}\ \bibnamefont
  {Lyapunov}},\ }\bibfield  {title} {\bibinfo {title} {Problème général de
  la stabilité du mouvement},\ }\href {https://doi.org/10.5802/afst.246}
  {\bibfield  {journal} {\bibinfo  {journal} {Annales de la faculté des
  sciences de Toulouse Mathématiques}\ }\textbf {\bibinfo {volume} {9}},\
  \bibinfo {pages} {203} (\bibinfo {year} {1907})}\BibitemShut {NoStop}\bibitem [{\citenamefont {Teschl}(2012)}]{Teschl2012}\BibitemOpen
  \bibfield  {author} {\bibinfo {author} {\bibfnamefont {G.}~\bibnamefont
  {Teschl}},\ }\href {https://doi.org/10.1090/gsm/140} {\emph {\bibinfo {title}
  {Ordinary Differential Equations and Dynamical Systems}}}\ (\bibinfo
  {publisher} {American Mathematical Society},\ \bibinfo {year}
  {2012})\BibitemShut {NoStop}\bibitem [{Note17()}]{Note17}\BibitemOpen
  \bibinfo {note} {To be more precise, let $\pm 1$ represent the two elements
  of $\protect \mathbb {Z}_2$ (viewed multiplicatively), then we define $\rho
  (t, -1, z(t)) = \protect \text {e}^{\protect \text {i}\omega t} z^*(t)$ and
  $\rho (t, +1, z(t)) = z(t)$.}\BibitemShut {Stop}\bibitem [{Note18()}]{Note18}\BibitemOpen
  \bibinfo {note} {In linear algebra, an exceptional point of order $n$
  corresponds to a Jordan block of size $n$. For instance, the canonical Jordan
  block of size three associated with the eigenvalue $\lambda $ has the form
  \begin {equation*} \begin {pmatrix} \lambda & 1 & 0 \\ 0 & \lambda & 1 \\ 0 &
  0 & \lambda \end {pmatrix}. \end {equation*} We refer to Refs.~\cite
  {Arnold1999,GonzalezTokman2013} for the equivalent notion in the case of
  covariant Lyapunov vectors.}\BibitemShut {Stop}\bibitem [{\citenamefont {Keim}\ \emph {et~al.}(2019)\citenamefont {Keim},
  \citenamefont {Paulsen}, \citenamefont {Zeravcic}, \citenamefont {Sastry},\
  and\ \citenamefont {Nagel}}]{Keim2019}\BibitemOpen
  \bibfield  {author} {\bibinfo {author} {\bibfnamefont {N.~C.}\ \bibnamefont
  {Keim}}, \bibinfo {author} {\bibfnamefont {J.~D.}\ \bibnamefont {Paulsen}},
  \bibinfo {author} {\bibfnamefont {Z.}~\bibnamefont {Zeravcic}}, \bibinfo
  {author} {\bibfnamefont {S.}~\bibnamefont {Sastry}},\ and\ \bibinfo {author}
  {\bibfnamefont {S.~R.}\ \bibnamefont {Nagel}},\ }\bibfield  {title} {\bibinfo
  {title} {Memory formation in matter},\ }\href
  {https://doi.org/10.1103/revmodphys.91.035002} {\bibfield  {journal}
  {\bibinfo  {journal} {Reviews of Modern Physics}\ }\textbf {\bibinfo {volume}
  {91}},\ \bibinfo {pages} {035002} (\bibinfo {year} {2019})}\BibitemShut
  {NoStop}\bibitem [{\citenamefont {Coffey}\ \emph {et~al.}(1996)\citenamefont {Coffey},
  \citenamefont {Kalmykov},\ and\ \citenamefont {Waldron}}]{Coffey1996}\BibitemOpen
  \bibfield  {author} {\bibinfo {author} {\bibfnamefont {W.}~\bibnamefont
  {Coffey}}, \bibinfo {author} {\bibfnamefont {Y.~P.}\ \bibnamefont
  {Kalmykov}},\ and\ \bibinfo {author} {\bibfnamefont {J.~T.}\ \bibnamefont
  {Waldron}},\ }\href@noop {} {\emph {\bibinfo {title} {The Langevin Equation -
  With Applications In Physics, Chemistry And Electrical Engineering}}}\
  (\bibinfo  {publisher} {World Scientific},\ \bibinfo {year}
  {1996})\BibitemShut {NoStop}\bibitem [{{\relax DLMF}()}]{NIST_DLMF}\BibitemOpen
  {\relax DLMF},\ \href {https://dlmf.nist.gov/} {\bibinfo {title} {{\it NIST
  Digital Library of Mathematical Functions}}},\ \bibinfo {howpublished}
  {\url{https://dlmf.nist.gov/}, Release 1.1.10 of 2023-06-15} (\bibinfo {year}
  {2023}),\ \bibinfo {note} {f.~W.~J. Olver, A.~B. {Olde Daalhuis}, D.~W.
  Lozier, B.~I. Schneider, R.~F. Boisvert, C.~W. Clark, B.~R. Miller, B.~V.
  Saunders, H.~S. Cohl, and M.~A. McClain, eds.}\BibitemShut {Stop}\bibitem [{\citenamefont {Kuznetsov}(2005)}]{Kuznetsov2005}\BibitemOpen
  \bibfield  {author} {\bibinfo {author} {\bibfnamefont {Y.~A.}\ \bibnamefont
  {Kuznetsov}},\ }\bibfield  {title} {\bibinfo {title} {Practical computation
  of normal forms on center manifolds at degenerate {B}ogdanov–{T}akens
  bifurcations},\ }\href {https://doi.org/10.1142/s0218127405014209} {\bibfield
   {journal} {\bibinfo  {journal} {International Journal of Bifurcation and
  Chaos}\ }\textbf {\bibinfo {volume} {15}},\ \bibinfo {pages} {3535–3546}
  (\bibinfo {year} {2005})}\BibitemShut {NoStop}\bibitem [{\citenamefont {Boccaletti}\ \emph {et~al.}(2002)\citenamefont
  {Boccaletti}, \citenamefont {Kurths}, \citenamefont {Osipov}, \citenamefont
  {Valladares},\ and\ \citenamefont {Zhou}}]{Boccaletti2002}\BibitemOpen
  \bibfield  {author} {\bibinfo {author} {\bibfnamefont {S.}~\bibnamefont
  {Boccaletti}}, \bibinfo {author} {\bibfnamefont {J.}~\bibnamefont {Kurths}},
  \bibinfo {author} {\bibfnamefont {G.}~\bibnamefont {Osipov}}, \bibinfo
  {author} {\bibfnamefont {D.}~\bibnamefont {Valladares}},\ and\ \bibinfo
  {author} {\bibfnamefont {C.}~\bibnamefont {Zhou}},\ }\bibfield  {title}
  {\bibinfo {title} {The synchronization of chaotic systems},\ }\href
  {https://doi.org/10.1016/s0370-1573(02)00137-0} {\bibfield  {journal}
  {\bibinfo  {journal} {Physics Reports}\ }\textbf {\bibinfo {volume} {366}},\
  \bibinfo {pages} {1–101} (\bibinfo {year} {2002})}\BibitemShut {NoStop}\bibitem [{\citenamefont {Pecora}\ and\ \citenamefont
  {Carroll}(1990)}]{Pecora1990}\BibitemOpen
  \bibfield  {author} {\bibinfo {author} {\bibfnamefont {L.~M.}\ \bibnamefont
  {Pecora}}\ and\ \bibinfo {author} {\bibfnamefont {T.~L.}\ \bibnamefont
  {Carroll}},\ }\bibfield  {title} {\bibinfo {title} {Synchronization in
  chaotic systems},\ }\href {https://doi.org/10.1103/physrevlett.64.821}
  {\bibfield  {journal} {\bibinfo  {journal} {Physical Review Letters}\
  }\textbf {\bibinfo {volume} {64}},\ \bibinfo {pages} {821–824} (\bibinfo
  {year} {1990})}\BibitemShut {NoStop}\bibitem [{\citenamefont {Grebogi}\ \emph {et~al.}(1982)\citenamefont
  {Grebogi}, \citenamefont {Ott},\ and\ \citenamefont {Yorke}}]{Grebogi1982}\BibitemOpen
  \bibfield  {author} {\bibinfo {author} {\bibfnamefont {C.}~\bibnamefont
  {Grebogi}}, \bibinfo {author} {\bibfnamefont {E.}~\bibnamefont {Ott}},\ and\
  \bibinfo {author} {\bibfnamefont {J.~A.}\ \bibnamefont {Yorke}},\ }\bibfield
  {title} {\bibinfo {title} {Chaotic attractors in crisis},\ }\href
  {https://doi.org/10.1103/physrevlett.48.1507} {\bibfield  {journal} {\bibinfo
   {journal} {Physical Review Letters}\ }\textbf {\bibinfo {volume} {48}},\
  \bibinfo {pages} {1507} (\bibinfo {year} {1982})}\BibitemShut {NoStop}\bibitem [{\citenamefont {Grebogi}\ \emph {et~al.}(1983)\citenamefont
  {Grebogi}, \citenamefont {Ott},\ and\ \citenamefont {Yorke}}]{Grebogi1983}\BibitemOpen
  \bibfield  {author} {\bibinfo {author} {\bibfnamefont {C.}~\bibnamefont
  {Grebogi}}, \bibinfo {author} {\bibfnamefont {E.}~\bibnamefont {Ott}},\ and\
  \bibinfo {author} {\bibfnamefont {J.~A.}\ \bibnamefont {Yorke}},\ }\bibfield
  {title} {\bibinfo {title} {Crises, sudden changes in chaotic attractors, and
  transient chaos},\ }\href {https://doi.org/10.1016/0167-2789(83)90126-4}
  {\bibfield  {journal} {\bibinfo  {journal} {Physica D: Nonlinear Phenomena}\
  }\textbf {\bibinfo {volume} {7}},\ \bibinfo {pages} {181} (\bibinfo {year}
  {1983})}\BibitemShut {NoStop}\bibitem [{\citenamefont {Grebogi}\ \emph {et~al.}(1987)\citenamefont
  {Grebogi}, \citenamefont {Ott},\ and\ \citenamefont {Yorke}}]{Grebogi1987}\BibitemOpen
  \bibfield  {author} {\bibinfo {author} {\bibfnamefont {C.}~\bibnamefont
  {Grebogi}}, \bibinfo {author} {\bibfnamefont {E.}~\bibnamefont {Ott}},\ and\
  \bibinfo {author} {\bibfnamefont {J.~A.}\ \bibnamefont {Yorke}},\ }\bibfield
  {title} {\bibinfo {title} {Chaos, strange attractors, and fractal basin
  boundaries in nonlinear dynamics},\ }\href
  {https://doi.org/10.1126/science.238.4827.632} {\bibfield  {journal}
  {\bibinfo  {journal} {Science}\ }\textbf {\bibinfo {volume} {238}},\ \bibinfo
  {pages} {632} (\bibinfo {year} {1987})}\BibitemShut {NoStop}\bibitem [{\citenamefont {Beims}\ and\ \citenamefont
  {Gallas}(2016)}]{Beims2016}\BibitemOpen
  \bibfield  {author} {\bibinfo {author} {\bibfnamefont {M.~W.}\ \bibnamefont
  {Beims}}\ and\ \bibinfo {author} {\bibfnamefont {J.~A.~C.}\ \bibnamefont
  {Gallas}},\ }\bibfield  {title} {\bibinfo {title} {Alignment of {L}yapunov
  vectors: A quantitative criterion to predict catastrophes?},\ }\bibfield
  {journal} {\bibinfo  {journal} {Scientific Reports}\ }\textbf {\bibinfo
  {volume} {6}},\ \href {https://doi.org/10.1038/srep37102} {10.1038/srep37102}
  (\bibinfo {year} {2016})\BibitemShut {NoStop}\bibitem [{\citenamefont {Tantet}\ \emph {et~al.}(2017)\citenamefont {Tantet},
  \citenamefont {Lucarini},\ and\ \citenamefont {Dijkstra}}]{Tantet2017}\BibitemOpen
  \bibfield  {author} {\bibinfo {author} {\bibfnamefont {A.}~\bibnamefont
  {Tantet}}, \bibinfo {author} {\bibfnamefont {V.}~\bibnamefont {Lucarini}},\
  and\ \bibinfo {author} {\bibfnamefont {H.~A.}\ \bibnamefont {Dijkstra}},\
  }\bibfield  {title} {\bibinfo {title} {Resonances in a chaotic attractor
  crisis of the {L}orenz flow},\ }\href
  {https://doi.org/10.1007/s10955-017-1938-0} {\bibfield  {journal} {\bibinfo
  {journal} {Journal of Statistical Physics}\ }\textbf {\bibinfo {volume}
  {170}},\ \bibinfo {pages} {584–616} (\bibinfo {year} {2017})}\BibitemShut
  {NoStop}\bibitem [{\citenamefont {Castejón}\ \emph {et~al.}(2013)\citenamefont
  {Castejón}, \citenamefont {Guillamon},\ and\ \citenamefont
  {Huguet}}]{Castejon2013}\BibitemOpen
  \bibfield  {author} {\bibinfo {author} {\bibfnamefont {O.}~\bibnamefont
  {Castejón}}, \bibinfo {author} {\bibfnamefont {A.}~\bibnamefont
  {Guillamon}},\ and\ \bibinfo {author} {\bibfnamefont {G.}~\bibnamefont
  {Huguet}},\ }\bibfield  {title} {\bibinfo {title} {Phase-amplitude response
  functions for transient-state stimuli},\ }\href
  {https://doi.org/10.1186/2190-8567-3-13} {\bibfield  {journal} {\bibinfo
  {journal} {The Journal of Mathematical Neuroscience}\ }\textbf {\bibinfo
  {volume} {3}},\ \bibinfo {pages} {13} (\bibinfo {year} {2013})}\BibitemShut
  {NoStop}\bibitem [{\citenamefont {Freire}\ \emph {et~al.}(2007)\citenamefont {Freire},
  \citenamefont {Gasull},\ and\ \citenamefont {Guillamon}}]{Freire2007}\BibitemOpen
  \bibfield  {author} {\bibinfo {author} {\bibfnamefont {E.}~\bibnamefont
  {Freire}}, \bibinfo {author} {\bibfnamefont {A.}~\bibnamefont {Gasull}},\
  and\ \bibinfo {author} {\bibfnamefont {A.}~\bibnamefont {Guillamon}},\
  }\bibfield  {title} {\bibinfo {title} {Limit cycles and lie symmetries},\
  }\href {https://doi.org/10.1016/j.bulsci.2006.03.015} {\bibfield  {journal}
  {\bibinfo  {journal} {Bulletin des Sciences Mathématiques}\ }\textbf
  {\bibinfo {volume} {131}},\ \bibinfo {pages} {501–517} (\bibinfo {year}
  {2007})}\BibitemShut {NoStop}\bibitem [{\citenamefont {Tantet}\ \emph {et~al.}(2020)\citenamefont {Tantet},
  \citenamefont {Chekroun}, \citenamefont {Dijkstra},\ and\ \citenamefont
  {Neelin}}]{Tantet2020}\BibitemOpen
  \bibfield  {author} {\bibinfo {author} {\bibfnamefont {A.}~\bibnamefont
  {Tantet}}, \bibinfo {author} {\bibfnamefont {M.~D.}\ \bibnamefont
  {Chekroun}}, \bibinfo {author} {\bibfnamefont {H.~A.}\ \bibnamefont
  {Dijkstra}},\ and\ \bibinfo {author} {\bibfnamefont {J.~D.}\ \bibnamefont
  {Neelin}},\ }\bibfield  {title} {\bibinfo {title} {Ruelle-pollicott
  resonances of stochastic systems in reduced state space. part ii: Stochastic
  hopf bifurcation},\ }\href {https://doi.org/10.1007/s10955-020-02526-y}
  {\bibfield  {journal} {\bibinfo  {journal} {Journal of Statistical Physics}\
  }\textbf {\bibinfo {volume} {179}},\ \bibinfo {pages} {1403–1448} (\bibinfo
  {year} {2020})}\BibitemShut {NoStop}\bibitem [{\citenamefont {Tantet}\ \emph {et~al.}(2019)\citenamefont {Tantet},
  \citenamefont {Chekroun}, \citenamefont {Neelin},\ and\ \citenamefont
  {Dijkstra}}]{Tantet2019}\BibitemOpen
  \bibfield  {author} {\bibinfo {author} {\bibfnamefont {A.}~\bibnamefont
  {Tantet}}, \bibinfo {author} {\bibfnamefont {M.~D.}\ \bibnamefont
  {Chekroun}}, \bibinfo {author} {\bibfnamefont {J.~D.}\ \bibnamefont
  {Neelin}},\ and\ \bibinfo {author} {\bibfnamefont {H.~A.}\ \bibnamefont
  {Dijkstra}},\ }\bibfield  {title} {\bibinfo {title} {Ruelle–pollicott
  resonances of stochastic systems in reduced state space. part iii:
  Application to the cane–zebiak model of the el niño–southern
  oscillation},\ }\href {https://doi.org/10.1007/s10955-019-02444-8} {\bibfield
   {journal} {\bibinfo  {journal} {Journal of Statistical Physics}\ }\textbf
  {\bibinfo {volume} {179}},\ \bibinfo {pages} {1449–1474} (\bibinfo {year}
  {2019})}\BibitemShut {NoStop}\bibitem [{\citenamefont {Chekroun}\ \emph {et~al.}(2020)\citenamefont
  {Chekroun}, \citenamefont {Tantet}, \citenamefont {Dijkstra},\ and\
  \citenamefont {Neelin}}]{Chekroun2020}\BibitemOpen
  \bibfield  {author} {\bibinfo {author} {\bibfnamefont {M.~D.}\ \bibnamefont
  {Chekroun}}, \bibinfo {author} {\bibfnamefont {A.}~\bibnamefont {Tantet}},
  \bibinfo {author} {\bibfnamefont {H.~A.}\ \bibnamefont {Dijkstra}},\ and\
  \bibinfo {author} {\bibfnamefont {J.~D.}\ \bibnamefont {Neelin}},\ }\bibfield
   {title} {\bibinfo {title} {Ruelle–pollicott resonances of stochastic
  systems in reduced state space. part i: Theory},\ }\href
  {https://doi.org/10.1007/s10955-020-02535-x} {\bibfield  {journal} {\bibinfo
  {journal} {Journal of Statistical Physics}\ }\textbf {\bibinfo {volume}
  {179}},\ \bibinfo {pages} {1366–1402} (\bibinfo {year} {2020})}\BibitemShut
  {NoStop}\bibitem [{\citenamefont {Shmakov}\ and\ \citenamefont
  {Littlewood}(2023)}]{Shmakov2023}\BibitemOpen
  \bibfield  {author} {\bibinfo {author} {\bibfnamefont {S.}~\bibnamefont
  {Shmakov}}\ and\ \bibinfo {author} {\bibfnamefont {P.~B.}\ \bibnamefont
  {Littlewood}},\ }\href@noop {} {\bibinfo {title} {Coalescence of limit cycles
  in the presence of noise}} (\bibinfo {year} {2023}),\ \Eprint
  {https://arxiv.org/abs/2306.09524} {arXiv:2306.09524} \BibitemShut {NoStop}\bibitem [{\citenamefont {Weissman}(1988)}]{Weissman1988}\BibitemOpen
  \bibfield  {author} {\bibinfo {author} {\bibfnamefont {M.~B.}\ \bibnamefont
  {Weissman}},\ }\bibfield  {title} {\bibinfo {title} {1/f noise and other
  slow, nonexponential kinetics in condensed matter},\ }\href
  {https://doi.org/10.1103/revmodphys.60.537} {\bibfield  {journal} {\bibinfo
  {journal} {Reviews of Modern Physics}\ }\textbf {\bibinfo {volume} {60}},\
  \bibinfo {pages} {537–571} (\bibinfo {year} {1988})}\BibitemShut {NoStop}\bibitem [{\citenamefont {Kogan}(1996)}]{Kogan1996}\BibitemOpen
  \bibfield  {author} {\bibinfo {author} {\bibfnamefont {S.}~\bibnamefont
  {Kogan}},\ }\href {https://doi.org/10.1017/cbo9780511551666} {\emph {\bibinfo
  {title} {Electronic Noise and Fluctuations in Solids}}}\ (\bibinfo
  {publisher} {Cambridge University Press},\ \bibinfo {year}
  {1996})\BibitemShut {NoStop}\end{thebibliography}
\end{document}